\documentclass[a4paper,fleqn,usenatbib,useAMS]{mnras}

\usepackage{enumitem}
\usepackage[para]{threeparttable}
\usepackage{graphicx}	
\usepackage{amsmath}	
\usepackage{amssymb}	
\usepackage{multicol}   
\usepackage{bm}		    
\usepackage{pdflscape}	

\newcommand{\be}{\begin{equation}}
\newcommand{\ee}{\end{equation}}
\newcommand{\fracb}[2]{\left(\frac{#1}{#2}\right)}

\defcitealias{Barthelmy+08}{1}
\defcitealias{Gogus+10}{2}
\defcitealias{GC08}{3}
\defcitealias{LT07}{4}
\defcitealias{Mazets+79}{5}
\defcitealias{Cline+80}{6}
\defcitealias{Park+12}{7}
\defcitealias{Kouveliotou+98}{8}
\defcitealias{Watcher+04}{9}
\defcitealias{Corbel+99}{10}
\defcitealias{Sarma+97}{11}
\defcitealias{Israel+2016}{12}
\defcitealias{AMI09}{13}
\defcitealias{Camilo+07}{14}
\defcitealias{Deller+12}{15}
\defcitealias{VG97}{16}
\defcitealias{Tendulkar13}{17}
\defcitealias{TL08}{18}
\defcitealias{FG81}{19}
\defcitealias{Tendulkar+13}{20}
\defcitealias{KF12}{21}
\defcitealias{GV98}{22}
\defcitealias{Gaensler+99}{23}
\defcitealias{Vasisht+00}{24}
\defcitealias{Torii+98}{25}
\defcitealias{Aharonian+08}{26}
\defcitealias{HG10}{27}
\defcitealias{TL12}{28}
\defcitealias{Levin+10}{29}
\defcitealias{Anderson+12}{30}
\defcitealias{Gotthelf+00}{31}
\defcitealias{LT08b}{32}
\defcitealias{Gelfand+14}{33}

\usepackage[T1]{fontenc}
\usepackage{ae,aecompl}

\usepackage{txfonts}

\title[Magnetar Wind Nebulae]{Learning About the Magnetar Swift J1834.9-0846 \\ from its Wind Nebula}

\author[Granot et al. 2016]{Jonathan Granot,$^{1}$\thanks{Contact e-mail:\href{mailto:granot@openu.ac.il}{granot@openu.ac.il}}
Ramandeep Gill,$^{1}$ George Younes,$^{2}$ Josef Gelfand,$^{3}$ Alice Harding,$^{4}$
\newauthor{Chryssa Kouveliotou,$^{2}$ and Matthew G. Baring$^{5}$}
\\
$^{1}$Department of Natural Sciences, The Open University of Israel, 1 University Road, P.O. Box 808, Raanana 4353701, Israel \\
$^{2}$Department of Physics, The George Washington University, Washington, DC 20052, USA \\
$^{3}$NYU Abu Dhabi, P.O. Box 903, New York, NY, 10276, USA \\
$^{4}$Astrophysics Science Division, NASA Goddard Space Flight Center, Greenbelt, MD 20771 \\
$^{5}$Department of Physics and Astronomy, Rice University, MS-108, P.O. Box 1892, Houston, TX 77251, USA
}

\date{Last updated; in original form}

\pubyear{2016}

\begin{document}
\label{firstpage}
\pagerange{\pageref{firstpage}--\pageref{lastpage}}
\maketitle

\begin{abstract}
The first wind nebula around a magnetar was recently discovered in X-rays around Swift~J1834.9$-$0846.
We study this magnetar's global energetics and the properties of its particle wind or outflows. 
At a distance of $\sim4\;$kpc, Swift~J1834.9$-$0846 is located at the center of the supernova remnant 
(SNR) W41 whose radius is $\sim 19\;$pc, an order of magnitude larger than that of the X-ray nebula ($\sim2\;$pc). 
The association with SNR W41 suggests a common age of $\sim5-100\;$kyr, 
while its spin-down age is $4.9$~kyr. A small natal kick velocity 
may partly explain why a wind nebula was detected around this magnetar but not around other magnetars, most of which appear 
to have larger kick velocities and may have exited their birth SNR.
We find that the GeV and TeV source detected by Fermi/LAT and H.E.S.S., respectively, of radius $\sim11\;$pc  
is most likely of hadronic origin.
The dynamics and internal structure of the nebula are examined analytically
to explain the nebula's current properties. Its size
may naturally correspond to the diffusion-dominated cooling length of the X-ray emitting $e^+e^-$ pairs. 
This may also account for the spectral softening of the X-ray emission from the nebula's inner to outer parts. 
Analysis of the X-ray synchrotron nebula implies that 
(i) the nebular magnetic field is $\gtrsim 11\;\mu$G (and likely $\lesssim30\;\mu$G), and 
(ii) the nebula is not powered predominantly by the magnetar's quiescent spin-down-powered MHD wind, 
but by other outflows that contribute most of its energy. The latter are most likely associated with the 
magnetar's bursting activity, and possibly dominated by outflows associated with its past giant flares.
The energy source for the required outflows cannot be the decay of the magnetar's dipole field alone, and is most likely the decay of its much stronger internal magnetic field.
\end{abstract}




\section{Introduction}
Pulsar wind nebulae (PWNe) can act as excellent calorimeters as they radiate over a broad energy 
range, from radio to TeV gamma-rays, and directly reflect the power injected by their central pulsars 
in the form of a relativistic MHD wind \citep[see review by e.g.][]{GS06}. Observations of the Crab nebula, 
a prototype of this class of objects, has provided invaluable insight into the physics of highly magnetized 
relativistic outflows and its interaction with the surrounding supernova remnant (SNR; in the case of the 
Crab the SNR is fragmented due to the Rayleigh-Taylor instability). The defining characteristics of a PWN 
is a centrally filled nebula -- sometimes 
referred to as a plerion, within the much larger SNR, along with power-law X-ray emission and flat radio spectrum 
(polarized in some cases). Its interior is filled with relativistically hot particles that cool both adiabatically as the nebula
expands, and radiatively by emitting synchrotron radiation in the relatively 
weak (typically between a few $\mu$G to $\sim$ mG, determined using various techniques,  
\citealp[see for e.g.][and references therein]{RGB12}) nebular magnetic field, and by 
inverse Compton scattering on soft CMB and/or infrared starlight photons. Its bolometric 
power is supplied by the loss of rotational kinetic energy ($\dot E_{\rm rot} = -I\Omega\dot\Omega = L_{\rm sd}$) of the 
spinning down pulsar \citep{Gold69} with a broad range in X-ray efficiency 
$\eta_X = L_{X,\rm PWN}/L_{\rm sd} \sim 10^{-5} - 10^{-1}$ \citep[e.g. see][for a catalog of PWNe properties]{KP08}. 

PWNe are typically observed down to 
$L_{\rm sd} \sim 10^{33}~{\rm erg~s^{-1}}$. 
As pulsars spin down to longer periods, $P$, their rotational energy drops as $\dot E_{\rm rot}\propto \dot{P}/P^{3}$, until they eventually become
too dim to be detected in X-rays.
Moreover, the pulsar's aging and spin down is also accompanied by a drop in the open magnetic-field line voltage $V_0 \propto |E_{\rm rot}|^{1/2}$ over which particles can potentially be accelerated. 
High angular resolution observations with \textit{Chandra} 
have further revealed that PWNe that are still embedded in their host SNRs thus forming a 
composite morphology, are typically associated with younger pulsars 
with high spin-down power $L_{\rm sd} \gtrsim 10^{36.3}~{\rm erg~s}^{-1}$, whereas older pulsars typically power bow-shock PWNe. This is likely since for a given natal kick velocity the pulsar overtakes its SNR at a finite time (shortly after the SNR's velocity drops below the pulsar's velocity), so that younger pulsars are still inside their SNRs, which helps trap their wind and form a PWN, while older pulsars have already overtaken their SNRs, and thus form a bow-shock instead of a PWN. Majority of the pulsars powering observed PWNe are known to be run-of-the-mill rotation powered pulsars with equatorial surface 
dipole magnetic fields $B_s\sim10^{12}-10^{13}~{\rm G}$, which begs the question: Does the incidence of
wind-nebulae continue up to the strongly magnetized neutron stars (NSs)?

Magnetars are slowly rotating ($P \sim 2 - 12~{\rm s}$) NSs with super-QED (i.e. above $B_Q = m_e^2c^3/e\hbar 
= 4.414\times10^{13}~{\rm G}$) surface fields $B_s \sim 10^{14}-10^{15}~{\rm G}$ as inferred from the 
magnetic dipole radiation (MDR) spin-down, given in Equation (\ref{eq:Beq}) \citep[see review by e.g.][]{TZW15}.  
Compared to rotation-powered pulsars with X-ray PWNe, the magnetars have similarly large period-derivatives 
$\dot P \sim 10^{-13} - 10^{-11}~{\rm s~s}^{-1}$, but due to slower rotation, having spin periods in the 
range $\sim 2 - 12~{\rm s}$, their spin-down power is significantly 
less, $L_{\rm sd}\sim 10^{30} - 10^{34}~{\rm erg~s}^{-1}$. Therefore, for comparable X-ray efficiencies $\eta_X$ they would be dimmer and harder to detect, despite their small average characteristic ages 
($\tau_c\sim10^3 - 10^5~{\rm yr}$, e.g. \citet{OK14}), indicative of their youth, which might suggest otherwise. 
The empirical limit on $L_{\rm sd}$ mentioned above 
for Crab-like PWNe alone makes observations of X-ray wind-nebulae around magnetars (hereafter Magnetar Wind Nebulae -- MWNe) challenging. 

In contrast with rotation-powered pulsars, magnetars have a high quiescent X-ray luminosity (from the neutron star surface and/or magnetosphere) that is in excess of their spin down power, 
$L_X \sim 10^{33} - 10^{36}~{\rm erg~s}^{-1} > L_{\rm sd}$. Moreover, magnetars display 
a range of bursting activity, from the more common short bursts (lasting $\sim 0.1\;$s with observed luminosities 
$L\sim10^{39} - 10^{41}~{\rm erg~s}^{-1}$) to the rare and highly super-Eddington giant flares (with an initial spike of luminosity 
$L\sim10^{44} - 10^{47}~{\rm erg~s}^{-1}$ over a fraction of a second, followed by a less luminous pulsating tail lasting hundreds of seconds). 
If the power in particle outflows is at least comparable to that in radiation, it is not inconceivable that the accumulated 
effect of many outbursts over the lifetime of magnetars, that haven't yet escaped from their host SNRs, 
can give rise to a synchrotron bubble of relativistic particles. However, it has proven difficult to 
find wind-nebulae around magnetars, perhaps at least partly because most magnetars 
($\sim 75\%$) lack a clear SNR association (see Table \ref{tab:mag-snr}). 
The status quo changed with the 
discovery of diffuse X-ray emission around one of the recently discovered magnetars \citep{Y+12}.

Swift J1834.9-0846 (referred to as Swift J1834 hereafter) was discovered on 2011 August 7 by 
\textit{Swift} and \textit{Fermi} when it went into outburst
\citep{DElia+2011,Guiriec+2011}. Later observations by \textit{RXTE} and \textit{Chandra} established its magnetar 
nature by measuring a spin period $P\simeq2.48~\rm{s}$ \citep[e.g][]{Gogus_Kouveliotou_2011} and period-derivative 
$\dot P\simeq7.96\times10^{-12}~{\rm s~s}^{-1}$ \citep{Kuiper_Hermsen_2011,Kargaltsev_2012}. This implies an equatorial 
surface dipole magnetic field strength $B_s\simeq1.16\times10^{14}f^{-1/2}~{\rm G}$ ($1.4\times10^{14}~{\rm G}$ 
for $f = 2/3$, see Equation~(\ref{eq:Lsd}) for the definition of $f$), spin-down power 
$L_{\rm sd}\simeq2.05\times10^{34}~\rm{erg~s}^{-1}$, and characteristic age $\tau_c\simeq4.9~{\rm kyr}$. 

Swift J1834 is positioned very close to the geometrical center of its host SNR W41, 
which might suggest that the magnetar has a small space velocity. Out of the total 
29 magnetars \citep[][]{OK14}, which includes 15 soft-gamma repeaters 
(SGR, including 4 candidate sources) and 14 anomalous X-ray pulsars 
(AXP, including 2 candidate sources), only 7, other than Swift J1834, have secure SNR 
associations. In all of these cases, the magnetar (or candidate source) 
is positioned very close to the center of the SNR, and has a low measured 
transverse velocity $<400~{\rm km~s}^{-1}$ \citep[e.g.][]{Gaensler+01}. We 
list the magnetar-SNR associations, along with the distance to the SNR, its 
angular radius ($\theta_{1/2}$), the location of the magnetar with respect to the SNR center and 
its off-center angle, the measured or inferred transverse velocities of magnetars, and whether they power 
an MWN in Table \ref{tab:mag-snr}. 

\begin{table*}
    \caption{Magnetar-SNR Associations}
    \centering
    \begin{threeparttable}
    \begin{tabular}{lccccccccc}
    \hline\hline
        Source & SNR & Distance & $\theta_{1/2}$ & Location & Off-Center & Association & $V_T$ & MWN & Refs.\\
         & & (kpc) & SNR & & Angle & Secure? & ($\rm km~s^{-1}$) & & \\
        \hline
        SGR 0501+4516 & HB9 & $0.8\pm0.4$ & $62.5'$ & E & $\sim80'$ & No & $(1.7 - 4.3)\times10^3$ & No & 
        \citetalias{Barthelmy+08}, \citetalias{Gogus+10}, \citetalias{GC08}, \citetalias{LT07}\\
        SGR 0526-66 & N49 & $\sim50$ & $\sim 35''$ & E & $22''$ & No & $\sim1100$ & No & 
        \citetalias{Mazets+79}, \citetalias{Cline+80}, \citetalias{Park+12}\\
        SGR 1627-41 & G337.0+0.1 & $11.0\pm0.3$ & $45''$ & E & $105''\pm26''$ & No & $\sim10^3$ & No & 
        \citetalias{Kouveliotou+98}, \citetalias{Watcher+04}, \citetalias{Corbel+99}, \citetalias{Sarma+97}\\
        SGR 1935+2154 & G57.2+0.8 & - & $6'$ & C & - & Yes & - & Maybe & 
        \citetalias{Israel+2016}, \citetalias{AMI09} \\
        1E 1547.0-5408 & G327.24-0.13 & $6\pm2$ & $2'$ & C & $\lesssim 13''$ & Yes & $280^{+130}_{-120}$ & No & 
        \citetalias{Camilo+07}, \citetalias{Deller+12}\\
        1E 1841-045 & Kes 73 & $\sim8.5$ & $2.5'$ & C & - & Yes & $\lesssim160$ & No & 
        \citetalias{VG97}, \citetalias{Tendulkar13}, \citetalias{TL08}\\
        1E 2259+586 & CTB 109 & $3.2\pm0.2$ & $18'$ & OC & $\sim4'$ & Yes & $\sim157$ & No & 
        \citetalias{FG81}, \citetalias{Tendulkar+13}, \citetalias{KF12}\\
        AX J1845.0-0258 & G29.6+0.1 & $\sim8.5$ & $2.5'$ & OC & $\lesssim 40''$ & Yes & - & No & 
        \citetalias{GV98}, \citetalias{Gaensler+99}, \citetalias{Vasisht+00}, \citetalias{Torii+98}\\
        CXOU J171405.7${}^{**}$ & CTB 37B & $\sim13.2$ & $2.5'$ & OC & $2.17'$ & Yes & $\sim10^3$ & No & 
         \citetalias{Aharonian+08}, \citetalias{HG10}, \citetalias{TL12}\\
        PSR J1622-4950 & G333.9+0.0 & $\sim 9$ & - & - & - & No & - & No & 
        \citetalias{Levin+10}, \citetalias{Anderson+12}\\
        PSR J1846-0258 & Kes 75 & $6.0^{+1.5}_{-0.9}$ & $1.75'$ & C & $< 1'$ & Yes & - & Yes${}^*$ & 
        \citetalias{Gotthelf+00}, \citetalias{LT08b}, \citetalias{Gelfand+14}\\
        \hline
    \end{tabular}
    \begin{tablenotes}
    $\theta_{1/2}$ is the angular radius of the SNR. E - source near the edge of the SNR, C - 
    source near the center of the SNR, OC - source slightly off-centered. ${}^*$PSR J1846-0258 
    is a 0.326 s rotation-powered pulsar, but with a high surface dipole field 
    $B_s = 4.9\times10^{13}$ G. The classification of this source as a magnetar, due to an 
    episode of bursting activity, is not yet clear. ${}^{**}$ Source name: CXOU J171405.7-381031. 
    Refs. - \citepalias{Barthelmy+08} \citet{Barthelmy+08};
    \citepalias{Gogus+10} \citet{Gogus+10}; \citepalias{GC08} \citet{GC08}; 
    \citepalias{LT07} \citet{LT07}; \citepalias{Mazets+79} \citet{Mazets+79}; 
    \citepalias{Cline+80} \citet{Cline+80}; \citepalias{Park+12} \citet{Park+12};
    \citepalias{Kouveliotou+98} \citet{Kouveliotou+98}; \citepalias{Watcher+04} \citet{Watcher+04}; 
    \citepalias{Corbel+99} \citet{Corbel+99}; \citepalias{Sarma+97} \citet{Sarma+97};
    \citepalias{Israel+2016} \citet{Israel+2016}; \citepalias{AMI09} \citet{AMI09}; \citepalias{VG97} \citet{VG97}; 
    \citepalias{Tendulkar13} \citet{Tendulkar13}; \citepalias{TL08} \citet{TL08}; 
    \citepalias{FG81} \citet{FG81}; \citepalias{Tendulkar+13} \citet{Tendulkar+13}; 
    \citepalias{KF12} \citet{KF12}; \citepalias{GV98} \citet{GV98};
    \citepalias{Gaensler+99} \citet{Gaensler+99}; \citepalias{Vasisht+00} \citet{Vasisht+00}; 
    \citetalias{Torii+98} \citet{Torii+98}; \citepalias{HG10} \citet{HG10}; 
    \citepalias{TL12} \citet{TL12}; \citepalias{Aharonian+08} \citet{Aharonian+08}; 
    \citepalias{Levin+10} \citet{Levin+10}; \citepalias{Anderson+12} \citet{Anderson+12};
    \citepalias{Gotthelf+00} \citet{Gotthelf+00}; \citepalias{LT08b} \citet{LT08b}; 
    \citepalias{Gelfand+14} \citet{Gelfand+14}; For further details, see the magnetar catalog: \url{http://www.physics.mcgill.ca/~pulsar/magnetar/main.html}
    \end{tablenotes}
    \end{threeparttable}
    \label{tab:mag-snr}
\end{table*}

\begin{figure}
    \centering
    \includegraphics[width=0.46\textwidth]{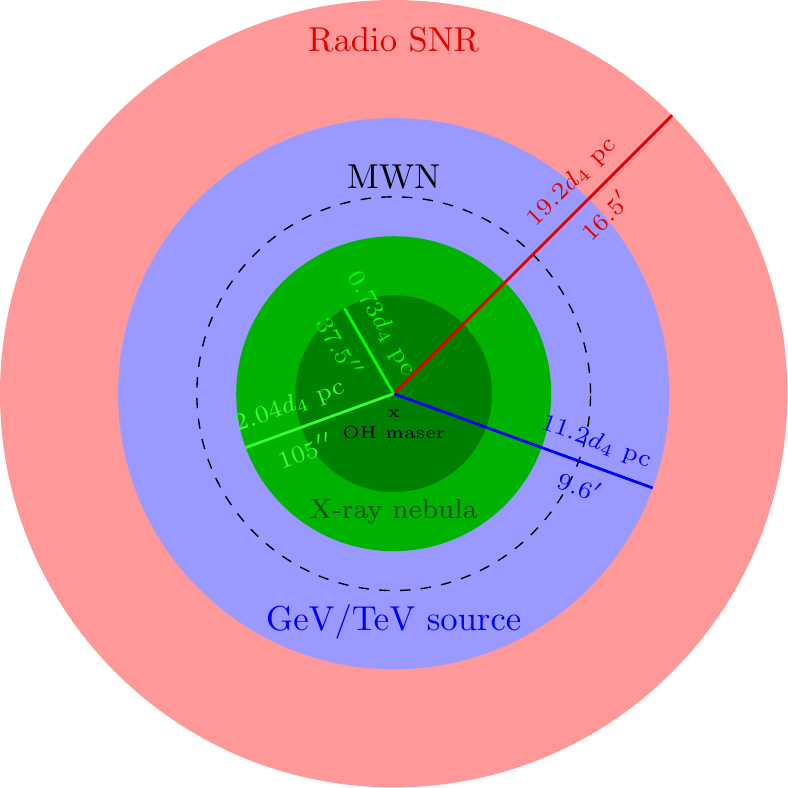}
    \caption{Physical setup of the system showing the radio 
    SNR W41, GeV/TeV source detected by H.E.S.S and 
    \textit{Fermi}, OH maser emission near the center of the 
    GeV/TeV region, and (inner and outer) X-ray bright region 
    of the magnetar wind nebula powered by Swift J1834. This 
    figure is not to scale, and for simplicity, the different 
    emission regions have been approximated here by circular 
    regions with approximate radial extents shown in the 
    figure; the true geometry is irregular.}
    \label{fig:setup}
\end{figure}

By analysing two deep \textit{XMM}-Newton observations post-outburst of Swift J1834 
and its extended X-ray emission, the existence of a first ever wind-nebula around a magnetar was 
confirmed by \citet{Y+16}. They have dismissed the dust-scattering halo interpretation \citep[e.g.][]{Esposito+13}, 
based on the fact that the X-ray flux of 
the diffuse emission has remained almost unchanged since the 2011 outburst and its spectrum is inconsistent with 
the much softer spectrum expected from scattering by dust. Figure~\ref{fig:setup} summarizes the measured sizes 
of the entire system. The spectral appearance of the X-ray nebula changes with distance from the central magnetar \citep{Y+16}, 
where the inner X-ray nebula, with angular size $37.5''$ and radial extent $R_{X,\rm in} = 0.73d_4\;$pc (for a distance of 
$d=4d_4\;$kpc to the source), is 
spectrally harder than the outer X-ray nebula ($105''$ and $R_X = 2.04d_4\;$pc). The size of the overlapping GeV/TeV source observed by Fermi/LAT and HESS is 
$9.6'$ or $R_{\rm GeV/TeV}=11.2d_4\;$pc, and the radio SNR extends to an angular size $16.5'$ with 
$R_{\rm SNR} = 19.2d_4\;$pc.

\begin{table*}
    \caption{Various symbols with their meanings that appear in this work.}
    \begin{minipage}[b]{0.45\linewidth}\centering
    \begin{tabular}{l|l}
        \hline
        Symbol & Meaning \\
        \hline
        $a$ & MWN radial expansion temporal power-law index \\
        $\alpha$ & Inverse of magnetic field decay power-law index \\
        $B$ & Nebular magnetic field \\
        $B_0$ & Initial surface dipole magnetic field \\
        $B_{\rm s}$ & Equatorial surface dipole magnetic field \\
        $B_{\rm int}$ & Internal magnetic field inside the magnetar \\
        $\beta_{\rm TS}$ & The flow's $v/c$ behind the wind termination shock\\
        $\chi$ & MWN compression ratio at reverse shock passage \\
        $d$ & Distance to the source from observer \\
        $E$ & Total energy in the MWN \\
        $E_0$ & Initial rotational kinetic energy injected by proto-NS \\
        $E_B$ & Magnetic energy of the X-ray nebula \\
        $E_{\rm B,dip}$ & Energy in magnetar's dipolar magnetic field\\
        $E_{\rm B,int}$ & Energy in magnetar's internal magnetic field \\
        $E_f$ & Total energy in the nebula after crushing \\
        $E_\gamma$ & Photon energy \\
        $E_i$ & Total energy in the nebula before crushing \\
        $E_{\rm inj}$ & Spin-down energy injected into MWN \\
        $E_p$ & Proton energy \\
        $E_{\rm rot}$ & Rotational kinetic energy \\
        $E_{\rm SN}$ & Supernova energy \\
        $\epsilon_e$ & Ratio of total energy in power-law electrons \\
         & to that in the matter component of MHD wind \\
        $\epsilon_X$ & Ratio of energy in X-ray radiating electrons \\
         & to total energy in power-law electrons \\
        $\eta_X$ & X-ray efficiency of the nebula \\
        $f$ & Parameter to distinguish between vacuum and \\
         & force-free spin-down law \\
        $g$ & Ratio of long-term mean energy injection rate \\
         & into the nebula by the magnetar to $L_{\rm sd}$ \\
        $\gamma_c$ & Lorentz factor of electrons cooling \\
         & at the dynamical time \\
        $\gamma_e$ & Lorentz factor of electrons \\
        $\gamma_{\rm max}$ & Maximal electron Lorentz factor, $eV_0/m_ec^2$ \\
        $\Gamma$ & Photon spectral index (whole nebula)\\
        $\Gamma_{\rm in}$ & Photon spectral index (inner nebula)\\
        $k$ & ISM density radial power-law index \\
        $L_{\rm sd}$ & Spin-down power \\
        $L_0$ & Initial spin-down power \\
        $l_{\rm adv}$ & Characteristic advection length scale \\
        $l_{\rm diff}$ & Characteristic diffusion length scale \\
        $L_{\rm GeV}$ & GeV luminosity \\
        $L_{\rm TeV}$ & TeV luminosity \\
        $L_X$ & (0.5-10 keV, XMM) X-ray luminosity of \\
         & the entire nebula\\
        $L_{X,\rm in}$ & (0.5-30 keV, XMM+NuSTAR) X-ray luminosity of \\
         & the inner nebula \\
        $L_{X,\rm tot}$ & (0.5-30 keV, XMM+NuSTAR) X-ray luminosity of \\
         & the entire nebula \\
        $M_{\rm ej}$ & SN ejecta mass \\ 
        $n$ & Braking index (assuming constant $B_s$) \\
        $n'$ & Measured braking index\\
        $n_{\rm ext}$ & Number density of ISM gas
        \end{tabular}
        \end{minipage}
        \hspace{0.5cm}
        \begin{minipage}[b]{0.45\linewidth}\centering
        \begin{tabular}{l|l}
        \hline
        Symbol & Meaning \\
        \hline
        $N_{HI}$ & Neutral hydrogen column density \\
        $n_p$ & Proton power law distribution \\
        $\nu_i$ & Characteristic synchrotron frequency \\
         & corresponding to electron lorentz factor $\gamma_i$ \\
        $\Omega$ & Spin angular frequency \\
        $\Omega_0$ & Initial spin angular frequency \\
        $s$ & Electron distribution power law index \\
        $P$ & Spin period \\
        $P_0$ & Initial spin period \\
        $p_i$ & Total pressure inside the nebula before crushing \\
        $p_f$ & Total pressure inside the nebula after crushing \\
        $p_S$ & Central pressure interior to a Sedov blast wave \\
        $r$ & Radial distance from the center of the nebula \\
        $R$ & Radius of MWN \\
        $R_0$ & Radius of MWN at time $t_0$ \\
        $r_{\rm adv}$ & Advection distance of nebular flow \\
        $R_b$ & Radius below which the nebular flow is in steady-state \\
        $R_c$ & Density core crossing radius of MWN outer shell \\
        $r_{\rm c,adv}$ & Max. advection distance of particles \\
         & over their synchrotron cooling times \\
        $r_{\rm c,diff}$ & Max. diffusion distance of particles \\
         & over their synchrotron cooling times \\
        $R_f$ & Radius of MWN after crushing \\
        $R_{\rm GeV/TeV}$ & Radius of the GeV/TeV energy region \\
        $R_i$ & Radius of MWN before crushing \\
        $R_{\rm NS}$ & Neutron star radius\\
        $R_{\rm SNR}$ & Radius of the SNR W41 \\
        $R_{\rm ST}$ & Radius of SNR at Sedov-Taylor onset \\
        $r_*$ & Radius beyond which diffusion dominates \\
         & over advection \\
        $R_{\rm TS,p}$ & Magnetars wind termination shock radius\\
        $R_X$ & Radius of the entire X-ray nebula \\
        $R_{X,\rm in}$ & Radius of the inner X-ray nebula \\
        $\rho_{\rm ext}$ & Mass density of the surrounding ISM \\
        $\sigma$ & Magnetization of nebular plasma \\
        $\sigma_e$ & Magnetization of power-law electrons \\
        $\tilde{t}$ & System's age $t$ normalized by $t_0$\\
        $\hat{t}$ & System's age $t$ normalized by $t_{\rm ST}$\\
        $t_0$ & Initial spin-down time \\
        $t_B$ & Characteristic initial magnetic field decay time \\
        $t_c$ & Density core crossing time of MWN outer shell \\
        $\tau_c$ & Characteristic spin-down age \\
        $t_{\rm SNR}$ & Age of the SNR \\
        $t_{\rm ST}$ & Sedov-Taylor onset time \\
        $t_{\rm syn}$ & Synchrotron cooling time of electrons \\
        $\theta_X$ & Mean angular size of X-ray nebula \\
        $V_0$ & Polar cap voltage difference \\
        $\xi$ & Ratio of the total energy in electrons to\\
         & the energy in the X-ray emitting electrons \\
        $\xi_{\rm in}$ & $\xi$ derived for the inner nebula \\
        $\zeta$ & Ratio of particle deflection length \\
         & to its Larmor radius
         \\ \vspace{-0.03cm}
    \end{tabular}
    \end{minipage}
    \label{tab:symbols}
\end{table*}

\subsubsection*{\rm \textbf{Plan of the Paper}}
The observation of a wind-nebula around Swift J1834 presents a rare opportunity to study the global 
energetics of magnetars and properties of their particle wind or outflows, both in quiescence and during outbursts.
In this work we consider in detail the implications of the discovery of this first-confirmed magnetar wind nebula (MWN). Here we briefly outline the structure of the paper along with our main line of reasoning for each part.

In \S~\ref{sec:SNR} we consider the implications of the association of Swift J1834 with the SNR. In particular,
the magnetar's systemic velocity on the plane of the sky is constrained to be at most tens of km/s. Its age is more uncertain
(\S~\ref{sec:age}) and depends mainly on the uncertain external density (and to a lesser extent on the somewhat uncertain SNR energy), 
but for reasonable parameter values it is in the range $5\;{\rm kyr}\lesssim t_{\rm SNR}\lesssim100\;$kyr. 
Reconciling the upper age estimate ($\sim100\;$kyr) of the radio SNR with the SGR's spin-down age of 
$\tau_c\simeq4.9~{\rm kyr}$ (\S~\ref{sec:n-tau_sd}) suggests either a braking index of $n\simeq1.1$
(corresponding to an initial surface dipole field well below its current value) or a braking index of
$n\approx 3$ but with a current value of $\dot{P}$ that is anomalously high, well above its long-term 
mean value (by a factor of $\sim20$). The evolution of the spin period and braking index for an 
evolving surface dipole field strength is considered in detail in Appendix~\ref{sec:appA}.

In \S~\ref{sec:EX} we consider the energetics of the X-ray nebula. First, we use its observed size ($R_X$) and luminosity ($L_X$)
to constrain the magnetic field strength $B$ within the X-ray nebula, assuming that the observed X-rays are synchrotron emission. 
From regular equipartition arguments we find $B\sim(4-30)\sigma_e^{2/7}\;\mu$G, where $\sigma_e = \frac{2}{3}\sigma/\epsilon_e$ is the ration of energy in the magnetic field and in the X-ray emitting power-law electron energy distribution that holds a fraction $\epsilon_e$ of the total energy in particles, and $\sigma$ is the magnetic to particle enthalpy density ratio in the MWN. For PWNe traditional 1D modeling typically implies $10^{-3}\lesssim\sigma\lesssim10^{-2}$ \citep[e.g.,][]{Kennel_Coroniti_1984,DeJager_Harding_1992,Hillas_1998,Meyer_2010}, while more recent 2D or 3D could allow for 
\textcolor{black}{slightly higher mean values of $10^{-2}\lesssim\sigma\lesssim10^{-1}$} in the nebula \citep[e.g.,][]{Mizuno_2011,Porth+2013,Porth+2014}. Second, by assuming that the maximal electron energy 
cannot exceed that corresponding to the maximal voltage across the magnetar's open magnetic field lines 
\citep[e.g.][]{DeJager_Harding_1992}, we derive a more robust limit (independent of $\sigma$ or the global 
electron energy distribution) of $B\gtrsim 11\;\mu$G, which corresponds to a lower limit on the MWN 
magnetic energy that scales with its volume (i.e. as the cube of its radius). Moreover, our results imply that 
for any value of $\sigma$, the long-term time averaged energy output from the SGR into the MWN significantly 
exceeds (by a factor of $g>3.1$) its current spin-down power $L_{\rm sd}$. The most likely candidate for such 
energy injection from the SGR into the MWN are the outflows associated with its bursting activity.

The MWN's dynamical evolution is outlined in \S~\ref{sec:Ead} and related to the energy injection by the magnetar's quiescent rotation powered MHD wind,
where some of the details are expanded upon in Appendix~\ref{sec:appB}. The MWN's dynamics and energy injection by the wind together determine the MWN's
current energy, considering adiabatic evolution of the electrons. The additional effects of radiative energy losses together with adiabatic cooling or 
heating are briefly considered in Appendix~\ref{sec:appC}, and neglecting them results in a fairly robust upper limit on the MWN's current energy. 

The internal structure within the MWN is considered in \S~\ref{sec:Lcool}, in order 
to calculate the resulting synchrotron cooling length. We find that there is an inner quasi-steady-state 
region (discussed in \S~\ref{sec:qss-nebula}) and an outer uniformly expanding region 
(\S~\ref{sec:non-steady-nebula}). This structure is discussed in the context of the observed spectral softening within the X-ray nebula with distance from the magnetar (\S~\ref{sec:softening}). Advection alone generally leads to a sharper spectral softening with radius than observed. Therefore, we also consider the effects of diffusion (\S~\ref{sec:diff}) and find that diffusion dominates over advection throughout most of the nebula. Moreover,  
the effects of diffusion can
likely account for the more gradual observed spectral softening and naturally explain the nebula's observed size.

In \S~\ref{sec:g-sigma} we show that the X-ray emitting $e^+e^-$ pairs are fast cooling, which allows us to write their detailed energy balance. This results in a lower limit  on $g = \langle\dot{E}\rangle/L_{\rm sd} \gtrsim 3.1(1+\sigma)/\sigma$, which is the ratio of the magnetar's long-term mean energy output in outflows (quiescent MHD wind + sporadic outbursts), $\langle\dot{E}\rangle$, to its spin-down luminosity.
This clearly implies that the MWN is not powered predominantly by 
the magnetar's spin-down-powered wind, and an alternative dominant energy source is required, most likely the decay of the magnetar's magnetic field. We show that the decay of its dipole 
field alone is not enough, and a significantly larger (by a factor of $\sim10^2-10^3$) energy reservoir is needed. The most plausible candidate is the magnetar's internal magnetic field, which has to be $\gtrsim 10-30$ times larger than its dipole field.
Finally, by assuming a maximum allowed initial internal field strength (on theoretical grounds) of $B_{\rm int,max}\sim10^{16}-10^{16.5}\;$G we obtain an upper limit on $g\lesssim 5\times(10^2-10^3)$.

In \S~\ref{sec:GeV/TeV} we consider the possible origin of the GeV/TeV emission observed by Fermi/LAT and H.E.S.S. In \S~\ref{sec:hadronic} we consider the possibility of an hadronic origin
for the GeV/TeV emission. While inverse-Compton emission by electrons in the MWN is energetically very challenging, we find a much more plausible and energetically
reasonable alternative to be the decay of neutral pions that are produced by the interaction of cosmic rays accelerated in the shock driven by the SNR into the 
external medium, with nuclei in the nearby giant molecular cloud.
In \S~\ref{sec:leptonic} we consider the possibility of inverse-Compton emission by relativistic electrons in the MWN
that upscatter seed photons from the NIR Galactic background and the CMB. For Galactic NIR seed photons we find it extremely difficult to account for the 
observed GeV/TeV luminosity. For CMB seed photons it is also very difficult to account for the GeV/TeV luminosity, and moreover, together with the power-law MWN electron energy distribution
(as reflected by their X-ray synchrotron emission) one would expect a much broader spectral peak at TeV energies, which is hard to reconcile with the 
Fermi/LAT and H.E.S.S. observations. Therefore, we find this option implausible. Finally, we also consider emission from relativistic non-thermal Bremsstrahlung from electrons accelerated at the SNR forward shock. 

Our conclusions are discussed in \S~\ref{sec:dis}. 
Our main findings are that the MWN is powered predominantly by outflows from the magnetar, whose main energy source is most likely  the decay of its strong internal magnetic field. 
These conclusions became possible because of the discovery of this first MWN. The outflows from the magnetar accumulate inside the MWN, which therefore serves as a calorimeter that helps us study the system's history.

\section{Implications of the association with the SNR}\label{sec:SNR}
The location of Swift J1834 at the very center of the SNR W41 strongly supports their association.
Moreover, it can also constrain the SGR's proper velocity on the plane of the sky, $v_{\perp,\rm SGR}$.
Its location can be constrained to be $\lesssim(0.05-0.1)R_{\rm SNR}$ from the center of the SNR,
which for an SNR/SGR age of $t_{\rm SNR}$ implies $v_{\perp,\rm SGR}\lesssim (30-60)d_4(t_{\rm SNR}/10^{4.5}\;{\rm yr})^{-1}\;{\rm km~s^{-1}}$.
Below we discuss how $t_{\rm SNR}$ may be estimated, and how it can be 
reconciled with the SGR's measured spin-down age.

\subsection{Age of SNR W41}\label{sec:age}
By using the Very Large Array Galactic Plane Survey (VGPS), \citet{Tian+2007} found 
strong evidence of HI emission at 1420 MHz associated with the SNR W41 in the 
radial velocity range of $53 - 63~{\rm km~s}^{-1}$. The SNR is also coincident with a giant molecular cloud (GMC) positioned just behind the SNR, as inferred from the $^{13}$CO 
emission lines, in the radial velocity range of $61 - 66~{\rm km~s}^{-1}$, that trace molecular hydrogen $\rm{H}_2$ \citep{Tian+2007}. For a mean radial velocity of 
$58~{\rm km~s}^{-1}$, galactic longitude of $l = 23.24^\circ$, and under the assumption of a flat rotation curve with $v_\odot \simeq 220~{\rm km~s}^{-1}$ 
at $d_\odot \simeq 8.5$ kpc, we find a galactocentric distance to SNR W41 of $d_G \simeq 5.1$ kpc,
which corresponds to a distance of $d \simeq 3.97$ kpc or $d \simeq 11.65$ kpc. The HI absorption spectrum of 
SNR W41 suggests a tangent point velocity $V_t\simeq 112~{\rm km~s}^{-1}$ and constrains W41 to be on the near 
side of the distance ambiguity \citep{LH08}. It has an average angular diameter of $33'$ \citep{Tian+2007} 
which gives a radial
extent of $R_{\rm SNR} \simeq 19.2d_4\;$pc, for a distance of $d=4d_4\;$kpc to the source. 
The HI column density maps from the VGPS suggest values in the range $1\lesssim N_{HI,21}\lesssim 2$ in the surrounding giant 
molecular cloud (GMC), where $N_{HI,21} = N_{HI}/(10^{21}\;{\rm cm}^{-2})$. 
Moreover, there is a significant
deficit of up to $\Delta N_{HI,21}\lesssim 1$ in the exact location of the SNR and with its exact shape,
which strongly suggests that it originates from the reduction in the $N_{HI}$ column density in the region 
now occupied by the SNR that was previously part of the surrounding GMC. Therefore, the maximum reduction in column density
due to the SNR can be used to estimate the corresponding mean number density in the region that it swept-up within the GMC,
$n_{HI} \sim \max(\Delta N_{HI})/(2R_{\rm SNR})\sim 8.4d_4^{-1}~{\rm cm}^{-3}$ (see also  \citealt{Tian+2007}).

Although the ISM particle density is dominated by the HI component, an estimate of the true total mass density $\rho_{\rm ext}$ should account 
for the presence of the GMC. Therefore, we express the mass density of the ISM using $\rho_{\rm ext} = \mu n_{HI}m_p$, where $m_p$ is the proton mass 
and $\mu$ is the mean number of nucleons per neutral hydrogen atom in the ISM. One might expect $\mu\approx 2$ when accounting for helium and 
metals, in addition to the fact that some of the hydrogen is ionized (HII) or molecular (H$_2$), 
which would give $n_{\rm ext}=\rho_{\rm ext}/m_p\approx 17(\mu/2)d_4^{-1}~{\rm cm}^{-3}$.
The age of the SNR can be estimated from the Sedov-Taylor self-similar solution
\begin{equation}
R(t) = 1.17\left(\frac{E_{\rm SN}t^2}{\rho_{\rm ext}}\right)^{1/5}\ ,
\label{eq:RST}
\end{equation}
which gives 
\begin{eqnarray}\nonumber
t_{\rm SNR} &=& 97E_{{\rm tot},51}^{-1/2}\left(\frac{n_{\rm ext}}{17\;{\rm cm^{-3}}}\right)^{1/2}d_4^{5/2}\;{\rm kyr}\ ,
\label{eq:tSNR}
\\ 
& =& 23.6E_{{\rm tot},51}^{-1/2}n_0^{1/2}d_4^{5/2}\;{\rm kyr}\ ,
\\ \nonumber
& =& 5.3E_{{\rm tot},52.3}^{-1/2}n_0^{1/2}d_4^{5/2}\;{\rm kyr}\ ,
\end{eqnarray}
where $n_0 = n_{\rm ext}/(1\;{\rm cm^{-3}})$ (this fiducial value of $n_{\rm ext}$ 
is chosen due to the inherent uncertainty in determining the true ISM density, which 
could be in the range $n_{\rm ext}\sim 0.1 - {\rm few\times}10~{\rm cm}^{-3}$, also see below), $E_{{\rm tot},51} = E_{\rm tot}/(10^{51}\;{\rm erg})$, and 
$E_{{\rm tot,52.3}} = E_{\rm tot}/(2\times10^{52}\;{\rm erg})$, where $E_{\rm tot}$ is the total energy injected into the system. The motivation for $E_{\rm tot}\sim2\times10^{52}\;$erg
comes from energy injection by a millisecond magnetar at early times (as explained in more detail below).
An estimate of $t_{\rm SNR}\sim 100\;$kyr was derived by \citet{Tian+2007} ($\sim 60\;$kyr assuming a Sedov-Taylor expansion, and $\sim 200\;$kyr accounting for radiative cooling of the SNR). Although according to \citet{Cioffi+88}, the SNR becomes 
radiative with its dynamical evolution described by the \textit{pressure-driven snowplow} (PDS) phase at 
$t_{\rm PDS} = 13.4E_{\rm SN,51}^{3/14}n_0^{-4/7}~{\rm kyr}$, the age derived above for the Sedov-Taylor phase serves as a 
robust lower limit.

The single model parameter with the largest uncertainty affecting the derived age is the density of the external medium $\rho_{\rm ext}$. 
Although other astronomical methods can be employed to estimate its true value, other than what is used in this work, 
they all can provide conflicting results \citep[see for e.g.][]{CS10}. 
For instance, the density of the shocked electrons behind the forward blast wave, that are emitting thermal X-rays, can be 
obtained from spectral fits that yield the volume emission measure, defined as $E.M. \approx n_e^2Vf_V/(4\pi d^2)$ where $n_e$ 
is the density of the shocked electrons, $f_V$ is the volume filling factor, $V$ is the volume and $d$ is the distance to 
the source. Alternatively, if the SNR forward blast wave is directly interacting with a neighbouring GMC, 
the possibility of which is confirmed by the observation of shock-excited OH (1720 MHz) maser emission \citep{Frail+96,Frail+2013}, then 
CR protons accelerated at the shock front can give rise to detectable GeV/TeV radiation. The high energy flux can then be 
used to constrain the density of the target protons (see \S \ref{sec:GeV/TeV}). However, care must be taken as for the interpretation 
of such a density estimate since it only reflects the volume averaged density. Molecular clouds are known to be clumpy with typically 
large density contrasts ($\rho_c/\rho_{\rm ic} > 10^3$) between the clumpy ($\rho_c$) and the inter-clump medium ($\rho_{ic}$), 
such that the average density is $\langle\rho\rangle = \rho_c[f_V+(\rho_{\rm ic}/\rho_c)(1-f_V)]$, where $f_V\sim0.02 - 0.08$ \citep[e.g.][]{Blitz93} is the volume 
filling factor of the clumps. For the SNR dynamics the relevant external density is $\rho_{\rm ext}\approx\rho_{\rm ic}$ (since dense clumps are hardly accelerated by the SNR shock and instead penetrate through it like bullets). Therefore, usage of the (generally larger) average density $\langle\rho\rangle$ for calculating the SNR dynamics might offset 
its size or age estimates considerably. In what follows, we continue to parametrize the external density in terms of $n_0 = n_{\rm ext}/(1\;{\rm cm^{-3}})$ where $n_{\rm ext}=\rho_{\rm ext}/m_p$.
\subsection{Characteristic age and braking index of Swift J1834}\label{sec:n-tau_sd}
The spin-down law is 
often expressed as a power law with a general `braking index' $n$ ($n = 3$ for magnetic dipole braking),
\begin{equation}\label{eq:ndef}
\dot{P} \propto P^{2-n}\ .
\end{equation}
For magnetic dipole braking this corresponds to the assumption that the strength of the dipole magnetic field and the moment of inertia remain constant over time. Integrating this equation over time gives the time evolution of the spin period
\begin{equation}
P(t) = P_0\left(1+\frac{t}{t_0}\right)^{\frac{1}{(n-1)}}\ ,
\end{equation}
where $P_0$ is the initial spin period, and $t_0$ is the initial spin-down time.
This provides the current age, $t$, in terms of $t_0$, $P$ and $\dot{P}$:
\begin{equation}\label{eq:true-age}
    t = \frac{P}{(n-1)\dot{P}}-t_0 = \frac{2\tau_c}{(n-1)}-t_0\ ,
\end{equation}
where $\tau_c\equiv P/2\dot{P}$ is the usual characteristic spin-down age. One finds the relation
\begin{equation}\label{eq:t_0_general}
    t_0 = \frac{P_0}{(n-1)\dot{P}}\fracb{P}{P_0}^{2-n}\ ,
\end{equation}
which provides $t_0$ given the current measured values of $P$ and $\dot{P}$, for assumed values of $n$ and $P_0$.

For $t \gg t_0$, the characteristic spin-down age can be obtained from the measured spin properties, and in the present case, for $n = 3$,
it is $t\approx \tau_c = 4.9\;$kyr.

If the true age of Swift~J1834 is significantly larger than $\tau_c$ (as suggested by the estimated age of the 
SNR $t_{\rm SNR}\sim100$ kyr) then one way in which this could be reconciled is if $n\approx 1$.
This could occur, e.g., if a particle-dominated wind opens up dipole magnetic field lines, thus increasing the resulting field strength at the light cylinder and enhancing the spin-down torque \citep[e.g.][]{Harding+99,Tong+13,Tong16}. 
However, while the likely energy source for driving such a wind is the magnetar's magnetic
field decay, it is not clear what mechanism would naturally produce such a particle-dominated wind. What is more, its required power needs to exceed the current
inferred spin-down power $L_{\rm sd}$ by a factor of $g \sim 50-5000$ \citep{Tong16}, 
so it would by far dominate the energy injection into the MWN.

Moreover, this has serious implications for the magnetic field $B_{\rm LC} = B(R_{\rm LC})$ at the light cylinder $R_{\rm LC}=c/\Omega$. If the spin-down torque is indeed due to rotational energy loss to Poynting flux of an outgoing MHD wind, then the spin-down would be the same as for magnetic dipole braking with an effective surface dipole field strength of $B_{\rm eff} = (R_{\rm LC}/R_{\rm NS})^3B_{\rm LC}\propto \sqrt{P\dot P}$ that satisfies
\begin{equation}
    \dot{P}\propto P^{2-n} \propto B_{\rm eff}^2P^{-1}\ ,
\end{equation}
and therefore
\begin{equation}\label{eq:B-evol}
B_{\rm eff}(t) = B_0\fracb{P}{P_0}^\frac{3-n}{2} = B_0\left(1+\frac{t}{t_0}\right)^{\frac{3-n}{2(n-1)}}\ .
\end{equation}
For the above mentioned model where $n\approx 1$ we can see that $B_{\rm eff}$ grows as a high power of time $t$. However, there is a limited dynamical range over which a particle wind can open up the dipole field lines, corresponding to a field-opening radius in the range
$R_{\rm NS}<r_{\rm open}<R_{\rm LC}$, or a factor of $R_{\rm LC}/R_{\rm NS}$ which is $\approx10^4$ for Swift~J1834. Therefore, such a behavior could last only up to a factor of $\sim (R_{\rm LC}/R_{\rm NS})^{2(n-1)/(3-n)}$ in time, between an age of $t_0$ and the current age $t$. For our case and $n\approx 1$ this corresponds to a current age satisfying $t/t_0\sim 10^{4(n-1)}$ or $\sim 2.6$ for $n\approx 1.1$ (which would correspond to a true age of $t\approx \tau_c 2/(n-1)\sim 100\;$kyr).

The current $P$ and $\dot{P}$ values of Swift J1834 imply a surface magnetic field $B_{\rm eff}(t) = 1.16\times10^{14}f^{-1/2}\;$G. 
A braking index $n=1.1$ would suggest a growing effective magnetic field $B_{\rm eff}$ over time, so that its value at birth should have been much lower,
$B_{\rm eff,0} = 6.2\times 10^{11}f^{-1/2}P_{0,-2}^{0.95}\;$G where $P_{0,-2} = P_0/(10^{-2}\;{\rm s})= P_0/(10\;{\rm ms})$. 
One way of interpreting this increase in $B_{\rm eff}$ is that it indeed reflects a growing
surface dipole field, which is not easy to account for on physical grounds 
\citep[see, however][]{MP96,Ho15,Marshall+16}. \textcolor{black}{Since the magnetars' bursting activity and high X-ray
luminosity $L_X>L_{\rm sd}$ is thought to be powered by the decay of their magnetic field, 
this would imply a gradual emergence of a much stronger internal magnetic field, from deep 
within the neutron star, out to its surface. This may potentially produce, at least temporarily, a net 
growth of the surface dipole magnetic field, if some of the strong internal toroidal field is converted to an 
external poloidal field faster than the latter decays}. Axisymmetric magnetic field stability analysis 
(see Eq. \ref{eq:Bint_max}) yields an internal field strength 
$B_{\rm int,0}\lesssim1.6\times10^{14}f^{-1/4}P_{0,-2}^{0.475}\;$G, which corresponds to 
the effective surface dipole field $B_{\rm eff,0}$. This estimate is consistent with the 
current surface dipole field, as inferred from $P$ and $\dot P$, and provides further 
support to the notion of internal field transport to the surface. Still, the underlying 
mechanism for growth of surface dipole field is still uncertain, however, we briefly mention possible 
channels below.

Now let us consider specifically the case of magnetic dipole braking, which corresponds to $n=3$ and for which the spin-down power is given by 
\citep[e.g.][]{Spitkovsky06}
\begin{equation}
L_{\rm sd} = f\frac{B_s^2 R_{\rm NS}^6 \Omega^4}{c^3}\ ,\quad 
f = \left\{\begin{array}{cr} \frac{2}{3}\sin^2\theta_B & {\rm in\ vacuum}\\ \\
1+\sin^2\theta_B & {\rm force\ free}\end{array}\right.
\end{equation}
where $I \simeq 10^{45}\;{\rm g\;cm^2}$, $R_{\rm NS}\simeq 10^6\;$cm and $\Omega = 2\pi/P$ are the neutron star's moment of inertia, radius and 
angular spin frequency, respectively, $\theta_B$ is the inclination angle between the magnetic dipole axis and the rotational axis, 
and $B_s$ is the equatorial surface dipole magnetic field strength, whose initial value is $B_0 = 10^{14}B_{14}\;$G.
By comparing the loss of rotational kinetic energy with spin-down power, $-I\Omega\dot\Omega = L_{\rm sd}$,  and initial spin-down time is given by
\begin{equation}\label{eq:t0}
t_0 = \frac{Ic^3}{2f\Omega_0^2 R_{\rm NS}^6 B_0^2} = 3.4\times 10^4\frac{P_{0,-3}^2}{f B_{14}^{2}}\;{\rm s}\ ,
\end{equation}
where $\Omega_0 = 2\pi/P_0$ is the initial angular spin frequency, and $P_{0,-3} = P_0/(10^{-3}\;{\rm s})= P_0/(1\;$ms). 
However, even in this case the spin-down age only reflects the current value of $\dot{P}$, which may be unusually higher than its long-term mean value, which may account for a true age $t\gg\tau_c$. Several other magnetars have shown large fluctuations in $\dot{P}$ before and after 
bursting episodes. For example, the average spin-down rate of SGR $1806-20$ increased by a factor of $\sim 6$ in the 12 yr period prior to the 
the giant flare in December 2004, with large fluctuations in $\dot P$ observed in the months leading up to the flare \citep{Woods+07,Y+15}. 
In the case of SGR $1900+14$, which emitted a giant flare in August 1998, the post-flare $\dot P$ was higher by a factor $\sim 4$ 
compared to the pre-flare long-time average spin-down rate \citep{Woods+02}. Finally, \citet{GK04} reported a factor of $\sim 12$ increase in the 
spin-down rate of AXP 1E $1048.1-5937$ over a timescale of weeks to months. Such a sudden increase (decrease) in $\dot{P}$ is likely due to the 
formation (destruction) of magnetospheric currents that enhance the magnetic field near the light cylinder to values above that due to the magnetar's 
dipole field component that is supported by currents inside the neutron star. In such a case the usual value of $B_s$ that is inferred from $P$ and $\dot{P}$ 
is an over-estimate of the true surface dipole field strength. \textcolor{black}{Such changes in $\dot{P}$ may also be associated with 
outflows accompanying the magnetar's bursting activity. Moreover, the cumulative effect of such irregular $\dot{P}$ changes may modify the spin-down histories of magnetars and make them correspondingly irregular and less predictable compared to regular pulsars.}


%
The current surface magnetic field is given by the familiar expression
\begin{equation}
B_s = \sqrt{\frac{Ic^3P\dot{P}}{4\pi^2fR_{\rm NS}^6}}= 2.61\times10^{19}R_6^{-3}\sqrt{\frac{I_{45}}{f}}\sqrt{P_{\rm s}\dot{P}}~{\rm G}\ ,
\label{eq:Beq}
\end{equation}
where $I=10^{45}I_{45}\;{\rm g\;cm^2}$, $R_{\rm NS}=10^6R_6\;$cm and $P_{\rm s} = P/(1\;{\rm s})$.

In Appendix~\ref{sec:appA} we show that for a more general equation for the magnetar's braking, of the form $\dot{P} \propto B_s^2P^{2-n}$,
where the magnetic field evolves in time as $B_s\propto t^{-1/\alpha}$ at times $t>t_B$, where $t_B$ is the 
characteristic field growth/decay timescale, the observable braking index is given by $n' = n+\Delta n$,
where the difference from $n$ that appears in this formula is $\Delta n = 2\dot B_sP/(B_s\dot P)$. For $\alpha <2$ the field decays rapidly enough 
that spin-down freezes out and the rotation period approaches a constant value at late times \citep{Colpi+2000,DallOsso2012}, in which case the measured
braking index would grow at late times as $n'\propto t^{(2-\alpha)/\alpha}$. For $\alpha>2$ the observed braking index approaches a constant value 
of $n'_\infty = (n\alpha-2)/(\alpha-2)$. 

For magnetic dipole braking that is preferred 
on theoretical grounds, $n = 3$, resulting in $n'_\infty = (3\alpha-2)/(\alpha-2)$ and $\alpha = (2n'_\infty-2)/(n'_\infty-3)$,
which is similar to Equation~(\ref{eq:B-evol}) when $n$ is replaced by $n'_\infty$. This may be able to account for the observed values of
$n'<3$, even for pure magnetic dipole braking. While this still requires $\alpha<0$, i.e. a growth in the surface magnetic dipole field strength, 
the true value of $\alpha$ may be lower than that inferred possibly from observations, $\alpha' = (2n'-2)/(n'-3)$, according to Equation~(\ref{eq:B-evol});
see, e.g., the upper left panel of Fig.~\ref{fig:n-prime}. 
Broadly similar ideas were recently discussed by \citet{RS16}. In regular pulsars a modest increase in the surface dipole field 
may potentially be caused by emergence of some of the stronger internal magnetic field from deep inside the NS 
\citep[e.g][]{MP96} or out of the crust \citep[e.g.][]{Ho15} and into the magnetosphere.

\subsection{The Current Age and Spin-Down Power}

The pulsar spins down as it injects energy into its environment largely in the form of an MHD wind, also causing its spin-down 
luminosity to decay over time from its initial value $L_0$ after an initial spin down time $t_0$
\begin{equation}\label{eq:Lsd}
    L_{\rm sd} = L_0\left(1+\frac{t}{t_0}\right)^{-m}
\approx L_0\times\left\{\begin{array}{cr} 1\quad & t<t_0\ ,\\ \\ 
(t/t_0)^{-m}\quad
& t>t_0\ .\end{array}\right.
\end{equation}
where $n$ is the braking index and $m = (n+1)/(n-1)$. 
For $n=3$ and $m=2$ we have
\begin{equation}
L_0 = 5.78\times10^{47}fB_{14}^2P_{0,-3}^{-4}\;{\rm erg\;s^{-1}}\ .
\end{equation}
The late time expression for $L_{\rm sd}$ can also be written as
\begin{eqnarray}
    &&L_{\rm sd}(t>t_0)\approx L_{\rm sd}({\rm obs})\fracb{B_0}{B_s({\rm obs})}^{-2}\fracb{t}{\tau_c}^{-2}
    \\ \nonumber
    &&= 2.05\times10^{34}\fracb{B_0}{1.16\times10^{14}f^{-1/2}\,{\rm G}}^{-2}\fracb{t}{\tau_c}^{-2}\;\frac{\rm erg}{\rm s}\ .
\end{eqnarray}
This demonstrates that for the usual magnetic dipole braking scenario, which assumes a constant surface dipole field, the age must be equal to the characteristic spin-down age in order to reproduce the observed spin-down power.

An older age, $t>\tau_c$, may still be possible if one or more of the underlying assumptions of this popular scenario breaks down.
For instance, if there is a strong particle wind that opens-up the magnetic dipole field lines above some radius $r_{\rm open}$ (where $R_{\rm NS}<r_{\rm open}<R_{\rm LC}$) then this would correspond to $n=1$ in Eq.~(\ref{eq:ndef}) which would give
an exponential growth of $P$ and $t = (P/\dot{P})\ln(P/P_0) = 2\ln(P/P_0)\tau_c$,
which is larger than $\tau_c=4.9\;$kyr by a factor of $\sim 11-15.6$ for $P_0\sim 1-10\;$ms. 

Another possibility is that the current $\dot{P}$ is larger than its long-term mean value, $\langle\dot{P}\rangle$, so that the true age for dipole braking would be $t\approx P/2\langle\dot{P}\rangle = \tau_c\dot{P}/\langle\dot{P}\rangle$, i.e. a factor of $\dot{P}/\langle\dot{P}\rangle$ larger than $\tau_c$.

A decay of the magnetic dipole field would go in the opposite direction and result in $t<\tau_c$ (e.g. \citet{Colpi+2000,DallOsso2012}, also see Fig. \ref{fig:alpha-tB-plot}).

\section{Energetics of the X-Ray Emitting Synchrotron Nebula}\label{sec:EX}
Here we consider the energetics of the X-ray nebula itself, within the observed radius $R_X = 2.04d_4$ pc 
(for a mean angular radius of $\theta_X = 105''$), rather than the entire MWN with radius $R$. The nebula's X-ray luminosity is  $L_X = 2.5\times10^{33}d_4^2~{\rm erg~s}^{-1}$ in the $(h\nu_m,h\nu_M) = (0.5, 10)~{\rm keV}$ range, with a photon index $\Gamma = 2.2$.

The inner part of the X-ray nebula, of semi-minor and semi-major axis of $25"\times50"$
($(0.48\times0.97)d_4\;$pc) or $R_{X,{\rm in}}\approx 0.73d_4\;$pc, was detected by NuSTAR up to $30\;$keV.
The total X-ray luminosity in this inner nebula combining XMM-Newton and NuSTAR observations in the
$(h\nu_m,h\nu_M) = (0.5, 30)~{\rm keV}$ energy range is 
$L_{X,{\rm in}}\approx5.0\times10^{32}\;{\rm erg\;s^{-1}}$, with a photon index of $\Gamma_{\rm in}=1.41\pm0.12$.

The magnetic field in the X-ray emitting region can be estimated from the frequency integrated synchrotron luminosity within the relevant region,
\begin{eqnarray}\label{eq:nebularB}
B&&=\left(\frac{L_X\sigma_e}{\mathcal{A} V} \frac{\Gamma-2}{\Gamma-1.5}\frac{\nu_1^{1.5-\Gamma}-\nu_2^{1.5-\Gamma}}
{\nu_m^{2-\Gamma}-\nu_M^{2-\Gamma}}  \right)^{2/7} \\ \nonumber
&&\simeq\left\{\begin{array}{cr} 
4.0\xiֿ\sigma_e^{2/7}d_4^{-2/7}~\mu{\rm G} &\ \  {\rm (whole \ nebula)}\ ,\\ \\
5.0\xi_{\rm in}ֿ\sigma_e^{2/7}d_4^{-2/7}~\mu{\rm G} &\ \ {\rm (inner\ nebula)}\ .\end{array}\right.
\end{eqnarray}
Here $\xi^{7/2}$ is the ratio of the total energy in electrons to that in the electrons radiating in the
observed frequency range (given below in Eq.~(\ref{eq:xi})), 
while $\nu_1$ and $\nu_2$  are the characteristic synchrotron frequencies 
($\nu_{\rm syn}\simeq eB\gamma^2/2\pi m_ec$) corresponding to the boundary energies
($\gamma_1 m_ec^2$ and $\gamma_2 m_ec^2$) of the electron spectrum
 in Eq.~(\ref{eq:partdist}), $V = 4\pi R_X^3/3$ or $V_{\rm in} = 4\pi R_{X{\rm in}}/3$ is the volume filled with radiating
 plasma, $\sigma_e = E_B/E_e$ is the ratio of energies in the magnetic field ($E_B = VB^2/8\pi$) and in the electrons within the considered power-law energy distribution ($E_e = \int_{\gamma_1}^{\gamma_2}\gamma_em_ec^2(dN_e/d\gamma_e)d\gamma_e$, where $dN_e/d\gamma_e\propto\gamma_e^{-s}$ and $s=2\Gamma-1$) and $\mathcal{A}=2^{1/2}e^{7/2}/18 \pi^{1/2}m_e^{5/2}c^{9/2}$ 
 \citep[see e.g.][]{Pacholczyk1970}.
 
We note that the harder photon index in the inner part of the nebula suggests that 
it may better reflect the true value of power-law index $s$ of the un-cooled
electron energy distribution, $s = 2\Gamma_{\rm in}-1 = 1.82\pm0.24$.

For our assumption of a single power-law electron distribution 
 \begin{equation}\label{eq:xi}
     \xi(\nu_1,\nu_2) = \fracb{\nu_2^{1.5-\Gamma}-\nu_1^{1.5-\Gamma}}{\nu_M^{1.5-\Gamma}-\nu_m^{1.5-\Gamma}}^{2/7}\ .
 \end{equation}
In the above, the magnetic field with $\xi=1$ or $\xi_{\rm in}=1$ assumes that $(\nu_1,\nu_2) = (\nu_m,\nu_M)$.
However, there may be a break in the spectrum at energies lower than 0.5$\;$keV (typically seen in PWNe between teh radio and X-rays) corresponding to 
$\nu_1 < \nu_m$. Taking $\nu_1 = 10^{13}$ Hz, the typical break frequency observed in other PWNe, yields a higher magnetic field strength in the whole X-ray nebula, $B \simeq 27~\mu{\rm G}$. 
This corresponds to changing $\xi \approx 6.80$ in the above equation. We define $\xi_7$ and $\xi$
normalized by this value. 

The unknown value of $\nu_2$ is not important when $\Gamma > 1.5$, which 
corresponds to $s>2$ for slow electron cooling. In this case most of the energy resides in electrons near the minimal electron Lorentz factor $\gamma_1$ with a corresponding synchrotron frequency $\nu_1$, so that one can take $\nu_2=\nu_M$. For $s<2$ most of the energy resides in the highest energy electrons
near $\gamma_2$, and $\xi$ depends mainly on $\nu_2=\nu_{\rm syn}(\gamma_2)$, so that one can instead take $\nu_1=\nu_m$. For $s\approx 2$ both boundaries are important and should be taken into account. 
 

 Alternatively, a lower bound on the nebular magnetic field can be obtained by requiring that the implied
 synchrotron frequency of the maximum energy electrons in the injected distribution, $\gamma_2$, would be 
 at least as high as the largest observed frequency from the nebula. Electrons are accelerated at or near 
 the wind termination shock and their maximum energy is limited by the size $\lesssim R_{\rm TS}$ of this 
 region over which they are initially confined and accelerated. 
 Confinement of the particles within their acceleration region requires
 $R_L\simeq \gamma_e m_e c^2/eB(R_{\rm TS}) < R_{\rm TS}$, which implies a maximum energy of $\gamma_{e}m_ec^2<\gamma_{\rm max}m_ec^2 = eB(R_{\rm TS})R_{\rm TS}\approx
 eB(R_{\rm LC})R_{\rm LC}$. The last equality arises since $B\propto 1/r$ beyond $R_{\rm LC}$, and this relates the confinement condition near $R_{\rm TS}$ in the nebula to the potential difference near $R_{\rm LC}$ and across the open field lines emanating from the neutron star, near its polar cap.
 As shown by \citet{DeJager_Harding_1992}, $\gamma_{\rm max}m_ec^2$ is, to within a factor of order unity, similar to 
 the energy gained by electrons while dropping across the polar cap potential $V_0$, which is given by \citep{GJ69}
 \begin{equation}\label{eq:V0}
     V_0 = \frac{R_{\rm NS}^3\Omega^2B_s}{c^2} = \sqrt{\frac{L_{\rm sd}}{fc}}~,
 \end{equation}
 and can be expressed using the spin-down power. This yields an estimate of the maximum lorentz 
 factor of the injected 
 electrons
 \begin{equation}
     \gamma_{\rm max} = \frac{e}{m_ec^2}\sqrt{\frac{L_{\rm sd}}{fc}} \simeq 4.9\times10^8f^{-1/2}\ .
 \end{equation}
As argued by \citet{DeJager_Harding_1992}, synchrotron losses will typically limit the maximum Lorentz factor of 
accelerated electrons to less than $\gamma_{\rm max}$. Therefore, $\gamma_{\rm max}\geq\gamma_2$ should be regarded as an 
absolute upper limit on the Lorentz factor of electrons injected downstream at the termination shock.
 
Then, in order to account for all of the X-ray emission in the observed range, we need $\gamma_{\rm max} > \gamma_X$ in all of this range, where $\gamma_X$ 
 is the Lorentz factor of the X-ray emitting electrons.
 The most constraining condition comes from the upper end of the observed energy range, which is 30~keV from the NuSTAR detection of the inner X-ray nebula, 
 and therefore we use here $E_M = 30E_{M,30}~{\rm keV}$.
 The synchrotron photon energy is $E_{\rm syn}(\gamma_e)\approx \hbar eB\gamma_e^2/(m_ec)$ so that $\gamma_M\propto \sqrt{E_{M}/B}$.
 Thus, the nebular magnetic field is bounded from below by
 \begin{equation}\label{eq:Bmin}
     B > B_{\rm min} \equiv \frac{m_e^3c^6fE_X}{\hbar e^3L_{\rm sd}} \simeq 11.0fE_{M,30}~\mu{\rm G}
 \end{equation}
which further constrains the flow magnetization
\begin{eqnarray}\label{eq:sigma_e}
    \sigma_e &>& 35.1\,d_4\fracb{fE_{M,30}}{\xi}^\frac{7}{2} = 0.043\,d_4\fracb{fE_{M,30}}{\xi_7}^\frac{7}{2}\ ,\quad\quad\\ \nonumber
\sigma_{e,{\rm in}} &>& 15.4\,d_4\fracb{fE_{M,30}}{\xi_{\rm in}}^\frac{7}{2} = 0.055\,d_4\fracb{fE_{M,30}}{\xi_{\rm in}/5}^\frac{7}{2}\ .
\end{eqnarray}

The usual plasma magnetization parameter is $\sigma = B^2/4\pi w$ where $w$ is the total particle enthalpy density, where for relativistically hot plasma $w = (4/3)e\gg\rho c^2$ where $e$ is the internal energy density, so that $\sigma = \frac{3}{2}B^2/8\pi e$. This reduces to $\sigma = \frac{3}{2}E_B/E_e = \frac{3}{2}\sigma_e$ or $\sigma_e\to\frac{2}{3}\sigma$ if the electrons in the considered power-law energy distribution hold all of the energy in particles, $E_e = E_{m}$. If there is another population of electrons or energy in protons etc., so that these electrons hold only a fraction $\epsilon_e$ of the total energy in particles or matter, $E_e = \epsilon_e E_{m}$ with $\epsilon_e\leq1$, then $\sigma_e\to\frac{2}{3}\sigma/\epsilon_e$. 

Equation (\ref{eq:sigma_e}) implies that in order to accommodate values of $\sigma\approx\frac{3}{2}\sigma_e\epsilon_e\sim 10^{-3}-10^{-2}$ 
in the bulk of the nebula, which are suggested by observations \citep[e.g.][]{Chevalier04} and appear to be broadly consistent with recent simulations \citep[e.g.][]{Porth+2013}, one requires a $\xi$ 
of at least a few to several,
\begin{eqnarray}\label{eq:xi_min}
    \xi&>& 2.76\,fE_{M,30}\fracb{d_4}{\sigma_e}^\frac{2}{7}=16.1\,fE_{M,30}\fracb{\epsilon_e d_4}{\sigma_{-2.5}}^\frac{2}{7}\ ,\quad\\ \nonumber
    \xi_{\rm in}&>& 2.18\,fE_{M,30}\fracb{d_4}{\sigma_{e,{\rm in}}}^\frac{2}{7}=12.7\,fE_{M,30}\fracb{\epsilon_e d_4}{\sigma_{{\rm in},-2.5}}^\frac{2}{7}\ ,
\end{eqnarray}
where $\sigma_{e,-2.5}=\sigma_e/10^{-2.5}$. In the following we will express the relevant quantities both
in terms of $\xi$ and $\sigma_e$, as well as in terms of $B = 15B_{15\mu{\rm G}}\;\mu$G.

We note that our analysis assumes a uniform magnetic field in the X-ray nebula. However, it might be plausible that the magnetic field 
gradually decreases with the distance from the wind termination shock and is higher in its inner parts than in its outer parts.
Moreover, the NuSTAR detection up to $30\;$keV is in the inner nebula, and applying Eq.~(\ref{eq:Bmin}) to the outer nebula gives $B_{\rm out} > 3.7fE_{M,10}\;\mu$G given that the outer nebula is detected only up to $10\;$keV, which is 3 times lower than the corresponding lower limit on the magnetic field in the inner nebula, $B_{\rm in} > 11.0fE_{M,30}\;\mu$G. However, our main results are not significantly affected by the assumption of a uniform field in the nebula.

The magnetic energy in the X-ray nebula from Equation~(\ref{eq:nebularB}) is 
 \begin{eqnarray}\nonumber
E_B &&\simeq 6.4\times10^{44}\xi^2\sigma_e^{4/7}d_4^{17/7}\ {\rm erg}\ ,\\ \label{eq:EB0}
&&\simeq3.0\times10^{46}\xi_7^2\sigma_e^{4/7}d_4^{17/7}\ {\rm erg}\ ,\\ \nonumber
&&\simeq9.3\times10^{45}B_{15\mu{\rm G}}^{2}d_4^{3}\ {\rm erg}\ ,
\end{eqnarray}
where the synchrotron cooling time of electrons radiating at $E =2E_2\;$keV is 
\begin{eqnarray}\nonumber
t_{\rm syn} &&= \frac{6\pi m_ec}{\sigma_TB^2\gamma_e} 
= \frac{6\pi}{\sigma_T}\fracb{e\hbar m_ec}{B^3E_\gamma}^{1/2} \\ \nonumber
&&\simeq 7.56\xi^{-3/2}\sigma_e^{-3/7}E_2^{-1/2}d_4^{3/7}\;{\rm kyr}\ ,\\ \label{eq:tsyn}
&&\simeq 0.42\xi_7^{-3/2}\sigma_e^{-3/7}E_2^{-1/2}d_4^{3/7}\;{\rm kyr}\ ,\\ \nonumber
&&\simeq 1.02B_{15\mu{\rm G}}^{-3/2}E_2^{-1/2}\;{\rm kyr}\ .
\end{eqnarray}
 
 The total energy  in the MWN, $E = E_B + E_m = E_B+E_e/\epsilon_e$, can be estimated under the assumption that the pressure is uniform 
 in the region between $R_{\rm TS} < r < R$, where $R_X < R < R_{\rm SNR}$, due to the sub-sonic 
 expansion of the nebula
\begin{eqnarray}
E &&= \fracb{1+\sigma}{\sigma}E_B\fracb{R}{R_X}^3 \\ \nonumber
&&\simeq 1.4\times10^{46}\frac{(1+\sigma)}{\sigma^{3/7}\epsilon_e^{4/7}}\xi^2d_4^{17/7}\fracb{R}{3R_X}^3~{\rm erg}\ ,\\ \nonumber
&&\simeq 6.5\times10^{47}\frac{(1+\sigma)}{\sigma^{3/7}\epsilon_e^{4/7}}\xi_7^2d_4^{17/7}\fracb{R}{3R_X}^3~{\rm erg}\ ,\\ \nonumber
&&\simeq 2.5\times10^{47}\fracb{1+\sigma}{\sigma}B_{15\mu{\rm G}}^{2}d_4^{3}\fracb{R}{3R_X}^3~{\rm erg}\ .
\end{eqnarray}

\section{MWN Dynamics \& Adiabatic Thermal History}\label{sec:Ead}

Here we consider the dynamical evolution of the MWN, accounting for the energy 
injection by the magnetar's quiescent spin-down powered
MHD wind, the interaction with the SN ejecta and external medium, and adiabatic cooling or heating of the relativistic electrons in the MWN
after they are initially accelerated in the wind termination shock, at radius $R_{\rm TS}$. 

The cumulative injected energy up to a time $t$ can be expressed using the 
initial rotational energy 
\begin{equation}
E_0 =\frac{n\pm1}{2} L_0t_0 = \frac{1}{2}I\Omega_0^2 
\simeq 2\times10^{52}P_{0,-3}^{-2}~{\rm erg}\ ,
\end{equation}
where the + and $-$ signs are,respectively, for the broken power-law approximation and the smooth form of $L_{\rm sd}(t)$ in Eq.~(\ref{eq:Lsd}). The latter implies
\begin{equation}
    E_{\rm inj}(t) = \int_0^t L_{\rm sd}(t')dt' = E_0\left[1-\left(1+\frac{t}{t_0}\right)^{-\frac{2}{n-1}}\right]\ .
\end{equation}
while the former is used 
when deriving Eq.~(\ref{eq:EST}), and  implies
\begin{equation}
    E_{\rm inj}(t) = \frac{2E_0}{n+1}\times\left\{\begin{array}{cr} t/t_0\quad & t<t_0\ ,\\ \\ 
1+\frac{n-1}{2}\left[1-\fracb{t}{t_0}^{-\frac{2}{n-1}}\right]\quad
& t>t_0\ .\end{array}\right.
\end{equation}

Downstream of the termination shock ($r > R_{\rm TS}$), the injected electrons can be assumed to have a 
power law distribution
\begin{equation}
N(\gamma_e) \propto \gamma_e^{-s}\quad\quad\rm{for}\quad\quad\gamma_1<\gamma_e<\gamma_2
\label{eq:partdist}
\end{equation}
and they suffer adiabatic losses due to the expansion of the SNR, where their energy drops as 
$E_e \propto V^{1-\hat{\gamma}}$. For a relativistic gas, the adiabatic index is $\hat{\gamma} = 4/3$, 
and thus $E_e \propto R^{-1}$. Such an adiabatic index should also hold for a tangled magnetic field and for relativistic protons (or ions)
that are also likely to be present in the nebula. The electrons also suffer radiative losses, however this effect is 
negligible for the majority of them as they are slow cooling; for $s > 1$ most electrons have energies 
close to the lower end of the distribution since $\gamma_eN(\gamma_e)\propto\gamma_e^{-s+1}$. Therefore
we neglect any energy losses and assume the energy is conserved in the system. Under such conditions the total energy 
in the wind nebula changes adiabatically and when the nebula expands the energy in electrons injected at time $t_i$ 
decreases with time $t$, such that
\begin{equation}
E(t) = \int_0^t L(t_i)f_{\rm ad}(t_i) dt_i\ ,
\end{equation}
where $f_{\rm ad}(t_i) = R(t_i)/R(t)$ is the adiabatic energy dilution factor and it 
depends on the expansion history of the MWN at times $t = t_i$. 
The fate of the MWN in turn depends on the initial energy $E_0$ 
and how it compares to the energy of the SN explosion $E_{\rm SN} = 10^{51}E_{\rm SN,51}~{\rm erg}$. 
When the MWN expands the energy it loses is gained by the SNR via $pdV$ work, and if it contracts (when compressed by the reverse shock
going into the SNR) it gains energy at the expense of the SNR. Altogether, under our assumption of no energy losses from the system the
SNR's energy can be expressed in terms of the other energies in the system,
\begin{equation}
    E_{\rm SNR}(t) = E_{\rm SN} + E_{\rm inj}(t) - E(t)\ .
    \label{eq:Esnr}
\end{equation}

\begin{figure}
    \centering
    \includegraphics[width=0.46\textwidth]{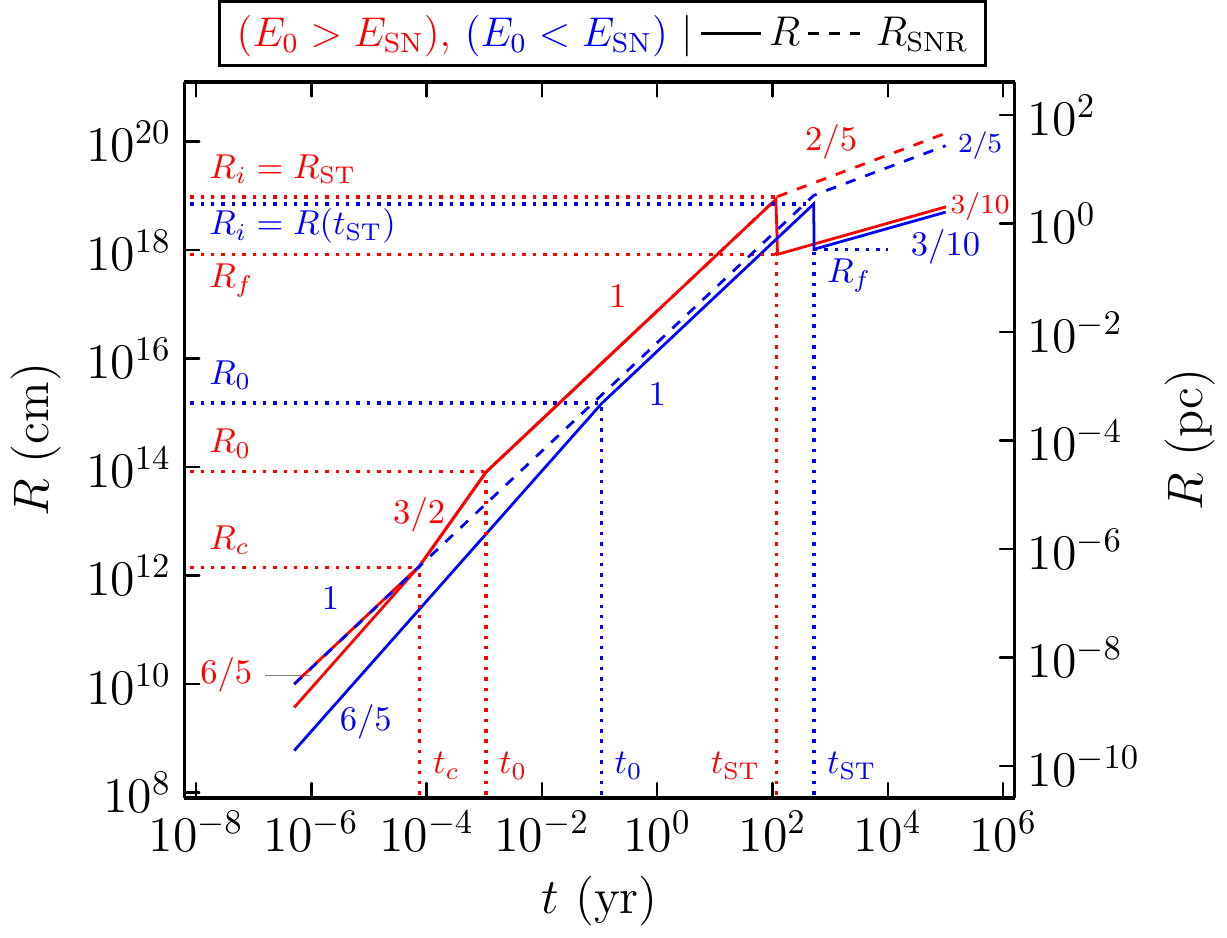}
    \caption{Schematic of the dynamical evolution of the MWN's radius (solid line) as a function of time in a constant density ISM ($k=0$), 
    with temporal power law indices specified for the different segments and color-coded for the two different cases. 
    These are: $E_0 > E_{\rm SN}$ ($P_{0,-3}=1$, red) and $E_0 < E_{\rm SN}$ ($P_{0,-3}=10$, blue). 
    The evolution of the SNR size is shown as a dashed line. The dotted lines 
    punctuate special times in the dynamical evolution 
    of the system, where $t_c$ is the density core crossing time by the outer edge of the MWN, 
    $t_0$ is the initial spin-down time of the magnetar after which its energy injection rate drops significantly, and $t_{\rm ST}$ is 
    the onset time for the Sedov-Taylor phase. All these times have their corresponding radii. $R_i$ and $R_f$ are the radii of the MWN 
    before and after it is crushed by the reverse shock at time $t = t_{\rm ST}$. See Table \ref{tab:Rtparams} and text 
    for details of these parameters. The following parameter values are assumed: $B_0 = 10^{14}~{\rm G}$, $n = 3$, $n_{\rm ext} = 1~{\rm cm}^{-3}$, 
    $M_{\rm ej} = 3M_\odot$, $f = 1$, $E_{\rm SN} = 10^{51}~{\rm erg}$.}
    \label{fig:Rplot}
\end{figure}

\begin{figure}
    \centering
    \includegraphics[width=0.46\textwidth]{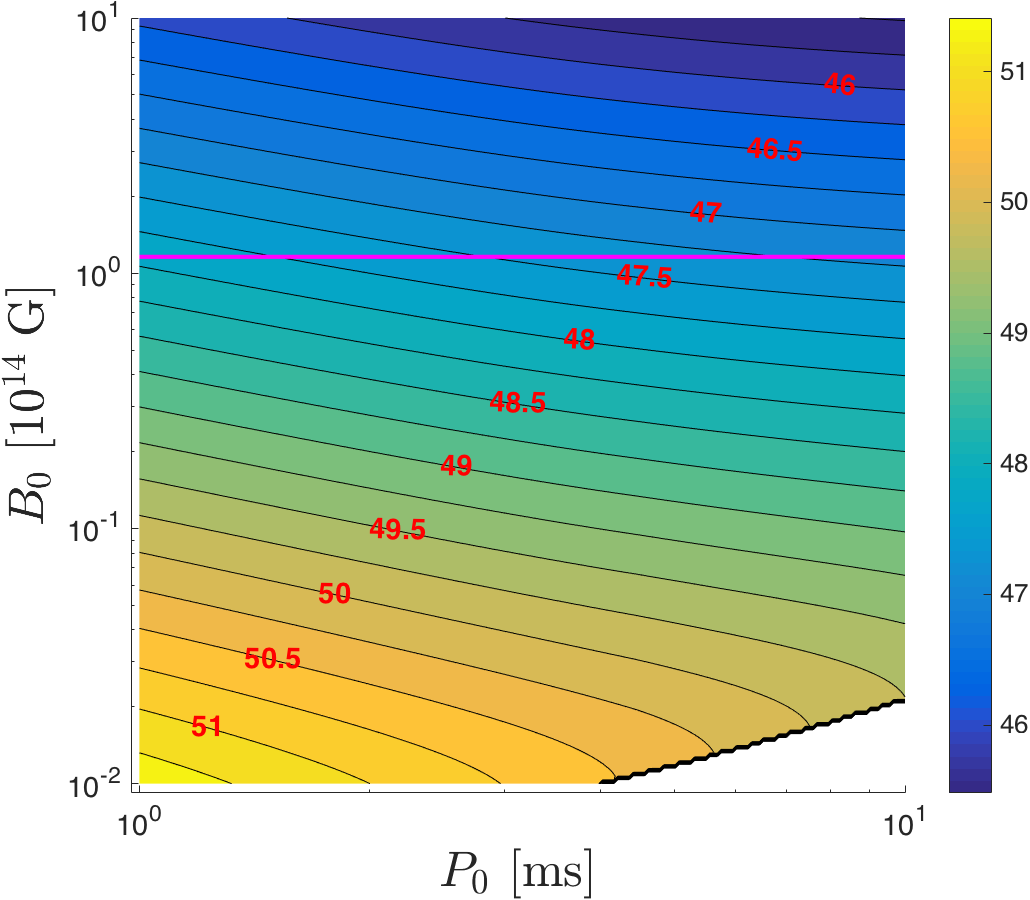}
    \vspace{0.3cm}
    \includegraphics[width=0.46\textwidth]{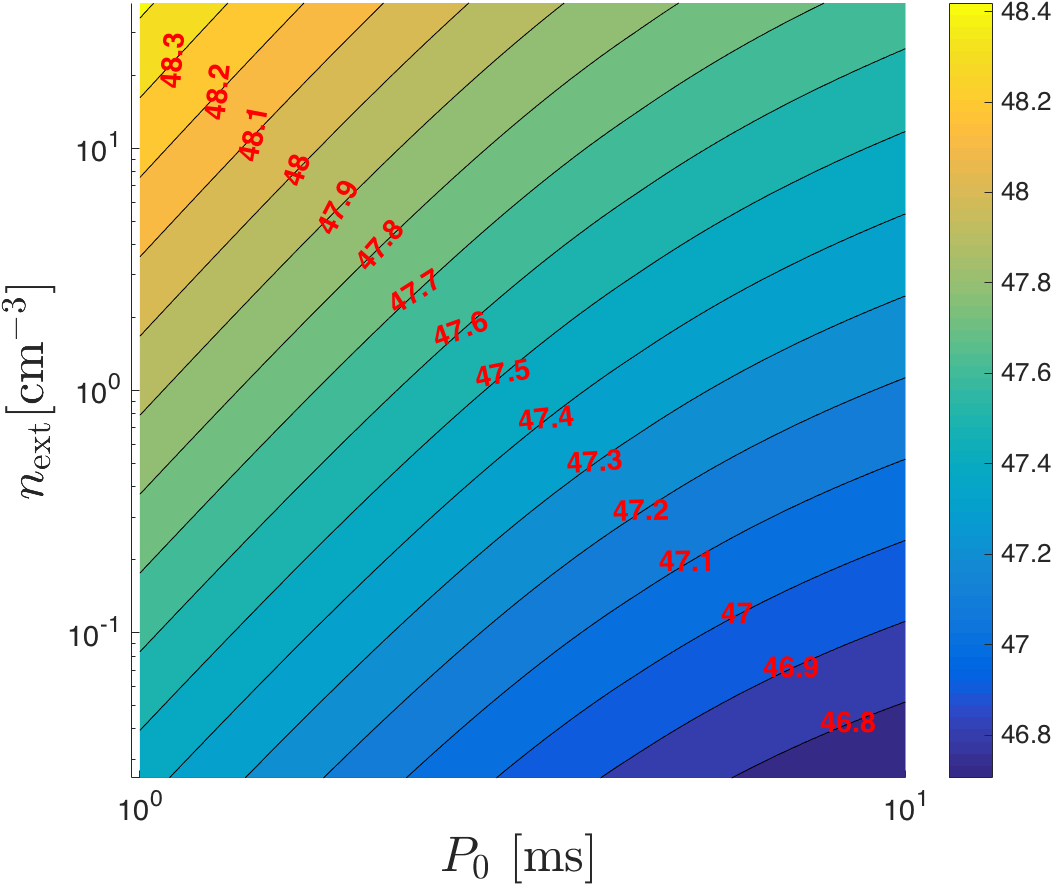}
    \caption{Contour plot of the total energy $\log_{10}(E_{\rm ST}~[{\rm erg}])$ in the MWN at time $t=t_{\rm ST}$ before 
    it is crushed by the reverse shock, for the following parameter values: $n = 3$,
    $M_{\rm ej} = 3M_\odot$, $f = 1$, $E_{\rm SN} = 10^{51}~{\rm erg}$.
    {\bf Top panel}: for 
    $n_{\rm ext} = 1~{\rm cm}^{-3}$ as a function of $B_0$ and $P_0$. The empty region in the bottom right is where 
    $t_0>0.5t_{\rm ST}$ so that our assumption of $t_0\ll t_{\rm ST}$ no longer holds. The horizontal magenta line corresponds to the inferred (current) surface dipole field (assuming it remained constant at its initial value, $B_0=B_s(t) = 1.16\times10^{14}f^{-1/2}\;$G).
    {\bf Bottom panel}: for $B_0=B_s(t) = 1.16\times10^{14}f^{-1/2}\;$G, as a function of $n_{\rm ext}$ and $P_0$.
    }
    \label{fig:EST_plot}
\end{figure}

We outline two cases in Appendix (\ref{sec:appB}) based on whether $E_0 \gtrsim E_{\rm SN}$, for which 
we choose the initial period $P_0 = 1~{\rm ms}$, or $E_0 \lesssim E_{\rm SN}$, which has $P_0 = 10~{\rm ms}$, 
while using a fiducial value for the initial surface magnetic field $B_0 = 10^{14}~{\rm G}$ in both cases, 
and give the size $R(t_i)$ of the MWN in different stages of its evolution. The self-similar expansion of the 
nebula, shown in Figure \ref{fig:Rplot}, can be generalized as 
\begin{equation}
    R(t) = R_\star\fracb{t}{t_\star}^a
    \label{eq:Rt}
\end{equation}
where $R_\star$ and $t_\star$ are the power law break points calculated in Appendix (\ref{sec:appB}) and 
summarized in Table \ref{tab:Rtparams}.
\begin{table*}
\caption{Parameters for the dynamical evolution of the MWN in Equation~(\ref{eq:Rt}) and assuming $n=3$ in two different scenarios, 
    where the magnetar injects more energy than that of the SN ($E_0 > E_{\rm SN}$) or less ($E_0 < E_{\rm SN}$). 
    Here $t_c$ is the density core crossing time, $t_0$ is the initial spin-down time of the magnetar, 
    and $t_{\rm ST}$ is the Sedov-Taylor phase onset time.}
    \centering
    \begin{tabular}{c|cccl}
        \hline\hline
         & $a$ & $R_\star~({\rm cm})$ & $t_\star~({\rm s})$ & \\
        \hline 
         & $6/5$ & $R_c = 1.4\times10^{12}P_{0,-3}^4E_{\rm SN,51}^{3/2}f^{-1}B_{14}^{-2}M_3^{-1/2}$ & 
         $t_c = 2.4\times10^3f^{-1}B_{14}^{-2}P_{0,-3}^4E_{\rm SN,51}$ & $t < t_c$ \\
        $E_0 > E_{\rm SN}$ & $3/2$ &  $R_c$ & $t_c$ & $t_c < t < t_0$ \\
         & $1$ & $R_0 = 8.3\times10^{13}f^{-1}B_{14}^{-2}P_{0,-3}M_3^{-1/2}$ & $t_0$ & 
         $t_0 < t < t_{\rm ST}$ \\
        \hline
        $E_0 < E_{\rm SN}$ & $6/5$ & $R_0 = 1.5\times10^{15}E_{\rm SN,51}^{3/10}M_3^{-1/2}f^{-1}B_{14}^{-2}P_{0,-2}^{8/5}$ & $t_0$ & $t < t_0$ \\
         & $1$ & $R_0$ & $t_0$ & $t_0 < t< t_{\rm ST}$ \\
        \hline 
    \end{tabular}
    \label{tab:Rtparams}
\end{table*}
Integration of the injected power, modified by adiabatic losses, then yields the total energy $E_{\rm ST} \equiv E(t=t_{\rm ST})$ 
in the nebula at the start of the Sedov-Taylor phase, shown in Figure \ref{fig:EST_plot}, with $\tilde t = t/t_0$,
\begin{equation}
E_{\rm ST} = \frac{2E_0}{(n+1)\tilde{t}_{\rm ST}}\times\left\{\begin{array}{ll}
    \multicolumn{2}{c}{{\rm For}~E_0 > E_{\rm SN}} \\ \\
    \frac{3}{55}\tilde{t}_c^{\,5/2}+\frac{2}{5}+\ln\tilde{t}_{\rm ST} & (m=2) \\ \\
    \frac{3}{55}\tilde{t}_c^{\,5/2}+\frac{2}{5}+
    \frac{1-\tilde{t}_{\rm ST}^{\,2-m}}{m-2} 
    & (m\neq2) \\ \\
    \multicolumn{2}{c}{{\rm For}~E_0 < E_{\rm SN}} \\ \\
    \frac{5}{11}+\ln\tilde t_{\rm ST} & (m=2) \\ \\
    \frac{5}{11}+\frac{1-\tilde{t}_{\rm ST}^{\,2-m}}{m-2} & (m\neq2)
\end{array}\right.
\label{eq:EST}
\end{equation}
Note that the two expressions for $E_0 > E_{\rm SN}$ and $E_0 < E_{\rm SN}$ are equal at $E_0 = E_{\rm SN}$ for $\tilde{t}_c = E_{\rm SN}/E_0$ 
(i.e. a factor of 1.32 smaller than in Equation~(\ref{eq:tc_over_to})),
which is therefore used when numerically evaluating $E_{\rm ST}$ in order to produce the figures.
We give approximate estimates for $E_{\rm ST}$ for the fiducial case of $m=2~(n=3)$ for the two cases, where we ignore 
the $\tilde t_c$ term as it is small when $E_0 > E_{\rm SN}$, which yields
\begin{equation}
    E_{\rm ST}=1.1\times10^{48}\fracb{\kappa_1}{5.99}f^{-1}B_{14}^{-2}M_3^{-5/6}E_{\rm tot,52.3}^{1/2}n_0^{1/3}~{\rm erg}
\end{equation}
with $\kappa_1 = \frac{2}{n+1}(2/5+\ln\tilde t_{\rm ST})$, and when $E_0 < E_{\rm SN}$
\begin{equation}
    E_{\rm ST} = 1.8\times10^{47}\fracb{\kappa_2}{4.47}f^{-1}B_{14}^{-2}M_3^{-5/6}E_{\rm SN,51}^{1/2}n_0^{1/3}~{\rm erg}
\end{equation}
with $\kappa_2 = \frac{2}{n+1}(5/11+\ln\tilde t_{\rm ST})$. Although the dependence on $P_0$ is not explicit in these estimates, 
there is a mild dependence on it through $\kappa_1$ and $\kappa_2$, as can be seen in Figure \ref{fig:EST_plot}.

\vspace{0.1cm}

\begin{figure}
    \centering
    \includegraphics[width=0.46\textwidth]{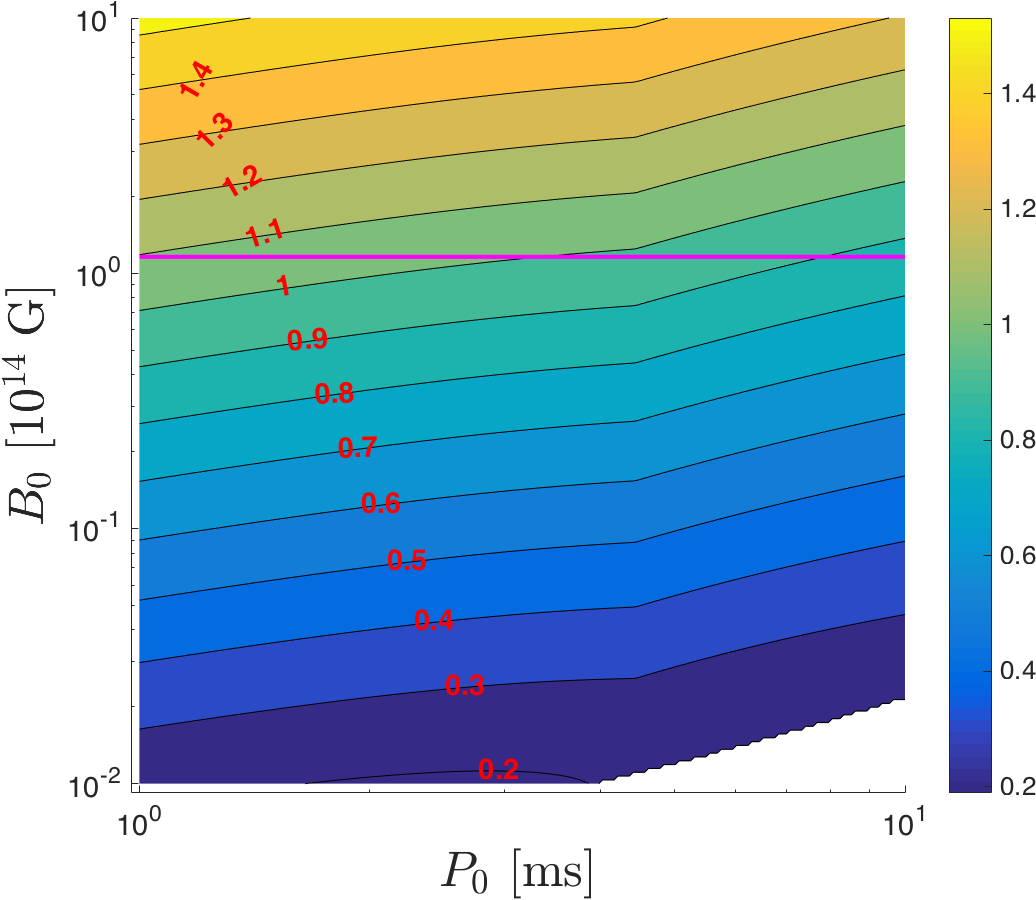}
    \vspace{0.3cm}
    \includegraphics[width=0.46\textwidth]{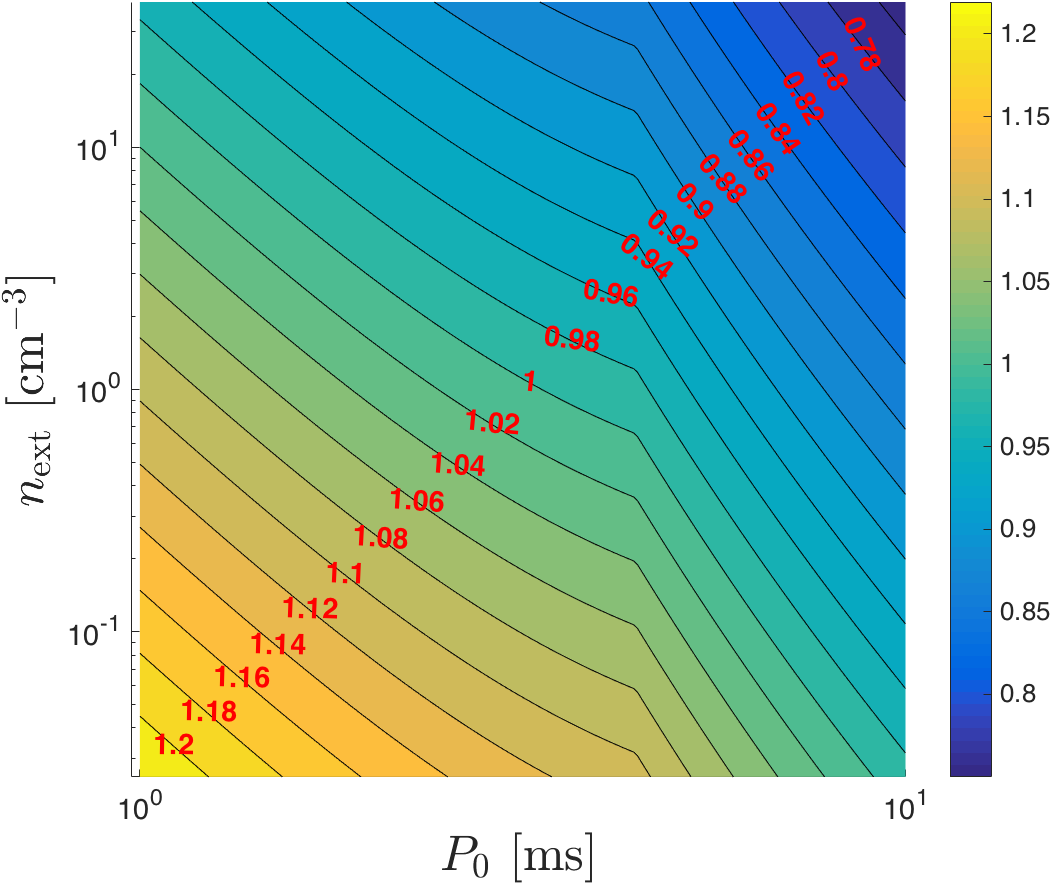}
    \caption{Contour plot of $\log_{10}(\chi)$ (where $\chi=R_i/R_f$ is given in Equation (\ref{eq:rirf})) for the 
    same parameter values as in the Figure~\ref{fig:EST_plot}. {\bf Top panel}: for $n_{\rm ext} = 1~{\rm cm}^{-3}$ 
    as a function of $B_0$ and $P_0$. The horizontal magenta line corresponds to the inferred (current) surface 
    dipole field (assuming it remained constant at its initial value, $B_0=B_s(t) = 1.16\times10^{14}f^{-1/2}\;$G).
    {\bf Bottom panel}: for $B_0=B_s(t) = 1.16\times10^{14}f^{-1/2}\;$G, as a function of $n_{\rm ext}$ and $P_0$.}
    \label{fig:rirf_plot}
\end{figure}

For $t > t_{\rm ST}$ the expansion of the SNR volume 
follows the Sedov-Taylor solution with $R_{\rm SNR} \propto t^{2/(5-k)}$. Here we have assumed a general 
density profile for the ISM $\rho_{\rm ext}(R) = AR^{-k}$ where $k = 0~(k = 2)$ applies to a constant density ISM 
(stellar wind). The onset of the Sedov-Taylor phase is marked by the equality of the swept up ISM mass, which 
starts to affect the dynamics, and that of the ejecta. It is also the approximate time when the MWN will 
interact strongly with the reverse shock and be compressed by the pressure behind the forward blast wave. 
This occurs at
\begin{eqnarray}
t_{\rm ST} &&= 519M_3^{5/6}E_{{\rm tot},51}^{-1/2}n_0^{-1/3}~{\rm yr} \\
&& = 116M_3^{5/6}E_{{\rm tot},52.3}^{-1/2}n_0^{-1/3}~{\rm yr}\nonumber
\label{eq:tst}
\end{eqnarray}
where $E_{\rm tot} = E_{\rm SN} + E_0$ is the total mechanical energy imparted to the blast wave, 
and $M_3 = M_{\rm ej}/3M_\odot$. The corresponding radius of the SNR is
\be
R_{\rm ST} = 3.07 M_3^{1/3}n_0^{-1/3}~{\rm pc}\ .
\label{eq:Rst}
\ee
In the following, we consider initial spin periods $P_0 = 10^{-3}P_{0,-3}~{\rm s}$ for which 
$E_{\rm tot} \simeq E_0 > E_{\rm SN}$, and $P_0 = 10^{-2}P_{0,-2}~{\rm s}$ with 
$E_{\rm tot} \simeq E_{\rm SN} > E_0$. The compression of the MWN by the 
reverse shock adiabatically heats the entire nebula, and the ratio of the 
final and initial energies is given by the ratio of the initial and final size of the nebula, shown in Figure \ref{fig:rirf_plot},
\begin{equation}
    \chi=\frac{E_f}{E_i} = \frac{R_i}{R_f} = \fracb{p_S}{p_i}^{1/4}~.
\end{equation}
Here the total energy in the nebula before it is crushed $E_i = E_{\rm ST}$, from Equation (\ref{eq:EST}).
The nebula compression ratio is determined by the ratio of the pressure behind the Sedov blast wave at $t = t_{\rm ST}$
\begin{equation}
    p_S \simeq 0.074\fracb{E_{\rm tot}}{R_{\rm ST}^3}~,
\end{equation}
approximated here by the central pressure \citep[e.g.][]{RC84,Shu92}, and the initial pressure in the nebula 
before it is crushed. Since $\sigma \ll 1$ throughout the MWN volume, the total energy and pressure are dominated 
by that of particles, such that the total initial pressure in the nebula is
\begin{equation}
    p_i = \frac{E_{\rm ST}}{4\pi R(t_{\rm ST})^3}\ .
\end{equation}
Thus $\chi = (0.074\times4\pi)^{1/4}(E_{\rm tot}/E_{\rm ST})^{1/4}[R(t_{\rm ST})/R_{\rm ST}]^{3/4}$. 
In the first case ($E_0 > E_{\rm SN}$), we assume that the size of the MWN $R_i \sim R_{\rm SNR} = R_{\rm ST}$ before it is 
compressed by the reverse shock, and in the second case ($E_0 < E_{\rm SN}$)
$R_i = R(t_{\rm ST}) = 6.3\times10^{18}[(n-1)/2]^{-1/5}M_3^{1/3}E_{\rm SN,51}^{-1/5}n_0^{-1/3}P_{0,-3}^{-2/5}~{\rm cm}$ from 
Equation (\ref{eq:Rt}). However, in order for $\chi$ to be continuous, we assume a slightly modified expression for 
$R_i\simeq R_{\rm ST}\times\min[1,(E_0/E_{\rm SN})^{1/5}]$, which is valid for both cases. Furthermore, taking 
$(0.074\times4\pi)^{1/4}\simeq1$, we obtain for the continuous approximation 
$\chi \simeq (E_{\rm tot}/E_{\rm ST})^{1/4}\min[1,(E_0/E_{\rm SN})^{3/20}]$, or
\begin{equation}
    \chi = \left\{\begin{array}{ll}
        \fracb{E_{\rm tot}}{E_{\rm ST}}^{1/4}\ , & E_0 > E_{\rm SN} \\
        \fracb{E_{\rm tot}}{E_{\rm ST}}^{1/4}\fracb{E_0}{E_{\rm SN}}^{3/20}\ , & 
        E_0 < E_{\rm SN}
    \end{array}\right.
    \label{eq:rirf}
\end{equation}
and when inserting in all the relevant scalings,
\begin{equation}
    \chi = \left\{\begin{array}{ll}
        11.6\,B_{14}^\frac{1}{2}M_3^\frac{5}{24}E_{\rm tot,52.3}^\frac{1}{8}\fracb{\kappa_1}{6.0f}^{-\frac{1}{4}}n_0^{-\frac{1}{12}}\ , & E_0 > E_{\rm SN} \\ \\
        6.76\,\frac{B_{14}^\frac{1}{2}M_3^\frac{5}{24}E_{\rm tot,51}^\frac{1}{8}}
        {E_{\rm SN,51}^\frac{3}{20}P_{\rm 0,-2}^\frac{3}{10}\fracb{\kappa_2}{4.47f}^\frac{1}{4}n_0^\frac{1}{12}}\ , 
        & E_0 < E_{\rm SN}
    \end{array}\right.
    \label{eq:chi2}
\end{equation}

\vspace{0.1cm}

After compression, the initial value of the nebula's equilibrium radius (around which the radius oscillates) is $R_f \simeq R_{\rm ST}(E_{\rm ST}/E_{\rm tot})^{1/4}\min[1,(E_0/E_{\rm SN}^{1/20})]$, or
\begin{equation}
    R_f = R_{\rm ST}\fracb{E_{\rm ST}}{E_{\rm tot}}^{1/4}\times\left\{\begin{array}{ll}
        1~, & E_0 > E_{\rm SN} \\
        \fracb{E_0}{E_{\rm SN}}^{1/20}~, & E_0 < E_{\rm SN}
    \end{array}\right.
    \label{eq:Rf}
\end{equation}
and with numerical values,
\begin{equation}
    R_f = \left\{\begin{array}{ll}
        \frac{0.26\;{\rm pc}}{\sqrt{B_{14}}}\fracb{\kappa_1}{6.0fn_0}^\frac{1}{4}\fracb{M_3}{E_{\rm tot,52.3}}^\frac{1}{8}\ , & E_0 > E_{\rm SN} \\ \\
        \frac{0.33\;{\rm pc}}{B_{14}^\frac{1}{2}E_{\rm SN,51}^\frac{1}{20}P_{\rm 0,-2}^\frac{1}{10}}\fracb{\kappa_2}{4.5fn_0}^\frac{1}{4}\fracb{M_3}{E_{\rm tot,51}}^\frac{1}{8}\ , & E_0 < E_{\rm SN}
    \end{array}\right.
\end{equation}

\begin{figure}
    \centering
    \includegraphics[width=0.46\textwidth]{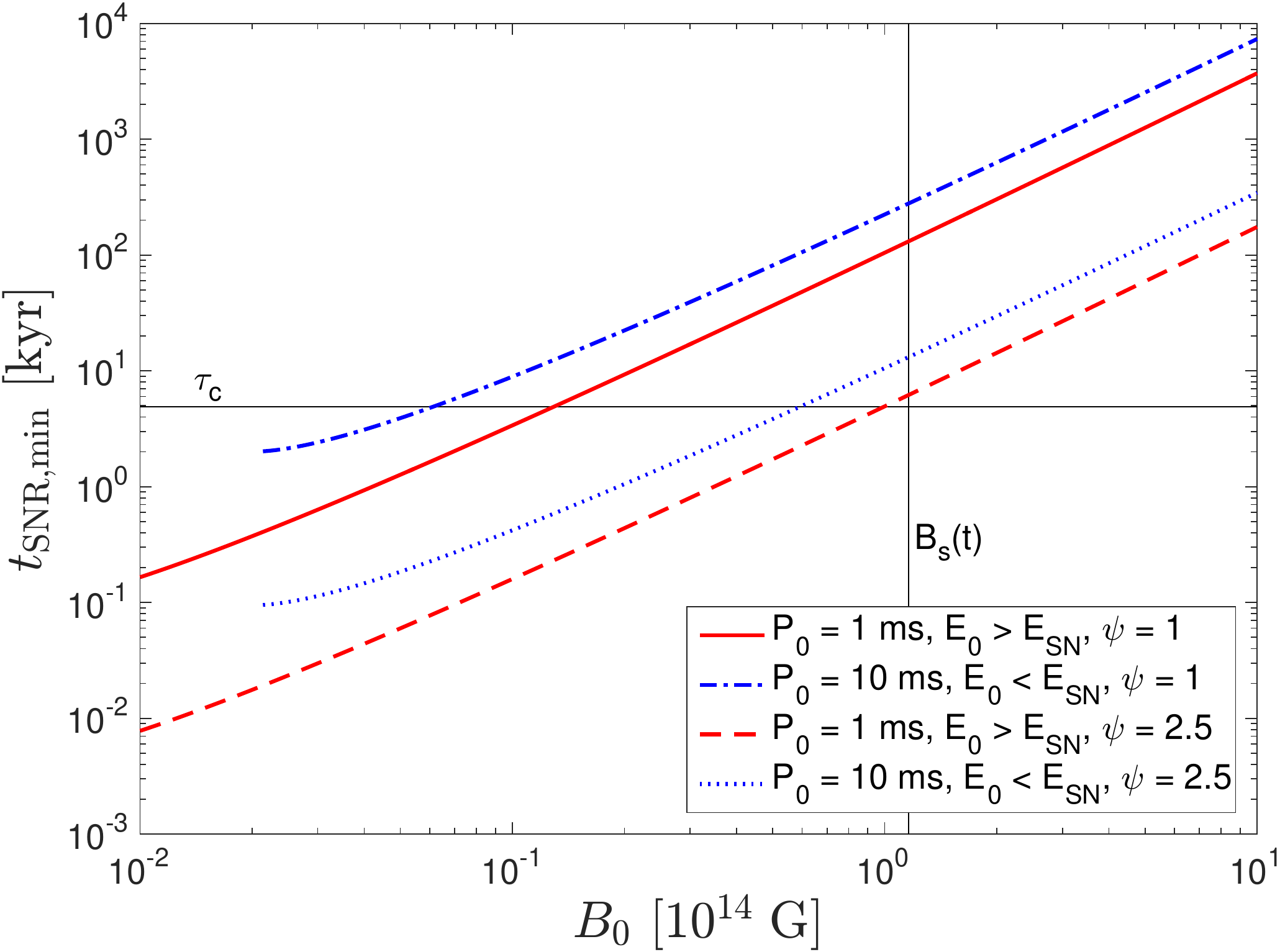}
    \caption{Minimum age of the system if the MWN is at least as large as the X-ray nebula, $R\geq R_X = 2.04d_4~{\rm pc}$, shown 
    as a function of the initial surface dipole magnetic field strength, for two dynamical scenarios and two values of $\psi$. For a fixed $B_0$ value, the age estimate depends only weakly on $P_0$. All parameters (except $\psi$) are the same as given in the caption of Figure \ref{fig:EST_plot}. The thin solid horizontal line indicates the value of the characteristic spin-down age. The values of the characteristic spin-down age $\tau_c$ and the current inferred surface dipole filed $B_s(t)$ are indicated, by the think black horizontal and vertical lines, respectively.}
    \label{fig:txB0}
\end{figure}
\vspace{0.1cm}

In reality, and as seen in hydrodynamic simulations 
\citep[e.g.][]{BCF01,Swaluw+01}, the compressed nebula goes through 
a reverberation phase where its size oscillates before achieving pressure equilibrium. Here we assume that the MWN 
is compressed to $R_f$ and then it begins its slow expansion outwards. 
Pressure equilibrium dictates that the MWN grows as $R(t) \propto p_S^{-1/4}$, where the internal pressure in the Sedov phase 
scales with time as $p_S \propto t^{-6/(5-k)}$, which yields
\begin{equation}
    R(t) = \psi(t) R_{\rm eq}(t) = \psi(t) R_f\fracb{t}{t_{\rm ST}}^{3/2(5-k)}\ ,
    \label{eq:Rt_ad}
\end{equation}
where we have introduced the parameter $\psi=2\psi_2$ below in order to account for the (order unity fractional) oscillations 
of $R$ around its equilibrium value $R_{\rm eq}$ after the crushing \citep{BCF01,Swaluw+01}. A fiducial value of 
$\psi=2$ is used since during the order unity fractional fluctuations in $R$ it spends most of its time near its 
maximal value (within a single oscillation), making $\psi$ close to its own maximal value more likely at a given 
snapshot of the system (corresponding to the present epoch). The current size of the MWN $R(t_{\rm SNR})$ must be 
larger than that of the X-ray bright region $R_X = 2.04d_4~{\rm pc}$, which yields an independent constraint on 
the age of the system,
\begin{equation}
    t_{\rm SNR} \gtrsim t_{\rm ST}\fracb{R_X}{\psi(t_{\rm SNR}) R_f}^{2(5-k)/3}\ .
\end{equation}
which equates to
\begin{equation}
    t_{\rm SNR} \gtrsim \left\{
    \begin{array}{ll}
        \frac{10.3\,{\rm kyr}\;d_4^{\frac{10}{3}}B_{14}^\frac{5}{3}M_3^\frac{5}{12}n_0^{\frac{1}{2}}}
        {E_{\rm tot,52.3}^\frac{1}{12}\fracb{\kappa_1}{6.0f}^{\frac{5}{6}}\psi_2^\frac{10}{3}}
        ~, & E_0 > E_{\rm SN} \\ \\
        \frac{22.2\,{\rm kyr}\;B_{14}^\frac{5}{3}M_3^\frac{5}{12}E_{\rm SN,51}^\frac{1}{6}d_4^{\frac{10}{3}}P_{0,-2}^{\frac{1}{3}}n_0^{\frac{1}{2}}}
        {E_{\rm tot,51}^\frac{1}{12}\fracb{\kappa_2}{4.47f}^{\frac{5}{6}}\psi_2^\frac{10}{3}}
        ~, & E_0 < E_{\rm SN} 
    \end{array}
    \right.
\end{equation}
This constraint is shown in Figure~\ref{fig:txB0} as a function of the initial magnetic field strength, for our two scenarios as well as for $\psi = 1$ and $\psi=2$. 

The total energy is then reduced adiabatically as the MWN re-expands, and if $t_0 \ll t_{\rm ST}$ then the energy injection from the magnetar can be ignored, such that
\begin{equation}
    E(R) = \fracb{R_f}{R}\chi E_{\rm ST}~.
    \label{eq:ER}
\end{equation}
On the other hand, if the magnetar continues to inject energy at late times then the total energy in the MWN 
at $t>t_{\rm ST}$ is given by
\begin{equation}\label{eq:Et}
    E(t) = \frac{\chi E_{\rm ST}}{\hat{t}^{a}\psi(t)}  + \frac{E_0\tilde{t}_{\rm ST}^{\,1-m}}{\hat{t}^{a}\bar{\psi}(t)}\times
    \left\{\begin{array}{ll}
     \frac{1-\hat{t}^{\,1+a-m}}{m-a-1}~, & m\neq1+a\ , \\ \\
     \ln\hat{t}~, & m=1+a\ ,
  \end{array}\right.
\end{equation}
where $a=3/[2(5-k)]$, $\hat{t}\equiv t/t_{\rm ST}$ and
\begin{eqnarray}
    \bar{\psi}(t) &=& \frac{\psi(t)\int_{t_{\rm ST}}^{t}dt_it_i^{a-m}}{\int_{t_{\rm ST}}^{t}dt_it_i^{a-m}\psi(t_i)} 
    \\ \nonumber
    &=&
    \psi(t)\frac{1-\hat{t}^{\,1+a-m}}{m-a-1}\left[\int_{1}^{\hat{t}}d\hat{t}_i\hat{t}_i^{a-m}\psi(\hat{t}_i)\right]^{-1}\ ,
\end{eqnarray}
is the effective weighted mean of the fractional change in $\psi$ between $t_{\rm ST}$ and $t$, which takes into account that the electrons are injected at $R(t_i)$ rather than $R_{\rm eq}(t_i)$ and therefore experience different adiabatic cooling (or heating) as the MWN evolves. The second term, which represents the energy injected by the wind at $t>t_{\rm ST}$
becomes particularly important for $m<1+a$ i.e. $m<1.3$ for $k=0$, in which case $E_{\rm ST}\simeq E_0\hat{t}_{\rm ST}^{1-m}/(2-m)$ and this term
eventually dominates at late times, $\hat{t}=t/t_{\rm ST}>[(1+a-m)\chi/(2-m)]^{1/(1+a-m)}$. Nevertheless, for $m>1.3$ and in particular for $m=2$
that corresponds to $n=3$ expected from magnetic dipole braking, the second term is not very imprtant and can usually be neglected. For $m=2$ and $n=3$,
the first term is larger than the second term by at least a factor of $0.7\chi\kappa$ where $\chi\sim5.7-9.8$ and $\kappa\sim8.9-12$ so that the second term
would at most introduce a small correction of a few percent, and one can neglect it to obtain a convenient expression,
\begin{equation}
    \frac{E(R_X)}{10^{48}\,{\rm erg}} \approx \left\{\begin{array}{ll}
        \frac{1.66}{B_{14}^2}\fracb{\kappa_1}{6.0fd_4}\sqrt{\frac{E_{\rm tot,52.3}}{M_3}}\ , & E_0 > E_{\rm SN} \\ \\
        \frac{0.200}{B_{14}^2}\frac{\fracb{\kappa_2}{4.47}}{fd_4E_{\rm SN,51}^\frac{1}{5}P_{\rm 0,-2}^\frac{2}{5}}
        \sqrt{\frac{E_{\rm tot,51}}{M_3}}\ , & E_0 < E_{\rm SN}
    \end{array}\right.
    \label{eq:Et1}
\end{equation}

\begin{figure}
    \centering
    \includegraphics[width=0.46\textwidth]{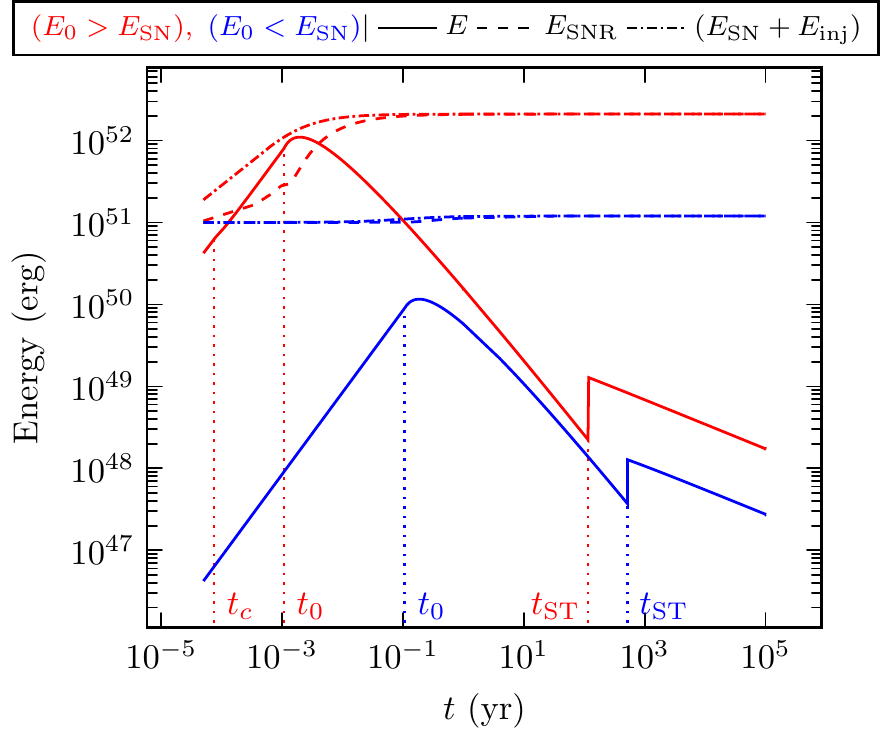}
    \caption{Total energy in the MWN as a function of its time.
    Two cases are shown: ($E_0 > E_{\rm SN}$, with $E_0 = 2\times10^{52}~{\rm erg}$, $P_0=1\;$ms, 
    and $B_{14} = 1$, solid red line) the magnetar injects more 
    energy than the kinetic energy of the SN ejecta, thus seriously affecting the dynamical evolution of the SNR + MWN system, and 
    ($E_0 < E_{\rm SN}$, with $E_0 = 2\times10^{50}~{\rm erg}$, $P_0=10\;$ms, and $B_{14} = 1$, solid blue line) for which the 
    magnetar is dynamically unimportant and the evolution of the 
    SNR + MWN system mirrors that of canonical PWNe observed around regular radio pulsars. The dashed lines show the 
    energy in the SNR from Equation (\ref{eq:Esnr}) and the dotted lines show the total energy in the system. 
    Following parameter values are assumed: $\psi(t) = 1$, $n = 3$, $n_{\rm ext} = 1~{\rm cm}^{-3}$, 
    $M_{\rm ej} = 3M_\odot$, $f = 1$, $E_{\rm SN} = 10^{51}~{\rm erg}$, and $k = 0$.}
    \label{fig:Etotneb_plot}
\end{figure}

\begin{figure*}
    \centering
    \includegraphics[width=0.85\textwidth]{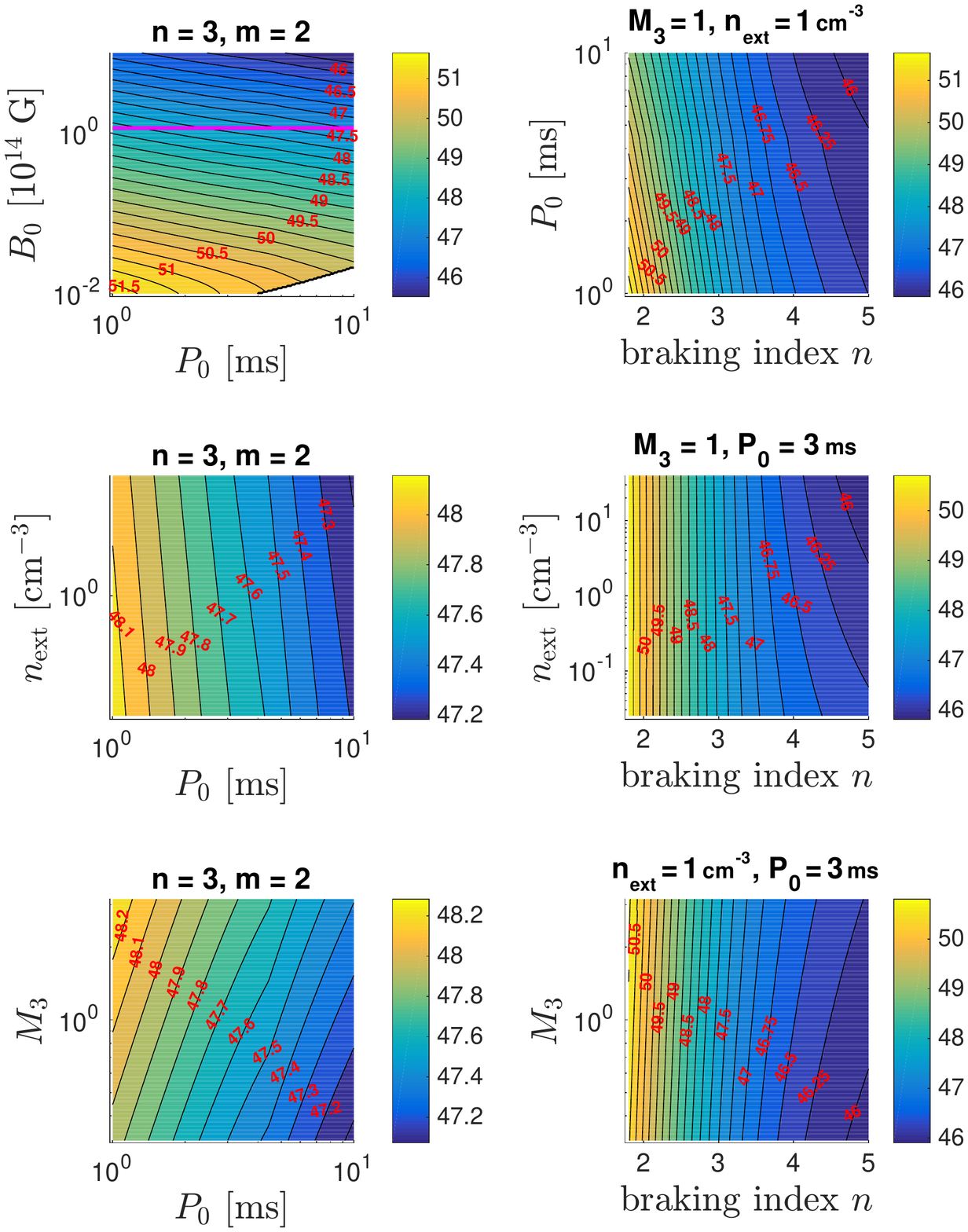}
    \caption{Contour plots of the log of the maximum total energy in the MWN,  $\log_{10}(E~[{\rm erg}])$ from Equation (\ref{eq:Et}), under the condition that $R_{\rm MWN}\geq R_X$. The {\bf left panels} are for magnetic dipole spin-down ($n=3$ and $m=2$), and show the dependence on the following parameters: {top panel} -- the initial spin period $P_0$ and the equatorial surface dipole magnetic field $B_0$, where the horizontal magenta line corresponds to its inferred (current) value (which for $n=3$ also corresponds to its initial value, $B_0=B_s(t) = 1.16\times10^{14}f^{-1/2}\;$G, which is also used in the two panels below this one); {\bf middle panel} -- $P_0$ and the external number density $n_{\rm ext}$; {\bf bottom panel} -- $P_0$ and the $M_3=M_{\rm ej}/3\,M_\odot$. The {\bf right panels} show the dependence on the braking index $n$ (defined by Eq.~(\ref{eq:ndef})), as well as $P_0$ ({\bf top panel}), $n_{\rm ext}$ ({\bf middle panel}), and $M_3$ ({\bf bottom panel}).
    The remaining default parameter values used here are 
    $n_{\rm ext} = 1~{\rm cm}^{-3}$, $M_{\rm ej} = 3M_\odot$, $P_0=3\;$ms, $E_{\rm SN} = 10^{51}~{\rm erg}$).}
    \label{fig:ER_plot}
\end{figure*}

The complete temporal evolution of the total energy in the MWN is shown in Figure 
\ref{fig:Etotneb_plot} for $\psi(t) = 1$; Accurate treatment for the dynamical evolution of $\psi(t)$ 
during the reverberation phase of the nebula at $t \gtrsim t_{\rm ST}$ is out of the scope of this work. 
In Figure \ref{fig:ER_plot}, the maximum 
energy in the MWN when its size is at least as large as the X-ray nebula is shown 
as a function of various parameters.

\vspace{0.3cm}

\section{MWN Internal Structure \& the Synchrotron Cooling Length}\label{sec:Lcool}
An electron injected with initial energy $\gamma_im_ec^2$ at the termination shock at time $t_i$ cools due to 
the adiabatic expansion of the flow and by emitting synchrotron radiation, as governed by
\begin{equation}
\frac{d\gamma_e}{dt} = -\frac{a}{t}\gamma_e -bB^2(t)\gamma_e^2\ ,
\label{eq:gamma_t}
\end{equation}
for $R\propto t^a$ as defined in Eq. (\ref{eq:Rt}), and where $b = \sigma_T/6\pi m_ec$. We assume that the magnetic field 
is spatially homogeneous but varies temporally due to the injection of magnetic energy by the central source and 
its adiabatic evolution driven by the expansion of the SNR volume,
\begin{equation}\label{eq:B(t)}
B(t) \approx \sqrt{\frac{6E_B(t)}{R^3(t)}}\ ,
\end{equation}
where the expression we have used for the MWN's volume is valid as long as its radius satisfies $R(t) \gg R_{\rm TS}$. 
The magnetic field strength when $R(t)=R_X$ is
\begin{equation}
    \frac{B(R_X)}{1\,\mu{\rm G}} \approx \left\{\begin{array}{ll}
        \frac{11.3}{B_{14}d_4^2}\sqrt{\frac{\kappa_1\sigma_{-2.5}}{6.0f(1+\sigma)}}
        \frac{E_{\rm tot,52.3}^\frac{1}{4}}{M_3^\frac{1}{4}}\ , & E_0 > E_{\rm SN} \\ \\
        \frac{3.92}{B_{14}d_4^2}\frac{\sqrt{\frac{\kappa_2\sigma_{-2.5}}{4.47f(1+\sigma)}}E_{\rm tot,51}^\frac{1}{4}}
        {E_{\rm SN,51}^\frac{1}{10}P_{\rm 0,-2}^\frac{1}{5}M_3^\frac{1}{4}}\ , & E_0 < E_{\rm SN}
    \end{array}\right.
    \label{eq:Bx}
\end{equation}
One can express the field as $B(t) = [\sigma/(1+\sigma)]^{1/2}B_{\rm max}(t)$, 
in terms of the maximal value of the field corresponding to all of the MWN energy
being in magnetic form ($\sigma\to\infty$),
\begin{equation}
    \frac{B_{\rm max}(R_X)}{1\,\mu{\rm G}} \approx \left\{\begin{array}{ll}
        \frac{201}{B_{14}d_4^2}\sqrt{\frac{\kappa_1}{6.0f}}
        \frac{E_{\rm tot,52.3}^\frac{1}{4}}{M_3^\frac{1}{4}}\ , & E_0 > E_{\rm SN} \\ \\
        \frac{69.6}{B_{14}d_4^2}\frac{\sqrt{\frac{\kappa_2}{4.47f}}E_{\rm tot,51}^\frac{1}{4}}
        {E_{\rm SN,51}^\frac{1}{10}P_{\rm 0,-2}^\frac{1}{5}M_3^\frac{1}{4}}\ , & E_0 < E_{\rm SN}
    \end{array}\right.
    \label{eq:Bx_max}
\end{equation}

\begin{figure}
    \centering
    \includegraphics{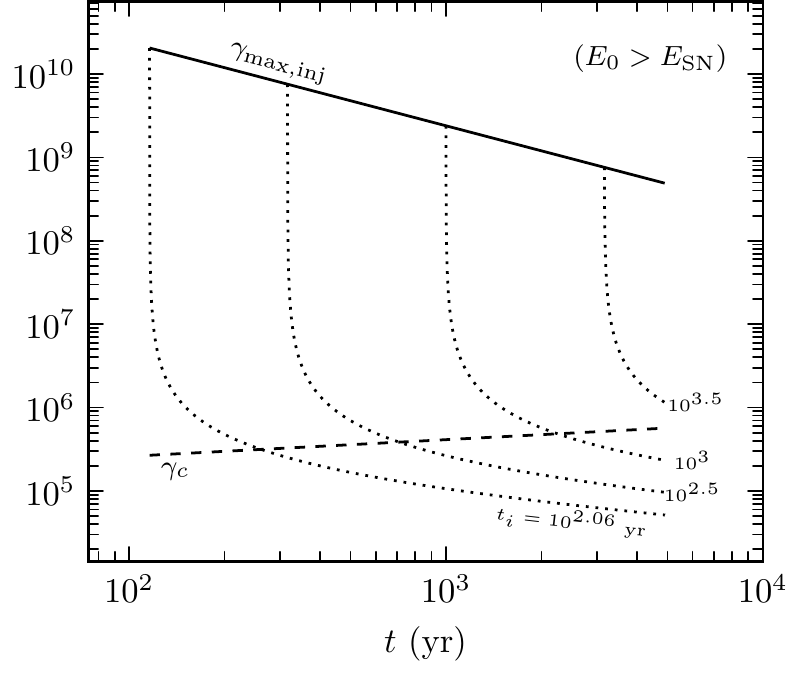}
    \vspace{0.2cm}\\
    \includegraphics{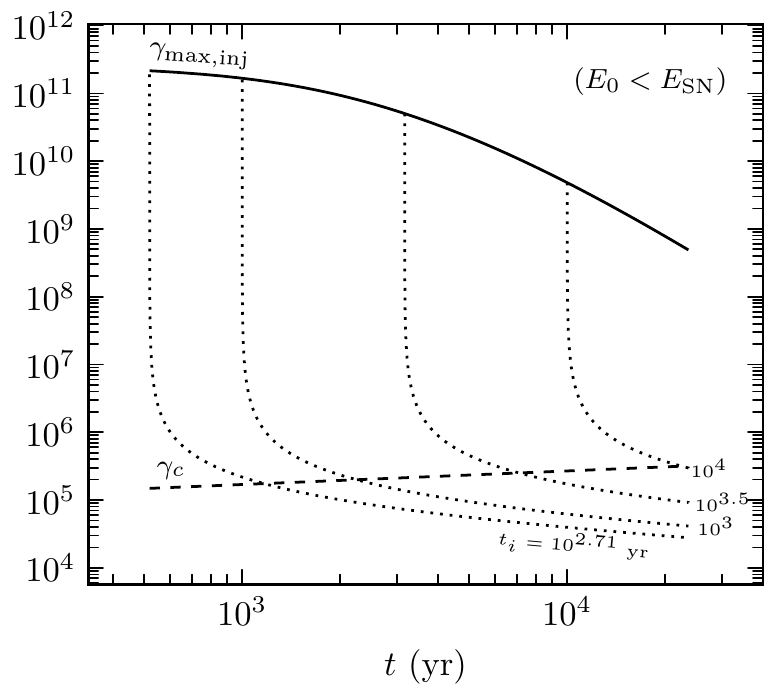}
    \caption{Temporal evolution of maximum electron Lorentz factor. $\gamma_{\rm max, inj}$ (solid) 
    is the maximum electron Lorentz factor injected at time $t_i$ at the termination shock 
    radius $R_{\rm TS,p}$. Its magnitude is governed by the maximum polar cap voltage 
    $V_0(t_i)$ which in turn depends on the surface magnetic field $B_s(t_i)$ and the 
    spin period $P(t_i)$, as given in Equation~(\ref{eq:gmax}). The dotted lines show 
    the cooling evolution of maximum energy electrons as a function of the dynamical 
    time up to the current time. $\gamma_c$ is the Lorentz factor of electrons that are 
    cooling at the dynamical time. (\textit{Top}):The magnetar is born as a fast rotator 
    with $P_0 = 1~{\rm ms}$ and high surface magnetic field $B_0 = 1.16\times10^{14}~{\rm G}$ 
    which remains constant over the lifetime of the system $t_{\rm SNR} = \tau_c = 4.9~{\rm kyr}$. 
    (\textit{Bottom}): The magnetar is born as a typical radio pulsar with $P_0 = 10~{\rm ms}$ 
    and surface field $B_0 = 10^{12}~{\rm G}$ but experiences field growth so that its current 
    field is $B_s = 1.16\times10^{14}~{\rm G}$ as inferred from its $P$ and $\dot P$ at 
    its model age $t = t_{\rm SNR} = 23.6~{\rm kyr}$. In both cases, the nebular field 
    strength at $t = t_{\rm SNR}$ is $B = 15\mu{\rm G}$, for which 
    $\sigma = 5.6\times10^{-3}$ when $E_0 > E_{\rm SN}$ and $\sigma = 4.7\times10^{-2}$ 
    when $E_0 < E_{\rm SN}$. See Appendix(\ref{sec:appA} \& \ref{sec:appC}) for more details.}
    \label{fig:gcool_plot}
\end{figure}

We solve Equation (\ref{eq:gamma_t}) in Appendix (\ref{sec:appC}) and show the temporal 
evolution of the maximum electron Lorentz factor $\gamma_{\rm max,inj}$ in Figure \ref{fig:gcool_plot} for 
both $E_0 > E_{\rm SN}$ and $E_0 < E_{\rm SN}$ cases. In the first case, we assume that 
the magnetar was born as a fast rotator with $P_0 \sim 1~{\rm ms}$ and surface magnetic 
field as inferred from the measured $P$ and $\dot P$, $B_0 \sim 1.16\times10^{14}~{\rm G}$
(for $f=1$), which remains constant over the age of the system 
$t_{\rm SNR} = \tau_c = 4.9~{\rm kyr}$. Alternatively, in the second case, we look at a field growth 
scenario where the nascent NS had a spin period $P_0\sim10~{\rm ms}$ and surface field 
$B_0 \sim 10^{12}~{\rm G}$ which then grows to the same strength as in the previous case over 
$t_{\rm SNR} = 26.3~{\rm kyr}$ (see Eq. (\ref{eq:tSNR}) for choice of age). Electrons injected 
at earlier times cool rapidly to lower Lorentz factors as they are in the fast-cooling regime when 
their $\gamma_e(t) > \gamma_c(t)=6\pi m_ec/\sigma_T B(t)^2t$, where $\gamma_c(t)$ is the Lorentz factor 
of electrons that are cooling at the dynamical time $t$. When $R(t)=R_X$ then (neglecting the oscillations)
\begin{equation}
    \frac{\gamma_c(R_X)}{10^6}
    \approx \left\{\begin{array}{ll}
        1.85\,\frac{(1+\sigma)}{\sigma_{-2.5}}\frac{B_{14}^\frac{1}{3}d_4^\frac{2}{3}{M_3}^\frac{1}{12}}
        {\fracb{\kappa_1}{6.0f}^\frac{1}{6}n_0^\frac{1}{2}E_{\rm tot,52.3}^\frac{5}{12}}\ , & E_0 > E_{\rm SN} \\ \\
        7.14\,\frac{(1+\sigma)}{\sigma_{-2.5}}\frac{B_{14}^\frac{1}{3}d_4^\frac{2}{3}E_{\rm SN,51}^\frac{1}{30}{M_3}^\frac{1}{12}P_{\rm 0,-2}^\frac{1}{15}}
        {\fracb{\kappa_1}{4.47f}^\frac{1}{6}n_0^\frac{1}{2}E_{\rm tot,51}^\frac{5}{12}}\ , & E_0 < E_{\rm SN}
    \end{array}\right.
    \label{eq:gamma_cx}
\end{equation}
is the Lorentz factor of electrons that are cooling at the dynamical time. However, majority 
of the particles in the MWN were injected at earlier times but they now have 
$\gamma_e(t) < \gamma_c(t)$, and thus are slow-cooling. Yet, freshly injected electrons over 
the last dynamical time, that are contributing to the X-ray emission, are presently fast-cooling.
\vspace{0.8cm}
\subsection{Quasi-steady state: the inner region}\label{sec:qss-nebula}
Assuming that the particle's overall motion is dominated by advection, the distance $r$ it travels over the time interval 
$\Delta t = t-t_i$ can be used to establish the cooling length of the X-ray emitting electrons. The flow speed downstream 
of the termination shock is governed by the magnetization of the wind. Observations of many PWNe 
suggest that the magnetization of the shocked pulsar wind is small, $\sigma \ll 1$, suggesting that the magnetic 
field is not dynamically important in the inner regions of the nebula where particle pressure dominates over magnetic 
pressure. Therefore, we use the hydrodynamic equations to describe the nebula's dynamics. The shocked wind in the nebula
is relativistically hot, with its energy density $e$, pressure $p$, and proper rest mass density $\rho$
satisfying $p=e/3\gg\rho c^2$. For such a relativistically hot fluid of proper number density $\tilde{n}$ 
 one obtains under spherical symmetry \citep{BM76},
\begin{eqnarray}\nonumber
    \frac{\partial}{\partial t}(\tilde{n}\gamma)+\frac{c}{r^2}\frac{\partial}{\partial r}(r^2\tilde{n}\gamma\beta) &=& 0\ ,\\ 
    \frac{d}{dt}\fracb{p}{\tilde{n}^{4/3}} &=& 0\ ,\\ \nonumber
    \frac{d}{dt}(p\gamma^4)&=&\gamma^2\frac{\partial p}{\partial t}\ ,
\end{eqnarray}
where $\frac{d}{dt} = \frac{\partial}{\partial t}+\beta c\frac{\partial}{\partial r}$. Approximating the flow in the nebula as a steady state, 
we have $\frac{\partial}{\partial t}\to 0$ and $\frac{d}{dt}\to\beta c\frac{d}{dr}$ so that
\begin{equation}
    \frac{d}{dr}(r^2\tilde{n}\gamma\beta) = 0\ ,\quad
    \frac{d}{dr}\fracb{p}{\tilde{n}^{4/3}} = 0\ ,\quad
    \frac{d}{dr}(p\gamma^4) = 0\ ,
\end{equation}
which imply that $r^2\tilde{n}\gamma\beta$, $\tilde{n}^{-4/3}p$, and $p\gamma^4$ are all uniform (i.e. independent of $r$) within the nebula.
Since the flow in the nebula is to a good approximation Newtonian ($\beta\lesssim 1/3$) we have $\gamma^4\approx 1$ and therefore
an approximately uniform pressure, $p\approx\;$const, and density, $\tilde{n}\approx\;$const. Finally, $r^2\beta\approx\;$const is also uniform, 
implying a velocity profile,
\begin{equation}
\beta(r) = \beta_{\rm TS}\fracb{r}{R_{\rm TS}}^{-2}\ ,
\label{eq:beta_r}
\end{equation}
where the normalization is obtained by matching the flow speed at the termination shock. 

In the low-$\sigma$ limit, MHD shock jump conditions for orthogonal shocks dictate that 
$\beta_{\rm TS} \approx \frac{1}{3}\sqrt{1+8\sigma}$ just downstream of the termination shock \citep{Kennel_Coroniti_1984}.
Furthermore, from magnetic flux conservation,
\begin{equation}
    \frac{d}{dr}(r\beta B) = 0\ ,
\end{equation}
where the magnetic field is assumed to be purely toroidal, its strength initially increases $B(r)\propto r$. 
This would suggest that the magnetic pressure dominates 
at large distances from the termination shock, such that the flow speed asymptotes to the terminal speed 
given by $\beta_\infty \approx \sigma(1+\sigma)^{-1}$. However, this is a serious artefact of the ideal MHD spherically 
symmetric model and has been shown to contradict observations and 3D numerical simulations. In fact, 3D numerical relativistic 
MHD simulations of \citet[][\textcolor{black}{see for e.g. their Fig. 5}]{Porth+2013} show that PWNe are mostly isobaric and filled with low magnetization \textcolor{black}{($10^{-2}\lesssim\sigma\lesssim10^{-1}$)} plasmas with particle 
pressure governing the bubble dynamics. Their simulations show that the magnetic field is efficiently dissipated due to its 
randomization throughout the volume of the PWN which significantly reduces the magnetization of the flow.

Self-consistently assuming a steady state with Equation~(\ref{eq:beta_r}) with $R_{\rm TS}=R_{\rm TS}(t_i)$ and $\beta_{\rm TS}=1/3$, which upon integration over time yields
\begin{equation}\label{eq:r1}
    r_{\rm adv}(t_i,t) = R_{\rm TS}(t_i)\left[1+\frac{c\Delta t}{R_{\rm TS}(t_i)}\right]^{1/3}\ ,
\end{equation}
where $\Delta t\equiv t-t_i$. For $r\gg R_{\rm TS}(t_i)$ this leads to a simple analytic scaling for the advection length and time,
$r_{\rm adv}\approx (R_{\rm TS}^2c\Delta t)^{1/3}\propto (\Delta t)^{1/3}$ and $\Delta t_{\rm adv}\approx r^3/cR_{\rm TS}^2\propto r^3$.

\subsection{The Non-steady outer region}\label{sec:non-steady-nebula}
The velocity boundary condition $\beta(r) = \beta_{\rm MWN}$ at $r = R(t)$, from Equation (\ref{eq:Rt}), gives the 
time evolution of the termination shock radius,
\begin{equation}
    R_{{\rm TS},v}(t) = \sqrt{\frac{a R(t)^3}{\beta_{\rm TS}ct}}\ .
\end{equation}
However, this heavily relies on the steady state assumption which is not valid throughout the nebula. 
A more realistic estimate for $R_{\rm TS}$ may be obtained through replacing
the steady-state assumption by a balance at $R_{\rm TS}$ between the winds ram pressure, $L_{\rm sd}/4\pi R_{\rm TS}^2c$, 
and the almost uniform pressure in the MWN, $p = e/3 = E(t)/3V(t) = E(t)/4\pi R(t)^3$, which yields
\begin{eqnarray}\label{eq:R_TSp}
    R_{{\rm TS},p}(t) &=& \sqrt{\frac{R(t)^3L_{\rm sd}(t)}{cE(t)}} \propto \psi(t)^2t^{(4a-m)/2}\\ \nonumber
                      &\approx& \left\{\begin{array}{ll}
    \frac{3.28\times10^{-3}\,{\rm pc}\;B_{14}d_4^2M_3^\frac{1}{4}}{\sqrt{\frac{\kappa_1}{6.0f}}E_{\rm tot,52.3}^\frac{1}{4}}\ , & E_0 > E_{\rm SN} \\ \\ \nonumber
    \frac{9.44\times10^{-3}\,{\rm pc}\;B_{14}d_4^2M_3^\frac{1}{4}E_{\rm SN,51}^\frac{1}{10}P_{\rm 0,-2}^\frac{1}{5}}
    {\sqrt{\frac{\kappa_2}{4.47f}}E_{\rm tot,51}^\frac{1}{4}}\ , & E_0 < E_{\rm SN}\end{array}\right.
\end{eqnarray}
where $R\propto 1/E\propto t^a$ and $L_{\rm sd}\propto t^{-m}$ was used to derive its scaling with the age 
of the system, $t$, and the numerical estimates are for $R=R_X$ and $a=0.3$.
The square of the ratio of these two estimates at equilibrium for this radius is
\begin{eqnarray}
    \fracb{R_{{\rm TS},p}(t)}{R_{{\rm TS},v}(t)}^2_{\rm eq} = \frac{tL_{\rm sd}(t)}{3aE(t)}\propto t^{1+a-m}\quad\quad\quad\quad\quad\quad\quad\quad\quad&\\ \nonumber
    \approx \left\{\begin{array}{ll}
    \frac{0.0452\,B_{14}^\frac{11}{3}d_4^\frac{13}{3}M_3^\frac{11}{12}n_0^\frac{1}{2}}
    {\fracb{\kappa_1}{6.0f}^\frac{11}{6}E_{\rm tot,52.3}^\frac{7}{12}}\ , & E_0 > E_{\rm SN} \\ \\ \nonumber
    \frac{0.806\,B_{14}^\frac{11}{3}d_4^\frac{13}{3}M_3^\frac{11}{12}E_{\rm SN,51}^\frac{11}{30}n_0^\frac{1}{2}P_{\rm 0,-2}^\frac{11}{15}}
    {\fracb{\kappa_2}{4.47f}^\frac{11}{6}E_{\rm tot,51}^\frac{7}{12}}\ , & E_0 < E_{\rm SN}\end{array}\right.
\end{eqnarray}
This shows that the approximation behind $R_{{\rm TS},v}$ is reasonable if the current energy
in the MWN, $E(t)$, is dominated by the energy injected by the magnetar's spin-down wind by the last dynamical time, $\sim L_{\rm sd}t$, 
but breaks down significantly when the MWN energy is dominated by energy injected at much earlier times that even after suffering 
adiabatic losses significantly exceeds the energy injected in the last dynamical time, $E(t)\gg tL_{\rm sd}(t)$.

\begin{figure}
    \includegraphics[width=\linewidth]{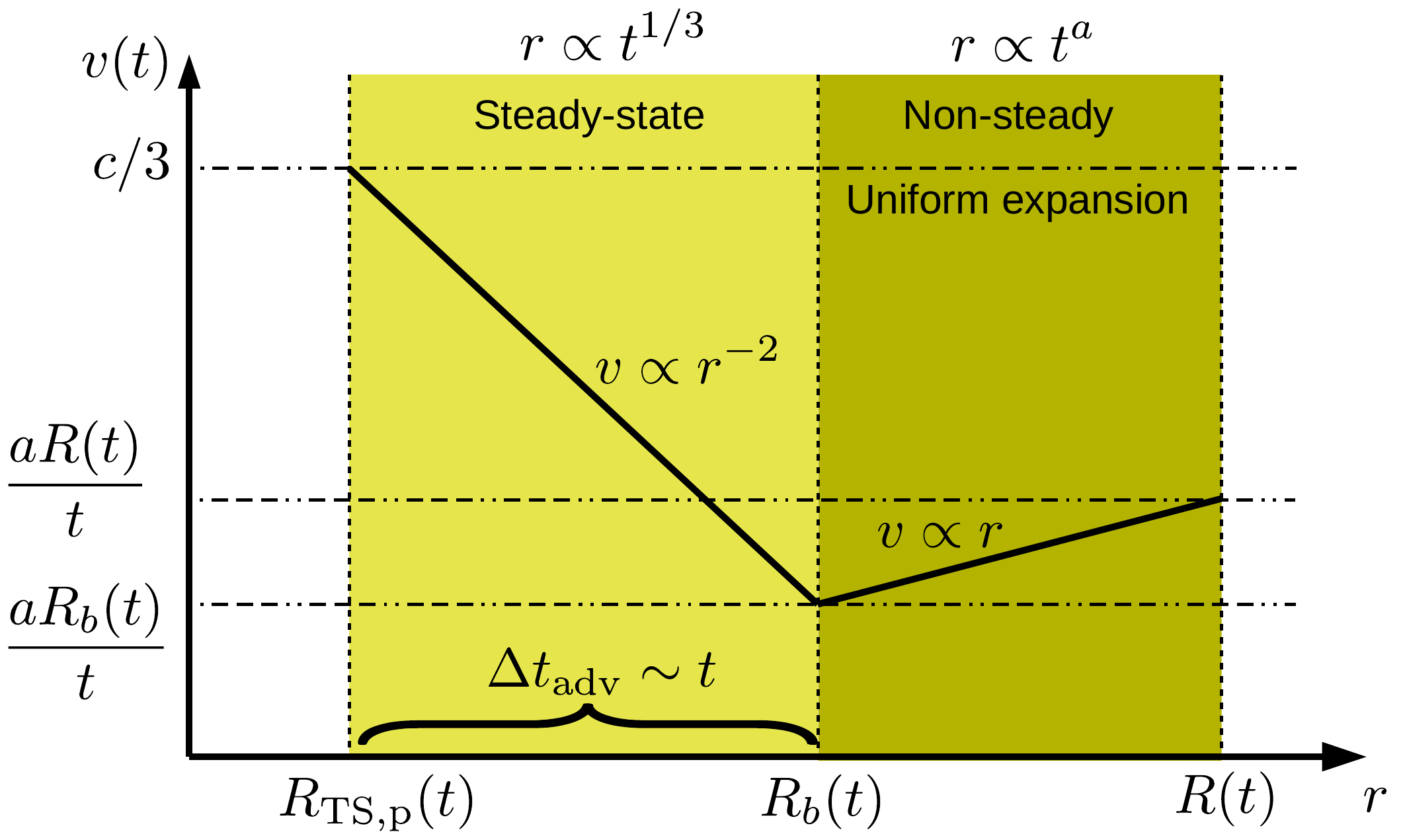}
    \\ \vspace{0.5cm}\\
    \includegraphics[width=\linewidth]{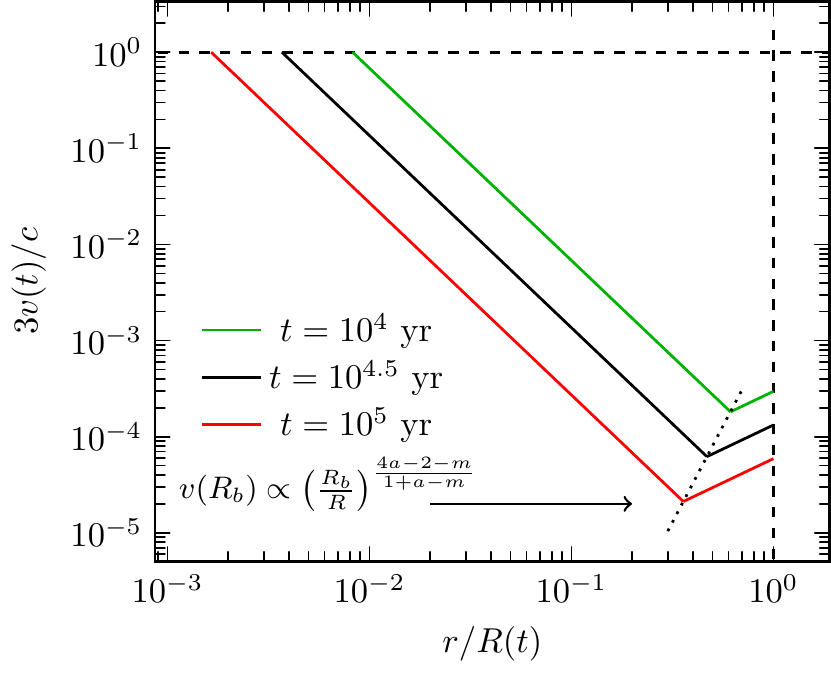}
    \caption{\textit{Top}: Velocity profile (log-log plot) of the advective flow in the nebula shown for both where it assumes a 
    steady-state and where this assumption fails. The flow is launched at the termination shock radius 
    $R_{\rm TS,p}(t)$ with velocity dropping quadratically with radius at a fixed temporal snapshot of the 
    nebula. The steady-state region terminates at radius $R_b(t)$ beyond which the flow expands 
    uniformly with radius out to the edge of the MWN at $R(t)$. \textit{Bottom}: Velocity profiles at three different 
    temporal snapshots, $t_0 < t_1 < t_2$ with fiducial values assumed for $m=2$ and $a=0.3$ (when $k=0$). The dotted line 
    shows the dependence of the velocity on $R_b$ at different snapshots in time.}
    \label{fig:steady_state_v}
\end{figure}

Nonetheless, near $R_{{\rm TS},p}$ where the flow was injected during the last dynamical time (roughly the timescale over which the nebula doubles its age or size)
one still expects the steady-state 
approximation of Equation~(\ref{eq:beta_r}) and (\ref{eq:r1}) to hold. This inner region extends out to a radius $R_b$ that 
can be estimated as follows. At $R_b<r<R$ we expect a uniform and isotropic expansion of the plasma, with a velociy
$v = dr/dt \propto r$ at any given time $t$. Matching the velocity at the MWN's outer boundary, $v(R)=\dot{R}=aR/t$ for $R\propto t^a$
implies a velocity $v = \dot{R}r/R = ar/t$ in this region. Therefore, it is natural to define $R_b$ through the continuity 
of the velocity between the inner (Equation~(\ref{eq:beta_r})) and outer regions, $v_{\rm TS}(R_{\rm TS}/R_b)^2=aR_b/t$, which for 
$v_{\rm TS}=c/3$ implies (at equilibrium)
\begin{eqnarray}\label{eq:R_b}
R_b(t) &= \fracb{R_{{\rm TS},p}^2ct}{3a}^\frac{1}{3} =  R(t)\fracb{tL_{\rm sd}(t)}{3aE(t)}^\frac{1}{3}\propto 
\psi(t)^\frac{4}{3}
t^\frac{1+4a-m}{3}\quad\quad  &\\ \nonumber
    &\approx \left\{\begin{array}{ll}
    \frac{0.336\,{\rm pc}\,B_{14}^\frac{11}{9}d_4^\frac{22}{9}M_3^\frac{11}{36}n_0^\frac{1}{6}}
    {\fracb{\kappa_1}{6.0f}^\frac{11}{18}E_{\rm tot,52.3}^\frac{7}{36}\psi_2^\frac{10}{9}}\ , & E_0 > E_{\rm SN} \\ \\ \nonumber
    \frac{0.887\,{\rm pc}\,B_{14}^\frac{11}{9}d_4^\frac{22}{9}M_3^\frac{11}{36}E_{\rm SN,51}^\frac{11}{90}n_0^\frac{1}{6}P_{\rm 0,-2}^\frac{11}{45}}
    {\fracb{\kappa_2}{4.47f}^\frac{11}{18}E_{\rm tot,51}^\frac{7}{36}\psi_2^\frac{10}{9}}\ , & E_0 < E_{\rm SN}\end{array}\right.
\end{eqnarray}
Altogether, the velocity profile as shown in Figure \ref{fig:steady_state_v} is given by
\begin{equation}\label{eq:vrt}
    v(r,t) = \left\{\begin{array}{ll}
        \frac{cR_{{\rm TS},p}^2(t)}{3r^2}=\frac{aR_b^3(t)}{tr^2}\ , & R_{{\rm TS},p}(t)<r<R_b(t) \\  \\
        \frac{ar}{t}\ , & R_b(t)<r<R(t)  \end{array}\right.
\end{equation}

We can verify that the advection time out to $R_b$ is close to the dynamical time, and indeed 
$\Delta t_{\rm adv}(R_b)\approx R_b^3/cR_{{\rm TS},p}^2=t/3a \sim t$.
As this time is estimated using the velocity derived in the steady state approximation,
which is valid for $\Delta t\ll t\Leftrightarrow t_i\approx t$, this implies that 
$\Delta t_{\rm adv}(R_b)\sim t_i$ or that the fluid element is advected to $R_b$ by the
time $t_b(t_i)=t_i+\Delta t_{\rm adv}(R_b) \sim 2t_i$. At a time $t$, the fluid that is
at $R_b(t)$ was injected at $t_i\sim t/2$, so that $r<R_b$ corresponds to 
$t_i\gtrsim t/2\Leftrightarrow \Delta t\lesssim t/2$ while $r>R_b$ corresponds to
$t_i\lesssim t/2\Leftrightarrow \Delta t\gtrsim t/2$. Once the fluid element enters the
outer region, $r>R_b$, its fractional radial location in the nebula, $r(t)/R(t)$ becomes constant, since both $r$ and $R$ scale as $t^a$. Therefore, in this picture the outer nebula is gradually filled from its inner boundary ($R_b(t)/R(t)\propto t^{(1+a-m)/3}$ 
decreases with time for $m>1+a$ that is needed in order to have such an outer region)
as it expands uniformly ($\Delta V\propto r^3\propto t^{3a}$ for any fluid element whose
radial coordinate is $r$) and isotropically. Adiabatic cooling becomes significant only
in the outer part of the nebula, over times that are at least comparable to the dynamical time.

The uniform pressure and energy density in the MWN imply that $E(<\!R_b)/E= (R_b/R)^3 = L_{\rm sd}t/3aE
= (R_{{\rm TS},p}/R_{{\rm TS},v})^2 \propto t^{1+a-m}$ and therefore 
$E(<\!R_b) \approx L_{\rm sd}t$, so that indeed the inner region, $R_{{\rm TS},p}(t)<r<R_b(t)$, where the steady-state flow approximation holds 
contains the energy and matter that was injected in the last dynamical time. The outer region, $R_b(t)<r<R(t)$, contains energy and matter 
that were injected before the last dynamical times and suffered significant adiabatic losses. Nonetheless, we find that for most of the 
relevant parameter space (namely $m = (n+1)/(n-1) > 1+a\to1.3$ or $n<(2+a)/a\to7.67$) it still dominates the total MWN energy $E(t)$.

Using Equation~(\ref{eq:vrt}) for $v(r<R_b,t)$ instead of Equation~(\ref{eq:beta_r}) accounts for the gradual evolution between the different instantaneous 
quasi-steady states of the flow in the inner nebula. It can be used to generalise the expression for $r_{\rm adv}(t_i,t)$ from Equation~(\ref{eq:r1}).
Denoting $\bar{t}\equiv t/t_i$ and by using the scaling $R_{{\rm TS},p}(t)=R_{{\rm TS},p}(t_i)\bar{t}^{(4a-m)/2}$ it can be integrated to obtain
\begin{eqnarray}\label{eq:radv}
    r_{\rm adv}
    = R_{{\rm TS},p}(t_i)\left(1+\frac{ct_i}{R_{{\rm TS},p}(t_i)}\left[\frac{\bar{t}^{\,1+4a-m}-1}{1+4a-m}\right]\right)^\frac{1}{3}\quad
    \\ \nonumber
    \quad= R_{{\rm TS},p}(t)\left(\bar{t}^\frac{3(m-4a)}{2}+\frac{ct}{R_{{\rm TS},p}(t)}\left[\frac{1-\bar{t}^{\,m-4a-1}}{1+4a-m}\right]\right)^\frac{1}{3}\;.
\end{eqnarray}
It can be seen that $r_{\rm adv}(t_i=t)=R_{\rm TS}(t)=R_{\rm TS}(t_i)$. 

The expression for $r_{\rm adv}(t_i,t)$ can be used to find the distance traveled by particles over their cooling times,
which is then used to express $\gamma_e = \gamma_e(r,t)$. 

For a uniform magnetic field in the nebula in steady state (neglecting adiabatic cooling due to the expansion 
of the nebula, which is a reasonable approximation if the cooling time is shorter than the dynamical times) the cooling time is 
$t_{\rm syn}=t_{c,0}/\gamma_e\propto E_{\rm syn}^{-1/2}$.
The cooling length $r_c$ is obtained by equating $t_{\rm syn}(\gamma_e)=\Delta t\approx r^3/cR_{{\rm TS},p}^2 $, which implies
a weak dependence of the cooling length on the observed synchrotron frequency,
\begin{eqnarray}\nonumber
    r_{c,{\rm adv}}&\approx \left(R_{{\rm TS},p}^2ct_{c,0}\right)^\frac{1}{3}\gamma_e^{-\frac{1}{3}} 
    = \left(R_{{\rm TS},p}^2ct_{c,0}\right)^\frac{1}{3}\fracb{E_{\rm syn}}{h\nu_0}^{-\frac{1}{6}}\\
         &=\left\{\begin{array}{ll}
        \frac{0.150\,{\rm pc}\;B_{14}^\frac{2}{3}d_4^\frac{4}{3}M_3^\frac{1}{6}}
        {\fracb{\kappa_1}{6.0f}^\frac{1}{3}E_{\rm tot,52.3}^\frac{1}{6}B_{15\mu{\rm G}}^\frac{1}{2}E_2^\frac{1}{6}}
        \ , & E_0 > E_{\rm SN} \\  \\ 
        \frac{0.303\,{\rm pc}\;B_{14}^\frac{2}{3}d_4^\frac{4}{3}M_3^\frac{1}{6}P_{\rm 0,-2}^\frac{2}{15}E_{\rm SN,51}^\frac{1}{15}}
        {\fracb{\kappa_2}{4.47f}^\frac{1}{3}E_{\rm tot,51}^\frac{1}{6}B_{15\mu{\rm G}}^\frac{1}{2}E_2^\frac{1}{6}}\ , & E_0 < E_{\rm SN}
        \end{array}\right.
\end{eqnarray}
where $t_{c,0}=6\pi m_ec/\sigma_TB^2$ and we have conveniently expressed $E_{\rm syn} =h\nu_0\gamma_e^2$ with $\nu_0 = eB/2\pi m_ec$. 
This implies that if the observed size is determined by the cooling length, i.e. by synchrotron burn-off,
then the energy dependence of the observed size should be rather weak.
Moreover, in our case the ratio $r_{c,{\rm adv}}$ is too small to account 
for the observed size of the X-ray nebula, $r_{c,{\rm adv}}/R_X\sim0.06-0.12\ll 1$.


\subsection{Comparison with Observations}\label{sec:softening}
The most relevant observation for the synchrotron cooling length is the spatial distribution of the 
spectral slope within the X-ray nebula. The photon index softens with distance from the magnetar \citep{Y+16},
from $\Gamma_{\rm in}=1.41\pm0.12$ in the inner ellipse of size (semi minor and major axes) $25''\times50''$ ($(0.48\times 0.97)d_4\;$pc) using XMM+NuSTAR data,
to $\Gamma_{\rm out}=2.5\pm0.2$ in the outer ellipse of size $80''\times130''$ ($(1.55\times 2.52)d_4\;$pc) using XMM data. It is still not clear whether this
softening, by $\Delta\Gamma=1.09\pm0.38$, reflects the MWN's intrinsic emission spectrum or is alternatively at least partly caused by a spatially varying 
absorption column $N_H$ through the nearby GMC. If the photon index of the inner ellipse is interpreted as representing that of the intrinsic 
emission spectrum of the uncooled shock accelerated electron distribution, $dN_e/d\gamma_e\propto\gamma_e^{-s}$, this leads to 
$F_\nu\propto\nu^{(1-s)/2}=\nu^{1-\Gamma_{\rm in}}$ and $s=2\Gamma_{\rm in}-1=1.82\pm0.24$, which is rather hard but with a rather large uncertainty. 
Values of $s\sim1.5$ or so are inferred in PWNe at lower energies, but usually in the X-ray energy range the inferred values are around $s\sim2-2.5$,
which is also consistent with the uncertainties. Moreover, the additional uncertainty due to the possible spatially varying
$N_H$ makes it even harder to draw any strong conclusions from the measured value of $\Gamma$  . 
  
In the following we will assume that the intrinsic spectral softening is at most similar to the observed one. The relatively gradual and modest degree 
of spectral softening ($\Delta\Gamma\lesssim1.1$ when the size of the region grows by a factor of $\sim3$ in terms of the distance from the magnetar) 
would be hard to reconcile with the weak dependence of the cooling length on the observed synchrotron photon energy, $r_{c,{\rm adv}}\propto E_{\rm syn}^{1/6}$
that we derived, if the observed size of the X-ray nebula, $R_X$, is indeed determined by synchrotron burn-off. One way around this is if $R_X$ is
instead limited by the sensitivity of our observation and the background, and it is somewhat smaller than the cooling length of the electrons emitting 
in the observed energy range, which results in a smaller and more gradual softening of the photon index with the distance from the magnetar.
However, this requires $r_{c,{\rm adv}}(E_X) > R_X$, while for typical parameters we obtain that $ r_{c,{\rm adv}}(E_X)\ll R_X$.

\subsection{The Role of Diffusion}\label{sec:diff}
This might still be reconciled with the observations if the effects of particle diffusion within the nebula are important and 
cannot be neglected as we did so far, when we considered only the particle advection with the bulk flow in the nebula. 
This would also tend to moderate the spectral softening with the distance from the
magnetar.

\begin{figure}
    \includegraphics[width=0.477\textwidth]{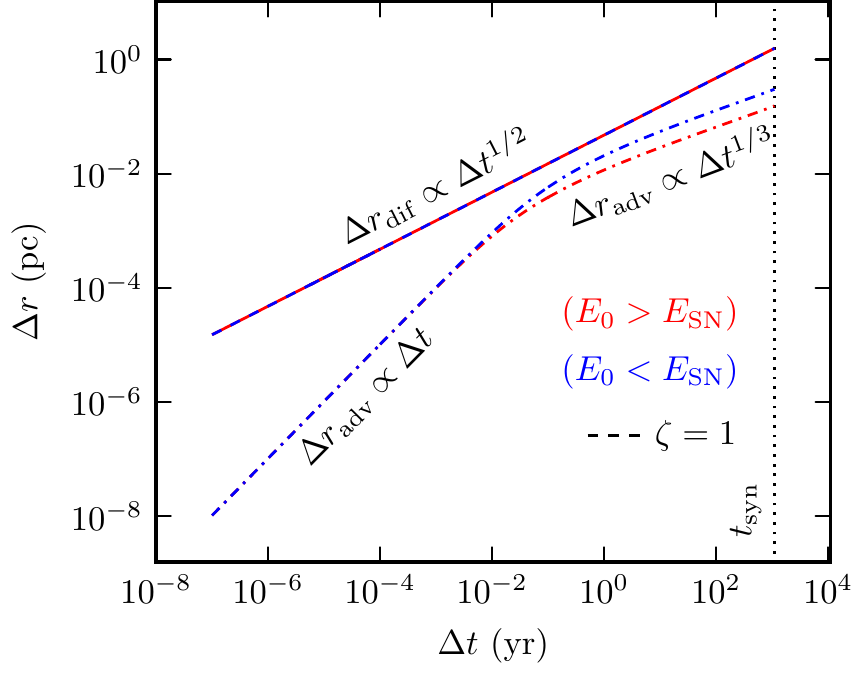} \\
    \quad \\
    \includegraphics[width=0.477\textwidth]{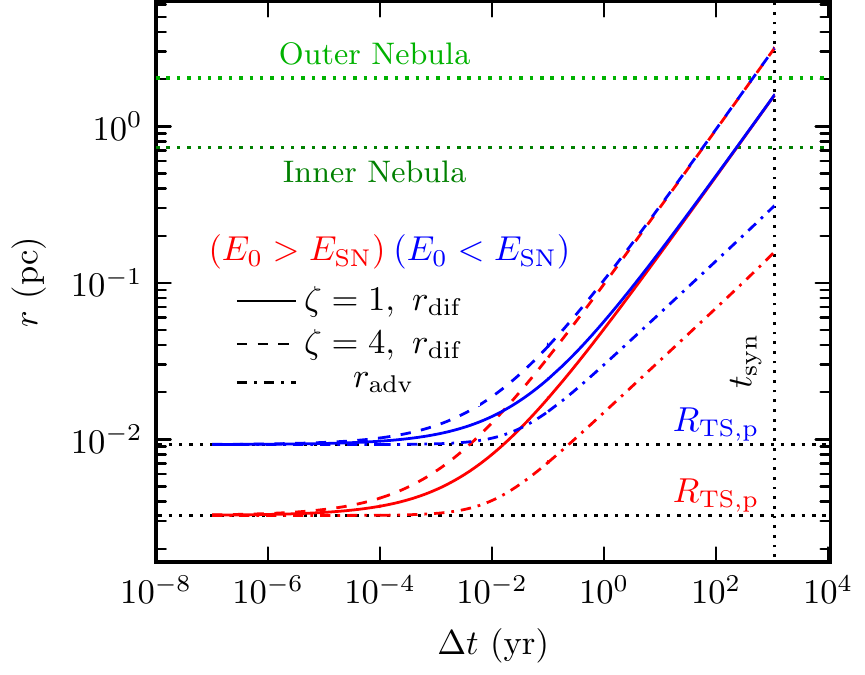}
    \caption{Diffusion and advection distance of fast cooling particles ($\gamma_e = 10^8 > \gamma_c(t)$) 
    over their cooling times $\Delta t = t-t_i \leq t_{\rm syn}$ (shown by the vertical dotted line) for a system age 
    $t = t_{\rm SNR} = 10.3~{\rm kyr}$ when $E_0 > E_{\rm SN}$ (red, for $\sigma=5.6\times10^{-3}$) and 
    $t_{\rm SNR} = 22.2~{\rm kyr}$ when $E_0 < E_{\rm SN}$ (blue, for $\sigma=4.7\times10^{-2}$). 
    \textbf{(Top)} - Radial distance from the termination shock $R_{\rm TS,p}$: 
    Diffusion length (dashed) $\Delta r_{\rm dif} = l_{\rm dif}$ 
    for $\zeta = 1$, and advection length (dot-dashed) $\Delta r_{\rm adv} = l_{\rm adv}$. 
    \textbf{(Bottom)} - Diffusion and advection (dot-dashed) distances  
    $r = R_{\rm TS,p}+\Delta r$. Diffusion length is shown for two cases: 
    (solid) $\zeta \equiv \lambda_{\rm def}/R_L = 1$ (Bohm diffusion) 
    with mean deflection length ($\lambda_{\rm def}$) equal to the Larmor radius ($R_L(\gamma_e)$), 
    and (dashed) $\zeta = 4$. The horizontal green dotted lines indicate the radial extents of the 
    inner and outer X-ray nebulae and the black dotted lines show the position of $R_{\rm TS,p}(t_{\rm SNR})$.}
    \label{fig:rdist}
\end{figure}

After injection at the termination shock ($r = R_{\rm TS,p}$), the particles start to both diffuse and advect downstream. We define their distance traveled over time $\Delta t$ due to advection as the advection length, $l_{\rm adv}$, and their typical distance traveled by diffusion as the diffusion length, $l_{\rm diff}$. For $c\Delta t\ll R_{\rm{TS},p}$ we have $l_{\rm adv} \approx c\Delta t/3$ as $c/3$ is the flow velocity just behind the wind termination shocks. For $c\Delta t\gg R_{\rm{TS},p}$ the advection length is given by $l_{\rm adv}\sim r(\Delta t)\approx(R_{{\rm TS},p}^2c\Delta t)^{1/3}$, while the diffusion length is\footnote{For diffusion in $i$ dimensions the
diffusion coefficient is $D=\lambda_{\rm def}v/i$ and the r.m.s. displacement is $\sqrt{\langle l^2\rangle} = \sqrt{2iD\Delta t}=\sqrt{2\lambda_{\rm def}v\Delta t}$, and in our case $v\approx c$.} 
$l_{\rm dif}\approx(2\lambda_{\rm def}c\Delta t)^{1/2}
\approx(2\lambda_{\rm def}R_{{\rm TS},p}^{-2}r^3)^{1/2}$
where $\lambda_{\rm def}(\gamma_e)$ is the deflection length of an electron of Lorentz factor 
$\gamma_e$.
The ratio of these two lengthscales is
\begin{equation}
    \frac{l_{\rm dif}}{l_{\rm adv}} \approx 
    \fracb{8\lambda_{\rm def}^3\,c\Delta t}{R_{{\rm TS},p}^4}^{1/6}
    \approx \frac{\sqrt{2\lambda_{\rm def}\,r}}{R_{{\rm TS},p}}\equiv\sqrt{\frac{r}{r_*}}\ ,
\end{equation}
and it grows as $\Delta t^{1/6}\propto r^{1/2}$ so that diffusion dominates at large radii $r>r_*$ where
\begin{equation}\label{eq:rstar1}
    r_*= \frac{R_{{\rm TS},p}^2}{2\lambda_{\rm def}} =
    \frac{R(t)^3L_{\rm sd}(t)}{2cE(t)\lambda_{\rm def}(\gamma_e,t)}\ .
\end{equation}
Since the advection and diffusion start just downstream of the shock, at $r=R_{{\rm TS},p}$, then $l_{\rm adv}$ or $l_{\rm dif}$ correspond to $\Delta r = r-R_{{\rm TS},p}$, rather than to $r$.
While for $c\Delta t\gg R_{\rm{TS},p}\Leftrightarrow r\gg R_{\rm{TS},p}$ we have $\Delta r\approx r$ and this distinction is not very important, in our case for the X-ray emitting electrons we have $\Delta r_*\lesssim R_{\rm{TS},p}$, which makes this distinction very important.
This can be seen when writing 
the numerical values in Eq.~(\ref{eq:rstar1}), for which one needs to specify the deflection length.

One generally expects $\lambda_{\rm def}$ to increase with $\gamma_e$ so that $r_*$ 
decreases with $\gamma_e$, i.e. increases for lower energy electrons that hence remain 
advection dominated out to a larger radius.
The deflection length should be at least comparable to the electron's Larmor radius, which for the X-ray emitting electrons 
($\gamma_e=\gamma_X$ for which $E_{\rm syn}(\gamma_X)=E_X = 2E_2\;$keV) is $R_L(\gamma_X)=\gamma_Xm_ec^2/eB$ or
\begin{equation}
    R_L(\gamma_X) = 3.95\times10^{-3}B_{15\mu{\rm G}}^{-3/2}E_2^{1/2}\;{\rm pc}\ .
\end{equation}
Using such a parameterization where $\zeta\equiv\lambda_{\rm def}/R_L\gtrsim 1$ (and $\zeta=1$ corresponds to Bohm diffusion), we find
\begin{equation}
\frac{\Delta r_*}{10^{-3}\,{\rm pc}}\approx \left\{\begin{array}{ll}
        \frac{2.36\,B_{14}^2d_4^4
        B_{15\mu{\rm G}}^\frac{3}{2}M_3^\frac{1}{2}}
        {\fracb{\kappa_1}{6.0f}E_{\rm tot,52.3}^\frac{1}{2}E_2^\frac{1}{2}\zeta}
        \ ,& E_0 > E_{\rm SN} \\  \\ 
        \frac{11.3\,B_{14}^2d_4^4B_{15\mu{\rm G}}^\frac{3}{2}
        M_3^\frac{1}{2}P_{\rm 0,-2}^\frac{2}{5}E_{\rm SN,51}^\frac{1}{5}}
        {\fracb{\kappa_2}{4.47f}E_{\rm tot,51}^\frac{1}{2}E_2^\frac{1}{2}\zeta}
        \ .& E_0 < E_{\rm SN}
        \end{array}\right.
\end{equation}
Comparison to Eq.~(\ref{eq:R_TSp}) shows that indeed for our case $\Delta r_*\lesssim R_{\rm{TS},p}$. Because of this, in this case there is no region where advection dominates over diffusion, i.e. diffusion dominates throughout the whole nebula (i.e. $r_* = R_{{\rm TS},p}$).
This can be seen in the upper panel of Figure~\ref{fig:rdist}.

Therefore, the expression we derived for the diffusion length, $l_{\rm dif}\approx (r^3/r_*)^{1/2}$, is no longer valid. 
Instead, the diffusion length is
$l_{\rm dif} = \Delta r_{\rm dif}\approx(2\lambda_{\rm def}c\Delta t)^{1/2}$. The effect of edvection can be neglected for the X-ray emitting electrons, so that their number density without accounting for the electron cooling approximately scales as $1/r$, following the solution for spherical steady-state diffusion from a steady source, $\tilde{n}(\gamma_e,r>r_*)\approx \tilde{n}(\gamma_e,r_*)\frac{r_*}{r}\to \tilde{n}(\gamma_e,R_{\rm{TS},p})\frac{R_{\rm{TS},p}}{r}$.
Accounting for electron cooling implies that $\tilde{n}(\gamma_e,r)$ starts dropping exponentially
with $r$ past the electron's cooling length, which is set by $r_{c,{\rm dif}}\approx[2\lambda_{\rm def}ct_{\rm syn}(\gamma_e)]^{1/2}$ or
\begin{equation}
    r_{c,{\rm dif}} \approx 1.57\,B_{15\mu{\rm G}}^{-3/2}\zeta^{1/2}\;{\rm pc}\ .
\end{equation}
Since we derived $B\gtrsim11\mu\rm G$ and one expects $\zeta\gtrsim 1$ then this lengthscale might potentially account for the observed size of the X-ray 
nebula ($R_X$), e.g. for $B_{15\mu{\rm G}}\sim 1$ and $\zeta\sim1-2$.
Projection effects imply that the surface brightness scales as $r\tilde{n}(r)$, which is roughly uniform at $r<r_{c,{\rm dif}}$.
An energy dependence of $\zeta=\zeta(\gamma_e)$ would introduce a corresponding energy dependence of $r_{c,{\rm dif}}$.
If $\zeta = \lambda_{\rm def}(\gamma_e)/R_L(\gamma_e)\gtrsim 1$ is of order unity near $\gamma_X$ but somewhat increases at lower $\gamma_e$
this might account for the mild spectral softening with $r$ that was observed (if it is indeed intrinsic and not due to a spatially varying $N_H$).
Improved spatially resolved measurement of the photon index could help pin down its origin and teach us more about the underlying phyiscs of the MWN.
Altogether these effects of particle diffusion (and advection in the inner nebula) could potentially account both for the observed size of the nebula 
and its spatially resolved spectrum, and a detailed fit to a more elaborate model along similar lines might help constrain the underlying physics 
(e.g. $\zeta(\gamma_e)$).

\section{The Quasi-Steady State Energy Balance of the X-ray Nebula}\label{sec:g-sigma}

As we have seen from Eqs.~(\ref{eq:tsyn}), the synchrotron cooling time of the X-ray emitting electrons is much smaller than the dynamical time, 
$t_{\rm syn}\ll t_{\rm SNR}$. Therefore, for modeling the X-ray nebula it is reasonable to use a steady-state approximation and neglect 
the expansion of the nebula over the relevant timescales. The relevant timescales here are mainly the advection ($\Delta t_{\rm adv}(R_X)$) 
and diffusion ($\Delta t_{\rm dif}(R_X)$) timescales from $R_{{\rm TS},p}$ to $R_X$. In steady state (i.e. for $r<R_b(t)$) the energy within that radius satisfies $E(<r)=\langle\dot{E}\rangle \Delta t_{\rm adv}(r)$,
where $\langle\dot{E}\rangle$ is the long-term mean (over times larger than our current
observations, up to the dynamical time) energy injection rate in outflows from the
magnetar into the MWN, and $\Delta t_{\rm adv}(r)$ is the time over which
the energy flows from $R_{{\rm TS},p}$ to $r$. We parameterize
$\langle\dot{E}\rangle\equiv gL_{\rm sd}$ where $L_{\rm sd}$ is the current spin down power. Thus defined, $g<1$ is possible if the current $L_{\rm sd}$ is above its long-term
mean value, while the contribution from burst-associated outflows does not compensate for
that. On the other hand, $g>1$ is possible if the long-term mean contribution from
burst-associated outflows exceeds the current $L_{\rm sd}$. 

\subsection{Magnetic Energy Balance}\label{sec:mag-bal}
We assume that a fraction 
$\sigma/(1+\sigma)$ of $\langle\dot{E}\rangle$ is injected in the form of magnetic fields,
such that $E_B(<r) = B^2r^3/6$ satisfies
\begin{equation}\label{eq:EB1}
\frac{1+\sigma}{\sigma}E_B(<r) = E(<r) = \langle\dot{E}\rangle \Delta t_{\rm adv}(r) = gL_{\rm sd}\Delta t_{\rm adv}(r)
\end{equation}
Using the lower limit on $B$ from Equation~(\ref{eq:Bmin}), $B>B_{\rm min}$, this can be written as
\begin{equation}
    \frac{g\sigma}{1+\sigma} = \frac{E_B}{L_{\rm sd}\Delta t_{\rm adv}(r)} > \frac{B_{\rm min}^2R_X^3}{6L_{\rm sd}\Delta t_{\rm adv}(r)}\ .
\end{equation}
This yields
\begin{eqnarray}\nonumber
    \frac{g\sigma}{1+\sigma} &>& \frac{\frac{1}{6}B_{\rm min}^2r^3}{L_{\rm sd}\Delta t_{\rm adv}\!(\!r\!)}=\frac{B_{\rm min}^2R(t)^3}{6E(t)}
    = \frac{E_{B,{\rm min}}(\!<\!R)}{E(t)}\\ \label{eq:g-sig1}
         &\approx&\left\{\begin{array}{ll}
        \frac{0.0030\,B_{14}^2d_4^4M_3^\frac{1}{2}f^3E_{30}^2}
        {\fracb{\kappa_1}{6.0}E_{\rm tot,52.3}^\frac{1}{2}}
        \ , & E_0 > E_{\rm SN} \\  \\ 
        \frac{0.0249\,B_{14}^2d_4^4M_3^\frac{1}{2}P_{\rm 0,-2}^\frac{2}{5}E_{\rm SN,51}^\frac{1}{5}f^3E_{30}^2}
        {\fracb{\kappa_2}{4.47}E_{\rm tot,51}^\frac{1}{2}}\ , & E_0 < E_{\rm SN}
        \end{array}\right.
\end{eqnarray}
where the numerical values are estimated for $R=R_X$.

A complimentary constraint can be obtained by using our modeling of the nebula dynamics 
and the implied magnetic field, e.g. as given by Equation~(\ref{eq:Bx}), which provides
$B(t) = [\sigma/(1+\sigma)]^{1/2}B_{\rm max}(t)$ where $B_{\rm max}$ is the upper limit on 
$B$ for which all of the nebula's energy $E(t)$ resides in its magnetic field. The condition 
$B>B_{\rm min}$ corresponds to $\sigma/(1+\sigma)>(B_{\rm min}/B_{\rm max})^2$ and leads to exactly the same constraint as Eq~(\ref{eq:g-sig1}) just without the factor $g$. Indeed,
the factor $g$ can be removed from Eq~(\ref{eq:g-sig1}) due to the following arguments. 
Taking into account that when the total energy in the nebula is dominated by injection well
before the last dynamical time, it also determines the pressure in the nebula 
(and its energy density, which determines $B_{\rm max}$), and the condition of pressure
equilibrium will determine the small fractional volume, $(R_b/R)^3$, occupied by the plasma injected over the last dynamical time. 
Varying $g$ will mainly change $R_b$ but as long as $(R_b/R)^3\approx gL_{\rm sd}t/E(t)<1$
it will hardly affect the energy density in the nebula. The result 
$\Delta t_{\rm adv}(R_b)\sim t_i$, which is also $\sim t$ for the plasma currently at $R_b$, will remain valid (averaging over a sufficiently long time
during which the sporadic outflows act together more coherently on the flow in the nebula) 
so that $\Delta t_{\rm adv}\approx r^3/cR_{{\rm TS},p}^2\to t(r/R_b)^3\sim r^3/cR_{{\rm TS},p}^2g$ since $R_b^3\propto L_{\rm sd}\to gL_{\rm sd}$. Also, recall that Equation~(\ref{eq:EB1}) that was used for deriving Equation~(\ref{eq:g-sig1})
relies on an estimate of the energy within a given volume, i.e. the energy density, which is
uniform in the nebula and largely independent of $g$ as long as $gL_{\rm sd}t/E(t)<1$.

\subsection{Energy Balance of X-ray Emitting Electrons}\label{sec:e-bal}
Since the cooling time of the X-ray emitting electrons is much smaller than the dynamical time, $t_{\rm syn}\sim 1\;$kyr, we can assume a steady state for their emission and take $\langle\dot{E}\rangle$ as the mean value over the time $t_{\rm syn}$ to obtain an equation for the energy balance of the X-ray emitting electrons,
\begin{equation}\label{eq:g-L_X0}
 \langle\dot{E}\rangle = gL_{\rm sd} = \frac{(1+\sigma)}{\epsilon_e\epsilon_X}L_{X,{\rm tot}}\ ,
\end{equation}
where $L_{X,{\rm tot}} = 2.74\times 10^{33}\;{\rm erg\;s^{-1}}$ is the luminosity in the whole X-ray nebula within the detected energy range, i.e. 0.5$\,$--$\,$10$\;$keV in the outer nebula and 0.5$\,$--$\,$30$\;$keV in the inner nebula.
Here a fraction $1/(1+\sigma)$ of the total energy injected into the nebula goes into particles, a fraction $\epsilon_e$ of the latter energy goes into the power-law electron (and positron) energy distribution responsible for the observed X-ray emission, and a fraction $\epsilon_X$ of the latter energy is radiated in the observed X-ray energy range (thus contributing to the observed X-ray luminosity $L_X$ between $\nu_m$ and $\nu_M$ corresponding to $\gamma_m<\gamma_e<\gamma_M$).

As long as $s>1$ and there are fewer high-energy electrons than low-energy electrons in the initial electron power-law energy distribution (without the effects of electron radiative cooling) then one can neglect the contribution to $L_X$ of electrons initially with $\gamma_{e}>\gamma_M$ that cool down into the contributing range $\gamma_m<\gamma_e<\gamma_M$ and deposit there a fraction $\sim\gamma_M/\gamma_{e,i}$
of their energy. For $\gamma_M\gg\gamma_m$ each such electron radiated and energy of $\approx\gamma_M m_ec^2$
within the observed range ($\nu_m<\nu<\nu_M$). Therefore, the increase in $L_X$ compared to the contribution of electrons initially in the range 
$\gamma_m<\gamma_e<\gamma_M$ is by a factor of $f_L$ given by
\begin{eqnarray}
    f_L-1 = \frac{\int_{\gamma_M}^{\gamma_2}d\gamma_e\frac{dN_e}{d\gamma_e}\gamma_M}{\int_{\gamma_m}^{\gamma_M}d\gamma_e\frac{dN_e}{d\gamma_e}\gamma_e}=
    \frac{\gamma_M\int_{\gamma_M}^{\gamma_2}d\gamma_e\gamma_e^{-s}}{\int_{\gamma_m}^{\gamma_M}d\gamma_e\gamma_e^{1-s}}\\ \nonumber = 
    \left\{\begin{array}{ll}
        \fracb{2-s}{s-1}\frac{\gamma_M(\gamma_M^{1-s}-\gamma_2^{1-s})}{(\gamma_M^{2-s}-\gamma_m^{2-s})}
        \ ,\quad & s\neq 1,\,2 \\ \\ 
        \frac{\ln(\gamma_2/\gamma_M)}{(1-\gamma_m/\gamma_M)}\ , & s=1\\ \\ 
        \frac{(1-\gamma_M/\gamma_2)}{\ln(\gamma_M/\gamma_m)}\ , & s=2
        \end{array}\right.
\end{eqnarray}
In our case $\Gamma_{\rm in}=1.41\pm0.12$ implies $s\gtrsim 1.82\pm 0.24$ and $\gamma_M/\gamma_m = (\nu_M/\nu_m)^{1/2}\approx7.746$ while for $h\nu_M=30\;$keV
\begin{equation}\label{eq:gM_over_g2}
    \frac{\gamma_M}{\gamma_2} >
    \frac{\gamma_M}{\gamma_{\rm max}} =
    \frac{0.86}{\sqrt{B_{15\mu{\rm G}}}}\ ,
\end{equation}
 implying $f_L-1\ll 1$ and $f_L\approx 1$, so this effect can be ignored. 
 
\begin{figure}
    \centering
    \includegraphics[width=0.473\textwidth]{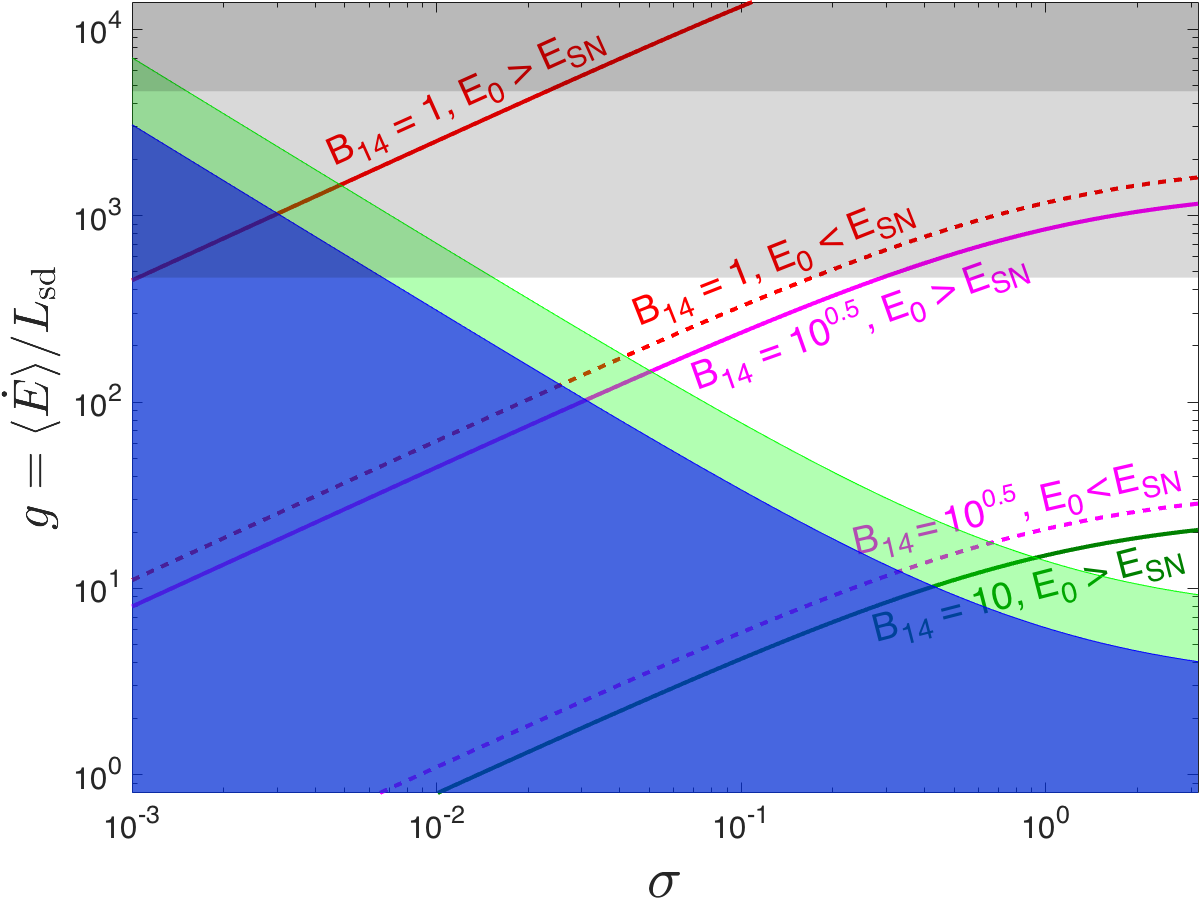}
    \caption{The excluded parameter space in the $g$-$\sigma$ plane, according to Equation~(\ref{eq:g-sigma}),
    is the shaded blue region on the bottom left. The shaded green region is excluded if one assumes that the magnetic field in the nebula does not increase with distance from the central magnetar. The gray shaded region
    is excluded according to Eq.~(\ref{eq:gmax}) for $t_{\rm SNR} = 10^{4.5}\;$yr and $B_{\rm int,max}=10^{16}\;$G (light gray) or $B_{\rm int,max}=10^{16.5}\;$G (darker gray). The thick solid and dashed lines represent the constraint from Eq.~(\ref{eq:g-sigma2}) for a few values of $B_0$ and for our two cases $E_0<E_{\rm SN}$ (dashed lines) and $E_0>E_{\rm SN}$ (solid lines).}
    \label{fig:g_sigma}
\end{figure}

The fraction of the energy of the electrons in the power-law component that initially radiates in the observed X-ray range ($\gamma_m<\gamma_e<\gamma_M$ ) is $\xi^{-7/2}$.
Therefore, we have $\epsilon_X = f_L\xi^{-7/2}\approx \xi^{-7/2}$ and Eq.~(\ref{eq:g-L_X0}) becomes
\begin{equation}\label{eq:g-L_X}
 g = \frac{L_{X,{\rm tot}}}{L_{\rm sd}}\frac{(1+\sigma)\xi^{7/2}}{\epsilon_e}
 =\frac{\eta_X(1+\sigma)\xi^{7/2}}{\epsilon_e}\ ,
\end{equation}
where $\eta_X = L_{X,{\rm tot}}/L_{\rm sd}\simeq 0.13d_4^2$. Now, using the lower limit on
 $\xi$ that conservatively assumes that the X-ray emission extends up to $30\;$keV only in the inner nebula where it is detected by NuSTAR up to such energies (i.e. $\xi_{\rm in}$ in Equation~(\ref{eq:xi_min})), this yields
\begin{equation}\label{eq:g-sigma}
    \frac{g\sigma}{1+\sigma} > 3.07d_4^3E_{M,30}^{7/2}f^{7/2}\ ,
\end{equation}
or
\begin{equation}\label{eq:g_min}
    g > g_{\rm min} = 3.07\frac{1+\sigma}{\sigma}d_4^3E_{M,30}^{7/2}f^{7/2}\ .
\end{equation}
Using the lower limit on $\xi$ that assumes the emission extends up to 30$\;$keV in the whole nebula (i.e. $\xi$ in Equation~(\ref{eq:xi_min})) or that the magnetic field in the nebula does not increase with distance from the central magnetar gives a numerical coefficient of $7.03$ in this equation. The resulting excluded and allowed regions in the $g$-$\sigma$ parameter space are shown in Figure~\ref{fig:g_sigma}.

Equating between the magnetic field in the inner nebula derived from its X-ray emission,
(Equation~(\ref{eq:nebularB})) and the nebular field derived from modeling its dynamics
(Equation~(\ref{eq:B(t)}) and (\ref{eq:Bx})) results in an expression for $\xi_{\rm in}$,
\begin{equation}\label{eq:xi-B}
  \xi_{\rm in} = \left\{\begin{array}{ll}
        \frac{39.8\,E_{\rm tot,52.3}^\frac{1}{4}\epsilon_e^\frac{2}{7}\sqrt{\frac{\kappa_1}{6.0f}}\,\sigma^\frac{3}{14}}
        {B_{14}d_4^\frac{12}{7}M_3^\frac{1}{4}\sqrt{1+\sigma}}
        \ , & E_0 > E_{\rm SN} \\  \\ 
        \frac{13.8\,E_{\rm tot,51}^\frac{1}{4}\epsilon_e^\frac{2}{7}\sqrt{\frac{\kappa_2}{4.47f}}\,\sigma^\frac{3}{14}}
        {B_{14}E_{\rm SN,51}^\frac{1}{10}d_4^\frac{12}{7}M_3^\frac{1}{4}P_{\rm 0,-2}^\frac{1}{5}\sqrt{1+\sigma}}\ . & E_0 < E_{\rm SN}
        \end{array}\right.
\end{equation}
Substituting this expression in Equation~(\ref{eq:g-L_X}) gives an expression for $g$,
\begin{eqnarray}\label{eq:g-sigma2}
  g &=& \left\{\begin{array}{ll}
        \frac{1422\,E_{\rm tot,52.3}^\frac{7}{8}\fracb{\kappa_1}{6.0f}^\frac{7}{4}}
        {B_{14.5}^\frac{7}{2}d_4^4M_3^\frac{7}{8}}\fracb{\sigma}{1+\sigma}^\frac{3}{4}
        \ , & E_0 > E_{\rm SN} 
        \\  \\ 
        \frac{35.03\,E_{\rm tot,51}^\frac{7}{8}\fracb{\kappa_2}{4.47f}^\frac{7}{4}}
        {B_{14.5}^\frac{7}{2}E_{\rm SN,51}^\frac{7}{20}d_4^4M_3^\frac{7}{8}M_3^\frac{7}{8}P_{\rm 0,-2}^\frac{7}{10}}\fracb{\sigma}{1+\sigma}^\frac{3}{4}\ , & E_0 < E_{\rm SN}
        \end{array}\right.
\end{eqnarray}
where $B_0 = 10^{14.5}B_{14.5}\,{\rm G}$. 

This constraint is also shown in Fig.~\ref{fig:g_sigma}.
It can be seen that for no evolution of the magnetic field, i.e. $B_0=B_s(t)=1.16\times10^{14}f^{-1/2}\;$G, $g$ is very high, arguably unrealistically so, although for our case $B_{14}=1$ and $E_0<E_{\rm SN}$ it is still possible
(e.g. with $10^{-1.5}\lesssim\sigma\lesssim10^{-1}$ and $10^2\lesssim g\lesssim10^{2.5}$). 

Since very high values of $g$ may be hard to produce physically, this may suggest
that $B_0>B_s(t)$ and the surface dipole field has decayed since the birth of the magnetar. Following the phenomenological study of \citet{DallOsso2012}, we shall adopt here as an illustrative example their preferred values of $\alpha\approx 3/2$ and $\tau_{d,i}=\alpha t_B\approx 1\;$kyr. Since a dipole field decay implies a true age younger than the characteristic spin-down age, which in our case $\tau_c=4.9\;$kyr is already barely compatible with the estimates of the SNR age $t_{\rm SNR}\sim5-100\;$kyr,
this would in turn suggest that the current $\dot{P}$ is anomolously large compared to it long-term mean value $\langle\dot{P}\rangle$, by a factor of $K = \dot{P}/\langle\dot{P}\rangle$. This would imply a new characteristic spin-down age of $\tau_{c,*}=K\tau_c$ and a true surface magnetic field $B_{s,*}=B_s K^{-1/2}$. For significant field decay $\tau_{c,*}\approx\tau_{d,i}(2-\alpha)^{-1}(B_0/B_{s,*})^2$,
which corresponds to
\begin{equation}
    B_0 = B_s\sqrt{\frac{(2-\alpha)\tau_c}{\tau_{d,i}}}
    \sim\frac{(1.15-1.82)\times10^{14}\;{\rm G}}{f^{1/2}(\tau_{d,i}/1\,{\rm kyr})}\ ,
\end{equation}
where the numerical values are for the favored values of $1.5\lesssim\alpha\lesssim1.8$ (not that $K$ has factored out here, and htat this expression is valid only for $t>$).
This suggests $1\lesssim B_{14}\lesssim 2$, which in turn suggests that $g\gtrsim10^{1.5}$. Moreover, this is a rather low $B_0$ for a magnetar. However, interestingly enough, this is rather similar to the value of $B_0\sim2\times10^{14}\;$G inferred for transient SGRs/AXPs \citep{DallOsso2012}. The corresponding true age would retain a dependence of $K$,
\begin{equation}
    t\approx\frac{(2-\alpha)^\frac{\alpha}{2}}{\alpha}
    \frac{\tau_c^\frac{\alpha}{2}K^\frac{\alpha}{2}}{\tau_{d,i}^\frac{\alpha-2}{2}}\sim (4.3-7.3)\fracb{\tau_{d,i}}{1\,{\rm kyr}}^\frac{2-\alpha}{2}\fracb{K}{10}^\frac{\alpha}{2}\;{\rm kyr}\ ,
\end{equation}
but $t\gg\tau_c$ would require rather extreme values of $K\gg 10$.

\subsection{An Alternative Energy Source is Required: the Decay of Magnetar's Magnetic Field}\label{sec:B-decay}

The natural channel of energy injection into the nebula, in 
addition to the rotationally powered MHD wind, is the 
decay of the initial super-QED magnetic field, which powers the 
sporadic bursting activity of the central magnetar and might also be 
responsible for the quasi-steady particle wind \citep[e.g][]{TB98}. 
As mentioned earlier, the internal (predominantly toroidal) magnetic 
field has $\lesssim 10^2$ times more energy than that in the dipole 
component. The feasibility of either field component for supplying the requisite 
energy to the nebula over the last dynamical time can be ascertained 
using simple arguments. 

\subsubsection{The Dipole Field Decay is Not Enough}
We first consider the decay of the dipole 
component and make the assumption that its decay from its initial 
value $B_{\rm dip,0} > B_s(t)$, where $B_s(t)$ is the surface 
dipole field inferred from its current $P$ and $\dot P$, is given 
by a power-law in time (see Eq.~(\ref{eq:B_evol})). The total energy of the dipole 
component is $E_{\rm B,dip}(t) = B_s^2(t)R_{\rm NS}^3/6$ and the power 
injected by its decay is 
\begin{equation}
\vert\dot E_{B,\rm dip}\vert = \frac{2}{\alpha}\frac{E_{B,\rm dip,0}}{t_B}\tau^{-1-\frac{2}{\alpha}} = 
\frac{2}{\alpha}\frac{E_{B,\rm dip}(t)}{t_B\tau}\ ,
\end{equation}
where $\tau=1+t/t_B$, $E_{B,\rm dip,0} = R_{\rm NS}^3B_{\rm dip,0}^2/6$, 
and $\alpha > 0$. For a given current age $t= t_{\rm SNR}$, the value of $t_B$ that maximizes this power is $t_B = 2t/\alpha$ corresponding to $\tau_{\rm max}=1+\alpha/2$ and
\begin{equation}\label{eq:dotE_B_max1}
\vert\dot E_{B,\rm dip}\vert_{\rm max} = \frac{2}{2+\alpha}\frac{E_{B,\rm dip}(t)}{t}\ ,
\end{equation}
which, for $B_s(t) = 1.16\times10^{14}f^{-1/2}~{\rm G}$ implied from the measured $P$ and $\dot{P}$ values, gives
\begin{equation}\label{eq:dotE_B_max2}
\frac{\vert\dot E_{B,\rm dip}\vert_{\rm max}}{gL_{\rm sd}} 
= 1.25\times10^{-3}g_{50}^{-1}f^{-1}t_{4.5}^{-1}\ ,
\end{equation}
for fiducial parameter values of $g = 50g_{50}$
(consistent with the lower limit on $g$ from Eq.~(\ref{eq:g-sigma}) for $\sigma < 1$), 
$t_{\rm SNR} = 10^{4.5}t_{4.5}~{\rm yr}$, 
and $\alpha = 3/2$ \citep[e.g., as $3/2\lesssim\alpha\lesssim 9/5$ was inferred by][]{DallOsso2012}. Therefore, the decay of the dipole field clearly fails to supply the needed power that is injected in the nebula, $\langle\dot{E}\rangle=gL_{\rm sd}$. 

\subsubsection{Decay of a Much Stronger Internal Field is Needed}
Supplying the needed $\langle\dot{E}\rangle=gL_{\rm sd}$ requires a significantly larger power, for which the most viable candidate is the decay of a much larger internal magnetic field within the magnetar, which dominates the total decay rate of the magnetic field, 
$\dot E_{B} = \dot E_{B,\rm dip} + 
\dot E_{B,\rm int} \simeq \dot E_{B,\rm int}$.
Doing a similar analysis for the decay of the internal field, and 
replacing $E_{B,\rm dip,0}\rightarrow E_{B,\rm int,0}$ and 
$B_{\rm dip,0}\rightarrow B_{\rm int,0}$, 
Eq.~(\ref{eq:dotE_B_max1}) together with the requirement that $\dot E_{B}\simeq \dot E_{B,\rm int}\geq\langle\dot{E}\rangle=gL_{\rm sd}$ (where the inequality accounts for some inefficiency in transferring the magnetic power to that supplying the nebular X-ray luminosity) and
therefore $\vert\dot E_{B,\rm int}\vert_{\rm max}\geq gL_{\rm sd}$, implies a lower limit on the current value of the internal field,
\begin{equation}\label{eq:Bint_min}
    B_{\rm int}(t) \geq 3.3\times10^{15}g_{50}^{1/2}t_{4.5}^{1/2}\;{\rm G}\ ,
\end{equation}
which is significantly larger than the inferred surface dipole field,
\begin{equation}
    \frac{B_{\rm int}(t)}{B_{\rm dip}(t)}>28.3f^{1/2}g_{50}^{1/2}t_{4.5}^{1/2}\ .
\end{equation}


A useful constraint on $g$ can be obtained by constraining the value of the 
internal field $B_{\rm int}$ from the stability criterion of magnetic fields 
in NSs. Numerical simulations by \citet{Braithwaite09} show that stable 
axisymmetric magnetic equilibrium is achieved when both poloidal and toroidal 
field components contribute, and most importantly, when the ratio of the 
total energy in the poloidal field component to the total magnetic energy is 
between $\mathcal{A}E_{\rm B,int}/E_G \simeq 10^{-3}\lesssim E_{\rm B,dip}/E_{\rm B,int} \lesssim 0.8$, 
where $\mathcal{A}\sim 10^3$ for NSs and $E_G\simeq3\times10^{53}~{\rm erg}$ 
is the gravitational binding energy (assuming uniform mass density with 
$M_{\rm NS} = 1.4M_\odot$ and $R_{\rm NS} = 10^6~{\rm cm}$). By using the lower limit 
on $E_{\rm B,dip}/E_{\rm B,int}$, which is relevant here, we find
\begin{eqnarray}\label{eq:Bint_max}
    B_{\rm int} \lesssim B_{\rm int,max} &&= 
    \fracb{6E_GB_{\rm dip,0}^2}{\mathcal{A}R_{\rm NS}^3}^{1/4} \nonumber \\
    &&= 6.6\times10^{15}B_{\rm dip,0,15}^{1/2}~{\rm G}~,
\end{eqnarray}
for a fiducial value of the initial surface dipole field 
$B_{\rm dip,0} = 10^{15}~{\rm G}$. The currently inferred surface dipole field 
is smaller than its initial value due to field decay over the age of the magnetar.

Combining equations (\ref{eq:Bint_min}) and (\ref{eq:Bint_max}) one obtains a lower limit on the initial dipole field of $B_{\rm 0,dip}\gtrsim 2.6\times10^{14}g_{50}t_{4.5}\;$G, which is problematic for scenarios in which the dipole field has grown significantly from its initial value to its present value ($B_s=1.16\times10^{14}f^{-1/2}\;$G).

Internal fields as high as $B_M = 10^{16}-10^{16.5}~{\rm G}$ are suggested 
by bursting activity in magnetars \citep[e.g.][]{TZW15} and are needed for powering 
their quiescent X-ray luminosities \citep[e.g.][]{DallOsso2012}. 
If this is the maximum internal field afforded by the NS then, 
Eq.~(\ref{eq:Bint_min}) yields (for $\alpha = 3/2$) 
an upper limit on $g$ (using Eq.~(\ref{eq:dotE_B_max1}) and the requirements 
that $\vert\dot E_{B,\rm dip}\vert_{\rm max}\geq gL_{\rm sd}$ and $B_{\rm int}(t)\leq B_{\rm int,max}$),
\begin{eqnarray}\label{eq:gmax}\nonumber
g \leq g_{\rm max} &&= 
\frac{2}{2+\alpha}\frac{R_{\rm NS}^3B_{\rm int,max}^2}{6L_{\rm sd}t}
\\ 
&&\simeq 2.0\times10^{2}B_{\rm dip,0,15}t_{4.5}^{-1} \nonumber \\
&&\simeq 4.65\times10^{2}B_{M,16}^2t_{4.5}^{-1} \\ \nonumber
&&\simeq 4.65\times10^{3}B_{M,16.5}^2t_{4.5}^{-1}
\end{eqnarray}
Due to the uncertainty in the value of the actual internal field, we 
show the above three upper limits on $g$ in Fig. \ref{fig:g_sigma}.

\subsection{The Radiative Efficiency and Electron Distribution}\label{sec:eps_X}
Finally, we note that Eq.~(\ref{eq:g-L_X}) can be rewritten as
\begin{equation}
    \epsilon_X\simeq\xi^{-7/2} = \frac{\eta_X(1+\sigma)}{\epsilon_e g}\simeq
    0.02\fracb{\epsilon_e}{0.13}^{-1}g_{50}^{-1}\ ,
\end{equation}
where $\epsilon_e\simeq0.13$ and $3.07<g\lesssim 5\times10^3$ would correspond to $2\times10^{-4}\lesssim\epsilon_X\lesssim 0.33$. It is hard to achieve a very low $\epsilon_X$ given the fairly hard inferred electron power-law index $s\simeq 1.82\pm0.24$ and given Eq.~(\ref{eq:gM_over_g2}), which seem to suggest $\epsilon_X\gtrsim 0.1$ or so.
This may be achieved for low values of $\epsilon_e$, which may be possible if most of the particle energy is either in electrons not taking part in the power-law energy distribution radiating in X-rays (e.g. an energetically dominant quasi-thermal energy component, below the observed power-law high-energy tail),
or if most of the energy in outflows fromt he magnetar is baryon rich with most of its energy in protons rather than $e^+e^-$ pairs. Finally, Eq.~(\ref{eq:xi_min}) can be rewritten as
\begin{equation}
\epsilon_X\simeq\xi^{-7/2} < 0.33\sigma\fracb{\epsilon_ed_4}{0.13}^{-1}f^{-7/2}E_{M,30}^{-7/2}\ ,
\end{equation}
which suggests that $\sigma$ cannot be very low unless $\epsilon_e$ is correspondingly low (and $g$ is correspondingly high according to Eq.~(\ref{eq:g_min})).

It is important to notice that once the total energy injection into the nebula is no longer dominated by the magnetar's spin-down power, i.e. $g>1$, then $\eta_X$ no longer represents the true radiative efficiency, $\eta_{X,{\rm true}}$. More generally, the injected power is larger by a factor of $g$, and therefore the overall radiating efficiency is smaller by the same factor,
\begin{equation}\label{eq:eta_Xtrue}
    \eta_{X,{\rm true}} = \frac{\eta_X}{g} = \frac{0.0026d_4^2}{g_{50}} < 0.042\frac{\sigma}{1+\sigma}
    d_4^{-1}E_{M,30}^{-\frac{7}{2}}f^{-\frac{7}{2}}\ .
\end{equation}
Note that this significantly lowers the requires radiative efficiency of the MWN, making it more compatible with efficiencies inferred for PWNe.

%

%


\section{The GeV and TeV Emission}\label{sec:GeV/TeV}
An extended TeV source with radius $9' - 10.2'$ (corresponding to a physical size of $(10.5 - 11.9)d_4$ pc) was discovered at the center 
of SNR W41 \citep{Aharonian+2006}. Additionally, the Fermi Large Area Telescope (LAT) found a high energy 
($E > 100$ MeV) extended source, similar in size to the TeV source, coincident with the same SNR 
\citep[e.g.][]{Nolan+2012}. Spectral analysis of the 
Fermi LAT data showed an approximately flat $E^2dN/dE\propto E^{2-\Gamma}$ spectrum with photon 
index $\Gamma \simeq 2.15$ and GeV luminosity $L_{\rm GeV} \simeq 1.45\times10^{35}d_4^2~{\rm erg~s}^{-1}$ 
in the $0.1 - 100$ GeV energy range \citep{Abramowski+2015}. The same work shows that the TeV extended region 
is slightly softer with $\Gamma\simeq2.6$ and has a ($1 - 30$) TeV luminosity 
$L_{\rm TeV} \simeq 1\times10^{34}d_4^2~{\rm erg~s}^{-1}$. 

There are three plausible scenarios that can explain GeV-TeV emission from SNRs, namely (i) hadronic emission by CR protons 
(${\rm CR}p + p \rightarrow p + p + \pi^0$) followed by the decay of neutral pions ($\pi^0\rightarrow2\gamma$), 
(ii) inverse-Compton scattering of soft seed photons (CMB and/or NIR Galactic background) by energetic electrons injected by the 
magnetar wind ($e + \gamma \rightarrow e + \gamma'$), and (iii) non-thermal bremsstrahlung emission from cosmic ray (CR) electrons 
directly accelerated by the SN forward blast wave (${\rm CR}e + p \rightarrow e + p + \gamma$). We examine all three cases of 
$\gamma$-ray production next and provide simple estimates of the energetics which are then used to ascertain the feasibility of such processes 
in the present case.

\subsection{Hadronic Emission}\label{sec:hadronic}
SNRs are thought to be the dominant contributors to the Galactic CR flux up to the ``knee" at $E = 10^{15}$ eV. 
The Galactic production rate of CRs can be explained if $\eta_{\rm CR}\sim 10\%$ of the SN energy 
($\eta_{\rm CR}$ is the cosmic-ray acceleration efficiency and $E_{\rm SN}\sim 10^{51}$ erg) 
goes into accelerating CRs at SN blast waves \citep[e.g.][]{Aharonian_2004}. 
Many middle-age SNRs ($t_{\rm SNR} \sim 10^5$ yr) found in dense environments hosting GMCs have been observed 
as bright GeV and TeV sources. In fact, SNRs interacting with GMCs, as inferred from the detection of OH masers
\citep[e.g.][]{Frail+2013}, constitute the dominant fraction of Galactic GeV SNRs \citep[e.g.][]{Thompson+2012}. 
An estimate of the GeV/TeV flux from the hadronic component can be obtained with simple arguments. At high energies 
($E_\gamma > 1~{\rm GeV}$), the $\gamma$-ray spectrum is spectrally similar to the parent distribution of CR protons 
\citep[e.g.][]{Aharonian_Atoyan_1996}. Then, for a power law distribution of CR protons
\begin{equation}
n_p(E_p) = K_pE_p^{-s_p}
\end{equation}
where the normalization $K_p$ is obtained by assuming that the CR energy density 
$U_{\rm CR}\sim 3\eta_{\rm CR}E_{\rm SN}/4\pi R_{\rm SNR}^3$ \citep[e.g.][]{Drury+1994}, the $\gamma$-ray 
photon emissivity [${\rm ph~s}^{-1}~{\rm cm}^{-3}~{\rm erg}^{-1}$] is \citep[e.g.][]{Aharonian_2004}
\begin{equation}
J_\gamma(E_\gamma) = \int_{E_{p,{\rm min}}}^\infty\frac{2cn_{\rm GMC}\sigma_{\rm pp}(E_p)\eta_An_p(E_p)}{\sqrt{(E_p-m_pc^2)^2\kappa_\pi^2-m_\pi^2c^4}}dE_p
\end{equation}
where the threshold proton energy for $\pi^0$ production is $E_{p,{\rm min}} = m_pc^2+
\kappa_\pi^{-1}(E_\gamma + m_\pi^2c^4/4E_\gamma)$, $m_p$ and $m_\pi$ are the proton and $\pi^0$ masses, $\eta_A \simeq 1.5$ includes the contribution of nuclei other than protons towards the production of $\gamma$-rays 
\citep{Dermer1986}, $\kappa_\pi=0.17$ is the mean fraction of proton's kinetic energy transferred to $\pi^0$-meson per collision, 
and $n_{\rm GMC}$ is the target proton number density (assumed uniform) in the GMC. The p-p inelastic collision cross-section is given by 
\citep[e.g.][]{CR04}
\begin{equation}
\sigma_{\rm pp}(E_p) \approx 30~\left[0.95+0.06\ln\fracb{E_p}{\rm GeV}\right]~{\rm mb}
\end{equation}
which is assumed to vanish below proton kinetic energy $E_p-m_pc^2 < 1~{\rm GeV}$. Since $\sigma_{\rm pp}$ only has a weak logarithmic dependence on proton 
energy, the $\gamma$-ray spectrum is expected to reproduce the spectrum of the parent proton population. The integrated $\gamma$-ray photon flux 
in the GeV-region for $s_p = 2.15$, $E_1 = 1~{\rm GeV}$ and $E_2 = 100~{\rm GeV}$ is
\begin{equation}
\Phi_\gamma = \int_{E_1}^{E_2}\frac{J_\gamma(E'_\gamma)V_{\rm TeV}}{4\pi d^2} dE_\gamma' 
\simeq 8.16\times10^{-10} A~{\rm cm}^{-2}~{\rm s}^{-1}
\end{equation}
where $A = \eta_{\rm CR} E_{\rm SN,51}n_{\rm GMC}d_4^{-2}$. 
This agrees with the photon flux measured by \textit{Fermi}\citep{Hess+2015} in the ($1-100$) GeV region 
$\Phi \approx 8.93\times10^{-9}~{\rm cm}^{-2}~{\rm s}^{-1}$ for $\eta_{\rm CR}\sim 0.1$ and 
$n_{\rm GMC} \sim 109~{\rm cm}^{-3}$. This is a reasonable estimate of the target proton number density in light 
of the fact that \citet{Tian+2007} find a number density of $\sim 10^3~{\rm cm}^{-3}$ for the GMC associated 
with W41 from ${}^{13}$CO observations. 

Above $\sim 1$ GeV, CR protons mainly lose energy to inelastic proton-proton collisional interaction over a 
characteristic timescale \citep[e.g.][]{Gabici+09}
\begin{equation}
    \tau_{\rm pp} = (c\eta_A\sigma_{\rm pp}n_{\rm GMC})^{-1} = 2.3\times10^5n_2^{-1}~{\rm yr}~,
\end{equation}
where $n_{\rm GMC} = 10^2n_2~{\rm cm}^{-3}$.
The propagation of CR protons, that are accelerated at the SN blast wave, into the ISM is governed by 
their energy-dependent diffusion, with diffusion coefficient \citep[e.g.][]{Gabici+09}
\begin{equation}
D(E_p,B_{\rm ISM}) = 10^{28}\fracb{E_p}{10 {\rm GeV}}^\frac{1}{2}\fracb{B_{\rm ISM}}{3\mu{\rm G}}^{-\frac{1}{2}}~{\rm cm}^2\,{\rm s}^{-1}~.
\end{equation}
Assuming the SNR W41 is directly in contact with the nearby GMC, as inferred from the OH maser emission, 
the penetration depth of CR protons into the GMC is
\begin{equation}
    \ell_{\rm p} \sim \sqrt{6D(E_p,B_{\rm ISM})\tau_{\rm pp}} = \frac{0.2\,{\rm kpc}}{\sqrt{n_2}}
    \fracb{E_p/10\,{\rm GeV}}{B_{\rm ISM}/5\,\mu{\rm G}}^\frac{1}{4}\ .
\end{equation}
The above estimate shows that the CR protons are capable of reaching distances 
much larger than the size of the SNR (before they lose most of their energy), for the assumed target proton density. As the protons cool only after reaching a large displacement $\ell_p$ and traversing an even larger distance (of $c\tau_{pp}\sim72n_2^{-1}$ kpc), then the relevant density is the mean value over the region they have traversed in, or the cumulative gramage (or path integral over $n_{\rm GMC}$) along their trajectory. Most of the $p$-$p$ collisions and corresponding energy loss that goes to producing GeV/TeV emission via neutral pion decay tends to occur in the regions where  $n_{\rm GMC}$ is largest, i.e. in dense clumps within the GMC. Since the CR proton illuminated high density clump is located behind the SNR \citep[e.g.][]{Tian+2007}, 
the smaller ($R_{\rm GeV/TeV} \ll \ell_p$) observed size of the GeV/TeV is consistent with this inference.

The proton energy distribution power law index $s_p \simeq 2$ agrees with that 
obtained from Fermi acceleration at the SN blast wave. The same applies to accelerated leptons and radio observations 
of SNR W41 find a photon index of $\Gamma_{\rm rad}\simeq 1.5 - 1.6$ \citep[e.g.][]{Kassim92}, which again yields a particle 
energy distribution power law index of $s_e = 2\Gamma_{\rm rad}-1 = 2.0 - 2.2$. However, at high energies the GeV/TeV 
spectrum shows a break around $\sim 100~{\rm GeV}$, with $\Gamma_{\rm TeV} \simeq 2.6$ in the TeV region. Such breaks are 
on average observed in other SNRs with associated GeV/TeV emission \citep[e.g.][]{HYW09} where the softening of the high 
energy spectrum can be attributed to (i) overlap of $\pi^0$-decay and non-thermal bremsstrahlung spectral components, or 
(ii) diffusion of CR protons in the neighbouring dense molecular clouds. In fact, in the case of SNR W41, using an energy 
dependent diffusion coefficient $D(E_p)\propto E_p^{0.6}$, \citet{LC12} were able to explain the GeV/TeV spectrum with 
hadronic emission and high target proton density $n_{\rm GMC} \gg 1~{\rm cm}^{-3}$. Lower energy protons that produce GeV 
photons experience smaller diffusion lengths as compared to the higher energy TeV producing protons. Thus, their energy 
distribution remains close to that of freshly accelerated protons, whereas TeV producing protons are subject to diffusive 
softening.

\subsection{Leptonic Emission}\label{sec:leptonic}
\subsubsection{Inverse Compton}
High energy electrons present in the MWN can also IC scatter softer seed photons 
with energy $E_s$ to much harder energies. The high energy $\gamma$-ray emission around Swift J1834 
is dominated by GeV photons with an order of magnitude larger radiated power than TeV photons. The $\nu F_\nu$ 
flux for the GeV emission peaks around a few GeV, then the Lorentz factor of the electrons needed to IC scatter 
softer seed photons such as the CMB, with mean energy $E_{\rm CMB} = 0.63~{\rm meV}$, and NIR Galactic background, 
with typical energy $E_{\rm NIR} \simeq 0.1~{\rm eV}$, must be
\begin{equation}
    \gamma_{\rm IC} = \fracb{3E_{\rm GeV}}{4E_s}^{1/2} =
    \left\{\begin{array}{ll}
        1.2\times10^5E_{2\rm GeV}^{1/2} & (E_s = E_{\rm NIR}) \\  \\ 
        1.5\times10^6E_{2\rm GeV}^{1/2} & (E_s = E_{\rm CMB})
        \end{array}\right.
\end{equation}
where we normalized the GeV photon energy $E_{\rm GeV} = 2E_{2\rm GeV}~{\rm GeV}$. 
Their inverse-Compton cooling time is $t_{\rm IC} = 3m_ec/(4\sigma_T\gamma_{\rm IC}U_s)$ where $U_s$ 
is the energy density of the relevant seed photon field, either $U_{\rm CMB} = 4.17\times10^{-13}~{\rm erg~cm}^{-3}$ 
or $U_{\rm NIR} \approx 10^{-13}~{\rm erg~cm}^{-3}$\citep{PS05}, while their energy is $E_e\equiv L_{\rm IC}t_{\rm IC}$ 
according to the definition of $t_{\rm IC}$, so that identifying $L_{\rm IC} = L_{\rm GeV}$ we obtain
\begin{equation}
    E_e = L _{\rm GeV}t_{\rm IC} =
    \left\{\begin{array}{ll}
        3.6\times10^{50}d_4^2E_{2\rm GeV}^{-1/2}~{\rm erg} & ({\rm NIR}) \\  \\ 
        6.9\times10^{48}d_4^2E_{2\rm GeV}^{-1/2}~{\rm erg} & ({\rm CMB})
        \end{array}\right.
\end{equation}
These requisite energies are larger than the total energy in the nebula (see Equation (\ref{eq:Et1})) in both scenarios considered 
in this work, where $E(R_{\rm GeV}) = E(R_X) R_X/R_{\rm GeV} = 0.18E(R_X) < E_e$ for an 
average GeV region size $R_{\rm GeV} = 11.2d_4~{\rm pc}$.

Also, X-ray emitting electrons that have Lorentz factors
\begin{equation}
    \gamma_X = 1.1\times10^8E_2^{1/2}B_{15\mu\rm G}^{-1/2}
\end{equation}
can be particularly powerful in boosting seed CMB photons to TeV energies
\begin{equation}
    E_\gamma = \frac{4}{3}\gamma_X^2E_{\rm CMB} = 9.8B_{15\mu\rm G}^{-1}E_2~{\rm TeV}
\end{equation}
where we have made use of the relation $E_X = \hbar eB\gamma_e^2/(m_ec)$. Note that this energy is still below the Klein-Nishina 
energy\footnote{This is the case up to $E_{\rm IC}\sim 310\;$TeV that corresponds to $\gamma_e=6.1\times 10^8$, which is slightly
larger than $\gamma_{\rm max}\simeq 4.9\times10^8f^{-1/2}$, so that Klein-Nishina effects are unimportant here for CMB seed photons.
$E_{\rm KN}\sim\gamma_X m_e c^2 = 55(E_2/B_{15\mu{\rm G}})^{1/2}\;$TeV. IC scattering of NIR Galactic background photons by the same
X-ray emitting electrons is Klein-Nishina suppressed for $\gamma_e>3.8\times10^6$.}. 
Given the hard photon index measured in X-rays in the inner nebula, $\Gamma_{\rm in}=1.41\pm0.12$ in the 0.5$-$30$\;$keV energy range,
one would expect a rising or at most flattish $\nu F_\nu$ slope at TeV energies, which is in contrast with the photon index measured by
H.E.S.S. of $\Gamma_{\rm TeV}=2.64\pm0.06$ in the 0.2$-$30$\;$TeV energy range \citep{Abramowski+2015}. 
Moreover, in this picture the size of the TeV source in this case should be much smaller than observed, as the electrons that inverse-Compton 
scatter the CMB photons into the H.E.S.S. energy range are fast cooling, and their cooling length is $\ll R_{\rm TeV}\approx 11d_4\;$pc, (see fig. \ref{fig:rdist}, bottom panel) and in particular it 
should be smaller than the GeV source size, while the two are measured to be similar.

\subsubsection{Non-Thermal Bremsstrahlung}
SNRs in the Sedov phase are capable of accelerating protons and electrons to TeV energies at the forward blast wave \citep[e.g.][]{Baring+99}. 
If the ISM density is high enough, then these primary CR electrons can become the dominant contributors to the GeV emission from SNRs 
that are in close proximity to GMCs. The cooling time of electrons emitting bremsstrahlung radiation only depends on the number 
density of target protons \citep[e.g.][]{Aharonian_2004}
\begin{equation}
    t_{\rm BR} = \gamma_e\fracb{d\gamma_e}{dt}^{-1} = \frac{X_0}{cm_pn_{\rm GMC}} = 4.5\times10^5n_2^{-1}~{\rm yr}~,
\end{equation}
where we have normalized $n_{\rm GMC} = 10^2n_2~{\rm cm}^{-3}$ (see \S\ref{sec:hadronic}), and $X_0 \approx 71~{\rm g~cm}^{-2}$ is the radiation 
length over which the electron loses all but $1/e$ of its energy to bremsstrahlung photons. Comparing this time to the 
synchrotron cooling time, we find that Bremsstrahlung losses are important for electrons below
\begin{equation}
    \gamma_{\rm BR} = 2.1\times10^6n_2\fracb{B_{\rm ISM}}{5\,\mu{\rm G}}^{-2}\ ,
\end{equation}
where the ISM magnetic field is only a few $\mu\rm G$. However, at energies below $\approx 700m_ec^2$ ionization losses 
become dominant. For $\gamma_e \gg 1$, the number distribution of CR electrons assumes a power law $N(E_e) = K_eE_e^{-s_e}$, 
where the normalization can be expressed using the electron-to-proton number ratio $\Phi_{ep} < 1$, such that for 
$E_{e,{\rm max}} \gg E_{e,{\rm min}}$, $s_e > 1$, and $s_p > 2$,
\begin{equation}
    K_e = \frac{(s_e-1)(s_p-2)\Phi_{ep}\eta_{\rm CR}E_{\rm SN}}{(s_p-1)}\fracb{E_{e,{\rm min}}^{s_e-1}}{E_{p,{\rm min}}}~,
\end{equation}
where $E_{e,{\rm min}}$ and $E_{p,{\rm min}}$ are respectively the minimum energies to which electrons and protons are accelerated by the 
blast wave. Just like the hadronic emission, bremsstrahlung photons retain the same spectral shape as the parent electron distribution. 
Then for $s_e = 2\simeq \Gamma_{\rm GeV}$ corresponding to the photon index of the GeV emission, and assuming that after one radiation 
length the electron energy $E_e$ is converted into photon energy $E_\gamma$, the emitted power in the GeV band 
($E_1 = 0.35~{\rm GeV}$ and $E_2 = 100~{\rm GeV}$) is \citep{Gaisser+98}
\begin{eqnarray}
L_{\rm GeV} &&\approx \int_{E_1}^{E_2}\frac{N(E_\gamma)E_\gamma}{t_{\rm BR}}dE_\gamma = \frac{K_e}{t_{\rm BR}}\ln\fracb{E_2}{E_1} \\
&&\approx 1.5\times10^{35}\fracb{\Phi_{ep}n_{\rm GMC}E_{e,{\rm min}}}{E_{p,{\rm min}}}~{\rm erg~s}^{-1} \nonumber
\end{eqnarray}
The term in the parenthesis can easily be of order unity, in which case the bremsstrahlung emission from CR electrons 
can power the GeV emission when $n_{\rm GMC} \gg 1$. However, a detailed treatment of particle acceleration and diffusion can shed more light on this 
possibility.

\section{Discussion}\label{sec:dis}
%
%
    

The confirmation of the first ever wind nebula around a magnetar in Swift~J1834.9$-$0846 has opened up a new avenue of investigation into the 
mysterious nature of magnetars. We have carefully analysed the properties 
of this MWN, along with its central magnetar and associated SNR (W41)
and GeV/TeV source, in order to improve our understanding of this system 
and the magnetar's activity pattern. Our main conclusions can be summarised 
as follows (while at the end of the discussion we elaborate more on some of 
these points):
\begin{itemize}[leftmargin=*]
    \item The X-ray nebula emission and energy budget:
    \begin{itemize}[leftmargin=0.5cm]
        \item Comparison of the energy in the magnetic field and in the X-ray synchrotron 
        emitting $e^+e^-$ pairs (with Lorentz factor $\gamma_X$) in the nebula implies a 
        nebular magnetic field $B\sim (5-30)\sigma_e^{2/7}\;\mu$G.  
        \item The maximum magnetar polar-cap voltage difference $V_0$ implies a maximum 
        electron Lorentz factor $\gamma_{\rm max}\lesssim10^{8.5}$, so that in order for    
        $E_{\rm syn}(\gamma_{\rm max})\gtrsim 30\;$keV the nebular magnetic field must be $B\gtrsim 11\;\mu$G.
        \item In such a $B$-field, $t_{\rm syn}(\gamma_X)\ll t_{\rm SNR}$ so we see X-rays from fast cooling electrons that cool radiatively in much less than the system's age.
        \item This allows us to write the detailed energy balance equation for the X-ray emitting electrons, which leads to a  constraint on $g = \langle\dot{E}\rangle/L_{\rm sd} > g_{\rm min} = 3.07\frac{1+\sigma}{\sigma}d_4^3E_{M,30}^{7/2}f^{7/2}$ (Eq.~(\ref{eq:g-sigma}) and Fig.~\ref{fig:g_sigma}) where $\langle\dot{E}\rangle$ is the magnetar's long-term mean energy output in outflows (quiescent MHD wind + sporadic outbursts).
        \item Altogether, the result that $g>g_{\rm min}>3.1$ 
        clearly implies that the MWN is not powered predominantly by 
        the magnetar's spin-down-powered wind, but instead requires 
        an additional energy source that contributes most of its energy.
        \item The most viable candidate energy source is the decay of the 
        magnetar's magnetic field. We show that the decay of its dipole 
        field alone is not enough, and a significantly larger (by a factor 
        of $\sim10^2-10^3$) energy reservoir is needed. The most natural 
        candidate for the latter is the magnetar's internal magnetic field, 
        which has to be $\gtrsim 30$ times larger than its dipole field.
        \item If the spin-down torque is dominated by a powerful steady 
        particle wind that opens up the dipole field lines thus increasing 
        the field at the light cylinder, then this gives $n\approx 1$ but 
        implies a lower true surface dipole field. However, this would in 
        turn imply an even more extreme ratio for 
        $E_{\rm B,dip}/E_{\rm B,int} = (B_{\rm dip}/B_{\rm int})^2$, 
        which might be susceptible to instabilities, thus arguing against 
        such a solution. 
        \item By assuming a maximum allowed initial internal field strength 
        (on theoretical grounds) of $B_{\rm int,max}\sim10^{16}-10^{16.5}\;$G 
        we obtain an upper limit on $g$ of $g\leq g_{\rm max} \sim 5\times(10^2-10^3)$ 
        (see Eq.~(\ref{eq:gmax})).
        \item The SNR age is inferred to be $t_{\rm SNR}\sim5-100\;$kyr, 
        where the main uncertainty arises from the external density. 
        However, the magnatar's spin-down 
        age is significantly lower, $\tau_c=4.9\;$kyr. Reconciling the two ages 
        requires $\langle L_{\rm sd}\rangle/L_{\rm sd}\sim \tau_c/t_{\rm SNR}\sim0.05-1$, 
        suggesting either a low breaking index ($1.1\lesssim n\lesssim 2$) or 
        a widely fluctuation  spin-down torque and thus $\dot{P}$ with 
        $\langle\dot{P}\rangle/\dot{P}\sim 0.05-1$.
    \end{itemize}
    
    \item The true X-ray radiative efficiency is the fraction of the total injected energy $\langle\dot{E}\rangle = gL_{\rm sd}$ that is radiated in X-rays,  and it is smaller than $\eta_X = L_{X,{\rm tot}}/L_{\rm sd}\simeq0.13$ (which assumes energy input only from the spin-down power) by a factor of $g$: $\eta_{X,{\rm true}}=\eta_X/g = 2.6\times10^{-3}g_{50}^{-1}<0.042$.  Therefore, while the naive estimate for the efficiency ($\eta_X\simeq0.13$) appears to be very high as compared to other PWNe, the true efficiency is considerably lower and consistent with that of PWNe. 
    
        \item The short cooling time of the X-ray synchrotron emitting shock-accelerated $e^+e^-$ 
    pairs implies that their diffusion dominates over their advection throughout almost all 
    of the MWN. Their diffusion-dominated cooling length approximately matches the observed 
    size of the MWN, which may naturally explain the spectral softenning between the inner 
    and outer parts of the X-ray nebula.

        \item It is very hard to explain the GeV/TeV emission as inverse-Compton emission from the MWN:
    \begin{itemize}[leftmargin=0.5cm]
        \item NIR Galactic background seed photons require a minimal energy in the GeV emitting electrons of 
        $E_e = 3.6\times10^{50}d_4^2E_{2\rm GeV}^{-1/2}~{\rm erg}$, which is very challenging energetically, unless the 
        initial surface field was $B_0 \lesssim 10^{12}~{\rm G}$, which would give a much longer initial spin down time $t_0$, where 
        most of the energy from the central pulsar would be injected until late times and, consequently, suffer less adiabatic losses. 
        \item For the CMB seed photons this minimal energy is lower, $\simeq 7\times10^{48}\;$erg (as both the photon energy density 
        $U_\gamma$ and the electron Lorentz factor $\gamma_e$ are higher). Even this energy requires a fairly large initial spin down 
        time ($t_0\gtrsim 10^{1.5}\;$yr) and short initial spin period ($P_{0,-3}\lesssim\;$a few), and in turn an initial dipole 
        field $B_0\lesssim 10^{12.5}\;$G, which is much less than the current value, and very low for a magnetar.  This would require
        a significant field {\bf growth}. While such a scenario is discussed in Appendix~\ref{sec:appA}, its physical plausibility is highly debated. 
        \item Moreover, even for CMB seed photons, because of the 
        wide power-law electron energy distribution implied when accounting also for the X-ray emission, the inverse-Compton spectral 
        peak should be much wider than observed.
    \end{itemize}
    \item The GeV/TeV emission is much more likely of hadronic origin \citep[e.g.][]{LC12}, 
    from interactions of cosmic rays accelerated at the SNR shock with the nearby GMC 
    (which has $n_{\rm GMC}\sim 10^3\;{\rm cm^{-3}}$). This scenario is also supported by 
    the detection of OH maser emission discovered at the center of the GeV/TeV region. 
    The energetics and spectrum of the emission, as well as its location and size, find a 
    much more plausible explanation in this scenario.

\end{itemize}

It is only natural to ask why similar MWNe were not detected so far around other magnetars -- is 
Swift~J1834 indeed {\bf unique} in this respect?
One possibility is that such MWNe exist around other magnetars but their detection requires more sensitive observations
due to their relatively low X-ray emission level. In particular, an extended X-ray emission was recently reported by
\citet{Israel+2016} around SGR~J1935+2154  and although they favor a dust scattering halo origin, they cannot rule out a MWN origin.
The latter option would represent the detection of a second MWN, which would double the current MWN population due to the currently 
very {\bf small number statistics}. This obviously stresses the large observational uncertainty at present on the fraction of magnetars that 
power a MWN.

Alternatively, MWNe might indeed be intrinsically relatively {\bf rare} 
and exist only around a reasonably small fraction of magnetars.
Let us therefore consider the possible physical characteristics that might 
impact the formation of a wind nebula around a magnetar, as well as its 
X-ray brightness. The factors that might determine the existence of a 
MWN around a magnetar and its X-ray brightness, and how they vary between 
different magnetars, may be broadly divided into two main classes: intrinsic 
magnetar properties and external environmental properties.

Important {\bf intrinsic magnetar properties} for this purpose are its initial spin {\bf period}, $P_0$, which determines its initial spin energy 
($E_0=\frac{1}{2}I\Omega_0^2=2\pi^2IP_0^{-2}$) as well as its initial surface {\bf dipole field} strength, $B_0$, and its evolution throughout its lifetime
that together determine the rotational energy loss rate, $\dot{E}_{\rm rot}$. The latter is essentially the energy injection rate into a MWN 
by the magnetar's quiescent rotation powered MHD wind, $L_{\rm sd}=|\dot{E}_{\rm rot}|$, into which its rotational energy is channeled. 
In addition, the magnetar's wind {\bf pair multiplicity} can be very high due to the high magnetic pair opacity in the inner 
magnetosphere, which enhances the energy in the wind component over radiation. Therefore, the high $e^+e^-$ pair injection rate 
into the MWN and its evolution over the magnetar's lifetime affect the MWN's radiation and radiative energy loss rate through 
its effect on the electron energy distribution.

The final intrinsic magnetar property worth mentioning in this context, which naturally leads us to the environmental effects or properties, 
is its natal {\bf kick velocity}, $v_{\rm SGR}$. For Swift~J1834 one can constrain its component on the plane of the sky, $v_{\rm\perp,SGR}$, through the 
fact that it is located at the center of the SNR W41. Its location is constrained to be $\lesssim(0.05-0.1)R_{\rm SNR}$ from the SNR's center,
which for an SNR/SGR age of $t_{\rm SNR}$ implies $v_{\rm\perp,SGR}\lesssim (30-60)d_4(t_{\rm SNR}/10^{4.5}\;{\rm yr})^{-1}\;{\rm km~s^{-1}}$.
It is quite reasonable that most magnetars have larger natal kick velocities, and therefore exit their host SNR at a fairly early stage 
(soon after the SNR's velocity drops below $v_{\rm SGR}$). In such a case once the magnetar exits its SNR then its wind is no longer confined by the SNR,
and it instead forms a bow shock structure due to its motion relative to the external medium. 

This appears to be the case for SGR~1806$-$20, from a detailed modeling of the radio nebula that was produced by its 24 Dec.~2004 giant flare
\citep{Granot+2006}. In that system the bright radio emission at $\sim 1\;$week after the giant flare is attributed to a collision between a 
mildly relativistic outflow ejected from the magnetar during the giant flare and the thin bow-shock structure that is produced by its 
quiescent wind and systemic motion relative to the external medium outside of its birth SNR. Its systemic velocity that is identified with 
its natal kick velocity was inferred to be $v_{\rm SGR}\sim 250n_0^{-1/2}\;{\rm km\;s^{-1}}$ \citep{Granot+2006}. Later, the proper velocity 
of SGR~1806$-$20 was measured through its near infrared (NIR) emission \citep{Tendulkar+2012} to be $v_{\rm\perp,SGR}=350\pm100\;{\rm km~s^{-1}}$ 
for an assumed distance of $d=9\pm2\;$kpc to this source, which corresponds to $v_{\rm\perp,SGR}\approx 580d_{15}\;{\rm km~s^{-1}}$ for a distance 
of $d=15d_{15}\;$kpc to this source that is well within the inferred range. Since $v_{\rm SGR}\geq v_{\rm\perp,SGR}$ and one generally expects 
$v_{\rm SGR}\gtrsim v_{\rm\perp,SGR}$, we can parametrize $v_{\rm SGR}=\kappa v_{\rm\perp,SGR}\approx 580\kappa d_{15}\;{\rm km~s^{-1}}$. 
When combined with the results of \citet{Granot+2006}, this would imply a density around SGR~1806$-$20 of 
$n_0\sim 0.2\kappa^{-2}d_{15}^{-2}~{\rm cm}^{-3}$, as well as an outflow kinetic energy $E_{\rm ej}\sim7\times10^{45}\kappa^{-2}d_{15}^{3}\;$erg, 
and mass $M_{\rm ej}\sim5\times10^{25}\kappa^{-2}d_{15}\;$g. \citet{Granot+2006} also obtain an independent limit on the ejected mass of  
$M_{\rm ej}\gtrsim10^{25}d_{15}^{0.5}\;$g, which implies $\kappa\lesssim2.2d_{15}^{0.25}$, $v_{\rm SGR}\lesssim 1300 d_{15}^{1.25}\;{\rm km~s^{-1}}$,
$n_0\gtrsim 0.1d_{15}^{-2.5}~{\rm cm}^{-3}$, and $E_{\rm ej}\gtrsim1.4\times10^{45}d_{15}^{2.5}\;$erg. For SGR~1900+14 \citet{Tendulkar+2012} measured in 
a similar manner a proper velocity of $v_{\rm\perp,SGR}=130\pm30\;{\rm km~s^{-1}}$  for an assumed distance of $d=12.5\pm1.7\;$kpc to this source. 

This demonstrates that {\bf environmental effects} can be very important in determining whether 
a MWN is formed or not, and tightly relate to the magnetar's natal kick velocity, $v_{\rm SGR}$ -- an intrinsic property. 
For a low $v_{\rm SGR}$ the magnetar remains within its parent SNR for a long time, and the SNR confines the magnetar's shocked MHD wind, 
thus enabling the production of a prominent MWN. On the other hand, for a high $v_{\rm SGR}$ the magnetar exits its parent SNR early on, 
and then its wind is no longer effectively confined after being shocked due to its interaction with the external medium, and instead it forms 
a bow shock structure through which the shocked wind flows at a fraction of its light crossing time.\footnote{Unless shear instabilities across
the contact discontinuity separating it from the shocked external medium significantly slow it down to speeds much smaller than its relativistic
sound speed.}
Therefore, there is no efficient accumulation of energy and the emission from the bow shock system tends to be dimmer and harder to detect 
than from a MWN for the same wind power. While a good part of the wind power goes into shocking the external medium in a bow shock system, 
the {\bf composition} of the latter is predominantly electrons and protons, and therefore both the overall radiative efficiency and the fraction 
of the energy that is radiated in the X-ray range, $\eta_X$, are typically lower than for a $e^+e^-$
pair plasma composition in the shocked magnetar MHD wind with a MWN. Moreover, the {\bf external density} affects the evolution of 
the SNR and MWN as it does for PWNe (in particular affecting their size at a given age or age corresponding to a given size),
as well as the size of the bow shock structure for a given systemic velocity and wind power. 

Another case in which it can be hard to confine a MWN is when the initial rotational energy exceeds the initial SN kinetic energy, 
$E_0 = \frac{1}{2}I\Omega_0^2 > E_{\rm SN}$. For a canonical $E_{\rm SN}\approx 10^{51}\;$erg, this corresponds to $P_0\lesssim 4-5\;$ms,
so it is expected to hold for the $\alpha$-$\omega$ dynamo scenario for the formation of a magnetar strength magnetic field in the newly born NS,
which requires $P_0\lesssim 3\;$ms \citep{DT92}. For such rapid initial rotation rates the initial spin-down time $t_0$ is very short for magnetar 
strength surface dipole fields (see Equation (\ref{eq:t0})). For $B_0\gtrsim 10^{15}\;$G it can correspond to the duration of a {\bf long GRB}, 
and produce a sufficiently large initial power, $L_0\approx E_0/t_0 > 10^{50}\;{\rm erg\;s^{-1}}$, which launches a relativistic {\bf jet} that can 
penetrate the stellar envelope and potentially power a long GRB at large distances. Moreover, in such a case the jet channels most of the rotational 
energy well outside of the stellar envelope \citep[e.g.][]{Bromberg+2014,Granot+2015,Bromberg+2016} 
and hence outside of the SNR shell that forms later on after the quasi-spherical supernova shock crosses the stellar envelope, 
while only a small fraction of the jet's power contributes to enhance the supernova explosion kinetic energy. While the channel 
initially cleared by the jet might get clogged at later times, enabling the formation of a confined wind nebula, by such a later time 
only a small fraction of the initial rotational energy is left in the magnetar,
so that the energy injected into such a MWN would be correspondingly smaller. 

For a lower $B_0$ and correspondingly higher $t_0$, the magnetar wind can potenitally 
power {\bf other transient events} such as ultra-long GRBs or ultra-luminous supernovae \citep[e.g.,][and references therein]{Metzger+2015}.
For sufficiently low $B_0$ and high $t_0$ the jet might eventually not be able to penetrate the stellar envelope, and all of the 
initial rotational energy may be initially channeled into a wind nebula. However, when $E_0>E_{\rm SN}$ the stellar envelope is initially
swept-up by the wind until the cumulative wind energy $L_0t$ exceeds the envelope's initial kinetic energy $E_{\rm SN}$ at $t_c$, 
and then it is accelerated until it acquires most of the wind energy $\sim E_0$ at $t\sim t_0$.
At $t_c<t<t_0$ the SNR shell is accelerated as $R_{\rm SNR}\propto t^{3/2}$ and is susceptible to a strong {\bf Rayleigh-Taylor instability}, 
which may {\bf fragment} the SNR shell, potentially to the extent that most of the shocked pulsar wind might be able to penetrate 
between the fragments and escape out of the SNR shell altogether.
While the shell might be mended at later times, most of the initial rotational energy might escape by then, 
and again this might result in a smaller injected energy that remains in a MWN. Nonetheless, since millisecond initial spin periods  
correspond to very high $E_0\simeq 2\times10^{52}P_{0,-3}^{-2}\;$erg, even a small fraction
of such an energy that might remain to power a MWN might still be sufficient for such a MWN to be detectable.

Energy injection by {\bf burst-associated outflows} is the most natural candidate. This may naturally occur if the mean energy in such outflows 
is comparable to the radiative energy observed from these bursts. For the latter, the distribution is compatible with $dN/dE\propto E^{-5/3}$ 
(where $N$ is the number of bursts and $E$ is their radiated energy) expected for self-organized criticality 
\citep[e.g.,][]{Cheng+1996,Gogus+1999,Gogus+2000,PK12}. 
This might suggest that the total energy output (which scales as $E^2dN/dE$) is dominated by the largest events - the rare giant flares 
(although it is possible that such giant flares may comprise a separate component that is not the high energy tail of the 
self-organized critical phenomenon and are isolated events). However, the uncertainty on the power-law index of $dN/dE$ 
could accommodate a flat $E^2dN/dE$ distribution or even one that slightly goes down with energy, so it is not clear whether 
the radiative energy output, let alone that energy output in the associated outflows, is dominated by a small number of 
giant flares, or by a much larger number of much weaker events. 
Such outflows can contribute to the high X-ray efficiency both directly (through X-ray radiation from the outflow itself), 
as well as indirectly by mechanically transferring a good part of their energy to the relativistically hot shocked 
$e^+e^-$ {\bf pair wind} already present in the MWN, which may be able to radiate this energy more efficiently into the 
X-ray energy range. The latter potenitally higher $\eta_X$ in the $e^+e^-$ shocked wind in the MWN compared to the outflow itself is motivated by 
evidence for a different composition of the outflow itself. A detailed modeling of the radio nebula produced by the outflow from the 24 Dec.~2004 giant 
flare from SGR~1806$-$20 \citep{Granot+2006} suggests that this outflow contained a significant mass in {\bf baryons}, as briefly mentioned above.

The MWN around Swift~J1834 can provide an estimate of the long-term mean energy output in outflows, 
$\langle\dot{E}\rangle\approx g_{50}\times10^{36}\;{\rm erg\;s^{-1}}$
which is fairly high. If $\langle\dot{E}\rangle$ is indeed dominated by outflows from giant flares, say each of mean energy $E=10^{45.5}E_{45.5}\;$erg, 
then this would correspond to a rate for such giant flares of one per $E/\langle\dot{E}\rangle=100g_{50}^{-1}E_{45.5}\;$yr, which is compatible with the rate estimates
of giant flares from the known SGR population given three giant flares observed so far from different SGRs. 
We note, however, that both the rate of giant flares and their mean energy can gradually change 
with the SGR's age, and that the three recorded giant flares are from SGRs that appear to be younger than Swift~J1834, while $\langle\dot{E}\rangle$ corresponds 
to the mean over a fraction of its current age and therefore should represent the current mean giant flare rate and energy in outflows. 
Therefore, detailed studied of MWNe could help constrain (e.g., through the parameter $g$) the mean energy output in such outflows for a given object
at close to its current age, thus effectively averaging over hundreds to thousands of years of its past activity. In this way the MWN acts as a calorimeter
that enables us to probe the history of the magnetar's activity and its energy budget.

The rich information that can be extracted from observations of a MWN provides strong motivation to search for additional MWNe around other known magnetars. We expect better prospect for detection of MWNe around magnetars that are still within their birth SNRs, though it is still worth looking for MWNe also around magnetars without a clear SNR association.
Detailed MWN observations are a very promising tool for in-depth studies of magnetar environments, evolutionary links, and past activity, which may shed light on the fundamental differences between magnetars and other types of NSs.

\section*{Acknowledgements}
\addcontentsline{toc}{section}{Acknowledgements}
We would like to thank Yuri Lyubarsky, Dale Frail, Lara Nava for very useful discussions that helped 
to improve the quality of this work, and George Pavlov for useful comments on the manuscript. 
We are very grateful to Oleg Kargaltsev for a thorough review of the 
article and insightful discussions on pulsar wind nebulae. J.G. and R.G. acknowledge support 
from the Israeli Science Foundation under Grant No.~719/14. R.G. is supported by an Outstanding Postdoctoral 
Researcher Fellowship at the Open University of Israel.




\begin{thebibliography}{99}
\bibitem[Aharonian(2004)]{Aharonian_2004} Aharonian, F. 2004, \textit{Very high energy cosmic gamma 
radiation : a crucial window on the extreme Universe}, World Scientific Publishing, River Edge, NJ.
\bibitem[Aharonian et al.(2006)]{Aharonian+2006} Aharonian, F. et al. (H.E.S.S Collaboration) 2006, \apj, 636, 777
\bibitem[Aharonian et al.(2008)]{Aharonian+08} Aharonian, F. et al. 2008, A\&A, 486, 829
\bibitem[Abramowski et al.(2015)]{Abramowski+2015} Abramowski, A., et al. (H.E.S.S. Collaboration) 2015, A\&A, 574, 27
\bibitem[Aharonian \& Atoyan(1996)]{Aharonian_Atoyan_1996} Aharonian, F. A. \& Atoyan, A. M. 1996, Astron. Astrophys. 309, 917
\bibitem[Anderson et al.(2012)]{Anderson+12} Anderson, G. E. et al. 2012, \apj, 751, 53
\bibitem[Atoyan et al.(1994)]{AAV94} Atoyan, A. M., Aharonian, F. A., \& V\"{o}lk, H. J. 1994, Phys. Rev. D, 52, 3265
\bibitem[Baring et al.(1999)]{Baring+99} Baring, M. G. et al. 1999, \apj, 513, 311
\bibitem[Barthelmy et al.(2008)]{Barthelmy+08} Barthelmy, S. D. et al. 2008, ATel, 1676, 1
\bibitem[Blandford \& McKee(1976)]{BM76}Blandford, R.~D., \& McKee, C.~F. 1976, Phys. Fluids, 19, 1130
\bibitem[Blitz(1993)]{Blitz93} Blitz, L. 1993, in Protostars and Planets III, ed. E. H. Levy \& J. I. Lunine 
(Tucson:Univ. of Arizona), 125
\bibitem[Blondin et al.(2001)]{BCF01} Blondin, J. M., Chevalier, R. A., \& Frierson, D. M. 2001, \apj, 563, 806
\bibitem[Braithwaite(2009)]{Braithwaite09} Braithwaite, J. 2009, MNRAS, 397, 763
\bibitem[Bromberg et al.(2014)]{Bromberg+2014}Bromberg, O., Granot, J., Lyubarsky, Y., \& Piran, T. 2014, MNRAS, 443, 1532
\bibitem[Bromberg \& Tchekhovskoy(2016)]{Bromberg+2016}Bromberg, O., \& Tchekhovskoy, A. 2016, MNRAS, 456, 1739
\bibitem[Camilo et al.(2007)]{Camilo+07} Camilo, F. et al. 2007, \apj, 666, L93
\bibitem[Castro \& Slane(2010)]{CS10} Castro, D. \& Slane, P. 2010, \apj, 717, 372
\bibitem[Cheng \& Romero(2004)]{CR04} Cheng, K. S. \& Romero, G. E. 2004, in Cosmic Gamma-Ray Sources, ed. K. S. Cheng \& G. E. Romero 
(Astrophys. Space Sci. Libr. 304; Dordrecht:Kluwer)
\bibitem[Cheng et al.(1996)]{Cheng+1996}Cheng, B., Epstein, R., Guyer, R., \& Young, A. C. 1996, Nature, 382, 518
\bibitem[Chevalier(2004)]{Chevalier04} Chevalier, R. A. 2004, AdSpR, 33, 456C
\bibitem[Cline et al.(1980)]{Cline+80} Cline, T. L. et al. 1980, \apj, 237, L1
\bibitem[Colpi et al.(2000)]{Colpi+2000} Colpi M., Geppert U., Page D., 2000, ApJ, 529, L2
\bibitem[Corbel et al. (1999)]{Corbel+99} Corbel, S. et al. 1999, \apj, 526, L29
\bibitem[Cioffi et al.(1988)]{Cioffi+88} Cioffi, D. F., McKee, C. F., \& Bertschinger, E. 1988, \apj, 334, 252
\bibitem[Dall'Osso, Granot \& Piran(2012)]{DallOsso2012} Dall'Osso, S., Granot, J. \& Piran, T. 2012, MNRAS, 422, 2878
\bibitem[De Jager \& Harding(1992)]{DeJager_Harding_1992}
De Jahar, O. C., \& Harding, A. K. 1992, ApJ, 396, 161
\bibitem[Deller et al.(2012)]{Deller+12} Deller, A. T. et al. 2012, \apj, 748, L1
\bibitem[D'Elia et al.(2011)]{DElia+2011} D'Elia et al. 2011, GRB Coordinates Network, 12253, 1
\bibitem[Dermer(1986)]{Dermer1986} Dermer, C. D. 1986, A\&A, 157, 223
\bibitem[Drury et al.(1994)]{Drury+1994} Drury, L. O'C., Aharonian, F. A., \& V\"{o}lk, H. J. 1994, Astron. Astrophy. 287, 959 
\bibitem[Duncan \& Thompson(1992)]{DT92} Duncan, R. C. \& Thompson, C. 1992, \apj, 392, L9
\bibitem[Esposito et al.(2013)]{Esposito+13} Esposito, P. et al. 2013, MNRAS, 429, 3123
\bibitem[Fahlman \& Gregory(1981)]{FG81} Fahlman, G. G. \& Gregory, P. C. 1981, Nature, 293, 202
\bibitem[Frail et al.(1996)]{Frail+96} Frail, D. A. et al. 1996, AJ, 111, 1651
\bibitem[Frail et al.(2013)]{Frail+2013} Frail, D. A., Claussen, M. J., \& M\'ehault, J. 2013, \apj, 773L, 19F
\bibitem[Gabici et al.(2009)]{Gabici+09} Gabici, S., Aharonian, F. A., \& Casanova, S. 2009, MNRAS, 396, 1629
\bibitem[Gaensler \& Chatterjee(2008)]{GC08} Gaensler, B. M. \& Chatterjee, S. 2008, GCN, 8149, 1
\bibitem[Gaensler \& Slane(2006)]{GS06} Gaensler, B. M. \& Slane, P. O. 2006, ARA\&A, 44, 17G
\bibitem[Gaensler et al.(1999)]{Gaensler+99} Gaensler, B. M. et al. 1999, \apj, 526, L37
\bibitem[Gaensler et al.(2001)]{Gaensler+01} Gaensler, B. M. et al. 2001, \apj, 559, 963
\bibitem[Gaisser et al.(1998)]{Gaisser+98} Gaisser, T. K., Protheroe, R. J., \& Stanev, T. 1998, \apj, 492, 219
\bibitem[Gavriil \& Kaspi(2004)]{GK04} Gavriil, F. P. \& Kaspi, V. M. 2004, \apj, 609, L67
\bibitem[Gelfand et al.(2014)]{Gelfand+14} Gelfand, J. D. et al. 2014, Astron. Nachr., 335, 318
\bibitem[Ginzburg \& Syrovatskii(1964)]{GS64} Ginzburg, V. L. \& Syrovatskii, S. I. 1964, 
\textit{The Origin of Cosmic Rays}, Pergamon, Oxford
\bibitem[G{\"o}{\u g}{\"u}{\c s} et al.(1999)]{Gogus+1999}G{\"o}{\u g}{\"u}{\c s}, E., Woods, P. M., Kouveliotou, C., et al. 1999, ApJ, 526, L93
\bibitem[G{\"o}{\u g}{\"u}{\c s} et al.(2000)]{Gogus+2000}G{\"o}{\u g}{\"u}{\c s}, E., Woods, P. M., Kouveliotou, C., et al. 2000, ApJ, 532, L121
\bibitem[Gold(1969)]{Gold69} Gold, T. 1969, Nature, 221, 25
\bibitem[Goldreich \& Julian(1969)]{GJ69} Goldreich, P. \& Julian, W. H. 1969, \apj, 157, 869G
\bibitem[Goldreich \& Reisenegger(1992)]{GR92} Goldreich, P. \& Reisenegger, A. 1992, \apj, 395, 250
\bibitem[G\"{o}\v{g}\"{u}\c{s} \& Kouveliotou(2011)]{Gogus_Kouveliotou_2011} G\"{o}\v{g}\"{u}\c{s}, E., \& Kouveliotou, C. 2011, ATel, 3542, 1
\bibitem[G\"{o}\v{g}\"{u}\c{s} et al.(2010)]{Gogus+10} G\"{o}\v{g}\"{u}\c{s}, E. et al. 2010, \apj, 722, 899
\bibitem[Gotthelf \& Vasisht(1998)]{GV98} Gotthelf, E. V. \& Vasisht, G. 1998, NewA, 3, 293
\bibitem[Gotthelf et al.(2000)]{Gotthelf+00} Gotthelf, E. V. et al. 2000, \apj, 542, L37
\bibitem[Guiriec et al.(2011)]{Guiriec+2011} Guiriec, S., Kouveliotou, C. \& van der Horst, A. J., 2011, GRB Coordinates Network, 12255, 1
\bibitem[Granot et al.(2006)]{Granot+2006} Granot, J., et al. 2006, ApJ, 638, 391
\bibitem[Granot et al.(2015)]{Granot+2015}Granot, J., Piran, T., Bromberg, O., Racusin, J.~L., \& Daigne, F. 2015, SSRv, 191, 471
\bibitem[Guilbert et al.(1983)]{GFR83} Guilbert, P. W., Fabian, A. C., \& Rees, M. J. 1983, MNRAS, 205, 593
\bibitem[Halpern \& Gotthelf(2010)]{HG10} Halpern, J. P. \& Gotthelf, E. V. 2010, \apj, 710, 941
\bibitem[Harding et al.(1999)]{Harding+99} Harding, A. K., Contopoulos, I., \& Kazanas, D. 1999, \apj, 525, L125
\bibitem[H.E.S.S. Collaboration(2015)]{Hess+2015} H.E.S.S. Collaboration. 2015, A\&A, 574A, 27H
\bibitem[Hewitt et al.(2009)]{HYW09} Hewitt, J. W., Yusef-Zadeh, H., \& Wardle, M. 2009, \apj, 706, L270
\bibitem[Hillas et al.(1998)]{Hillas_1998}
Hillas, A. M., et al. 1998, ApJ, 503, 744
\bibitem[Ho(2015)]{Ho15} Ho, W. C. G. 2015, MNRAS, 414, 2567
\bibitem[Hurley-Walker et al.(2009)]{AMI09} Hurley-Walker, N. et al. (AMI Consortium) 2009, MNRAS, 396, 365
\bibitem[Israel et al.(2016)]{Israel+2016}Israel, G.~L., et al., 2016, MNRAS, 457, 3448
\bibitem[Kargaltsev \& Pavlov(2008)]{KP08} Kargaltsev, O. \& Pavlov, G. G. 2008, 40 Years of Pulsars: Millisecond Pulsars, Magnetars and More, 
ed. {Bassa}, C. and {Wang}, Z. and {Cumming}, A. and {Kaspi}, V.~M., American Institute of Physics Conference Series, V. 983, 171
\bibitem[Kargaltsev et al.(2012)]{Kargaltsev_2012}
Kargaltsev, O. et al. 2012, ApJ, 748, 26
\bibitem[Kassim(1992)]{Kassim92} Kassim, N. E. 1992, AJ, 103, 943
\bibitem[Kennel \& Coroniti(1984)]{Kennel_Coroniti_1984} Kennel, C. F. \& Coroniti, F. V. 1984, \apj, 283, 694
\bibitem[Kothes \& Foster(2012)]{KF12} Kothes, R. \& Foster, T. 2012, \apj, 746, L4
\bibitem[Kouveliotou et al.(1998)]{Kouveliotou+98} Kouveliotou, C. et al. 1998, IAU, Circ. 694
\bibitem[Kuiper \& Hermsen(2011)]{Kuiper_Hermsen_2011} Kuiper, L, \& Hermsen, W. 2011, ATel, 3577, 1
\bibitem[Leahy \& Tian(2007)]{LT07} Leahy, D. A. \& Tian, W. W. 2007, \apj, A\&A, 461, 1013
\bibitem[Leahy \& Tian(2008a)]{LH08} Leahy, D. A. \& Tian, W. W. 2008, \apj, 135, 167
\bibitem[Leahy \& Tian(2008b)]{LT08b} Leahy, D. A. \& Tian, W. W. 2008, A\&A, 480, L25
\bibitem[Levin et al.(2010)]{Levin+10} Levin, L. et al. 2010, \apj, 721, L33
\bibitem[Li \& Chen(2012)]{LC12} Li, H. \& Chen, Y. 2012, MNRAS, 421, 935
\bibitem[Marshall et al.(2016)]{Marshall+16} Marshall, F. E. et al. 2016, ArXiv:1608.01901
\bibitem[Metzger et al.(2015)]{Metzger+2015}Metzger, B.~D., Margalit, B., Kasen, D., \& Quataert, E. 2015, MNRAS, 454, 3311
\bibitem[Meyer et al.(2010)]{Meyer_2010}
Meyer, M., Horns, D., \& Zechlin, H.-S. 2010, A\&A, 523, A2
\bibitem[Mezets et al.(1979)]{Mazets+79} Mazets, E. P. et al. 1979, Nature, 282, 587
\bibitem[Mizuno et al.(2011)]{Mizuno_2011}
Mizuno, Y., Lyubarsky, Y.,  Nishikawa, K.-I., \&  Hardee, P. E. 2011, ApJ, 728, 90
\bibitem[Muslimov \& Page(1996)]{MP96} Muslimov, A. \& Page, D. 1996, \apj, 458, 347
\bibitem[Nolan et al.(2012)]{Nolan+2012} Nolan P. L. et al. (Fermi Collaboration) 2012, ApJS, 199, 31
\bibitem[Olausen \& Kaspi(2014)]{OK14} Olausen, S. A. \& Kaspi, V. M. 2014, ApJS, 212, 6
\bibitem[Ostriker \& Gunn(1971)]{OG71} Ostriker, J. P. \& Gunn, J. E. 1971, \apj, 164, L95
\bibitem[Pacholczyk(1970)]{Pacholczyk1970} Pacholczyk, A. G. 1970
\bibitem[Park et al.(2012)]{Park+12} Park, S. et al. 2012, \apj, 748, 117
\bibitem[Porter \& Strong(2005)]{PS05} Porter, T. A. \& Strong, A. W. 2005, ICRC, 4, 77P
\bibitem[Porth, Komissarov, \& Keppens(2013)]{Porth+2013} Porth, O., Komissarov, S. S., 
\& Keppens, R. MNRAS, 2013, 431, L48
\bibitem[Porth, Komissarov, \& Keppens(2014)]{Porth+2014} Porth, O., Komissarov, S. S., 
\& Keppens, R. 2014, MNRAS, 438,278
\bibitem[Prieskorn \& Kaaret(2012)]{PK12}Prieskorn, Z., \& Kaaret, P. 2012, ApJ, 755, 1	
\bibitem[Rogers \& Safi-Harb(2016)]{RS16} Rogers, A. \& Safi-Harb, S. 2016, MNRAS, 457, 1180
\bibitem[Reynolds \& Chevalier(1984)]{RC84} Reynolds, S. P. \& Chevalier, R. A. 1984, \apj, 278, 630
\bibitem[Reynolds et al.(2012)]{RGB12} Reynolds, S, P., Gaensler, B. M., \& Bocchino, F. 2012, SSRv, 166, 231
\bibitem[Sarma et al.(1997)]{Sarma+97} Sarma, A. P. et al. 1997, \apj, 483, 335
\bibitem[Shu(1992)]{Shu92} Shu, F. H. 1992, The Physics of Astrophysics: vol II Gas Dynamics (University 
Science Books, Mill Valley)
\bibitem[Spitkovsky(2006)]{Spitkovsky06} Spitkovsky, A. 2006, \apj, 648, L51
\bibitem[Tendulkar(2013)]{Tendulkar13} Tendulkar, S. P. 2013, IAUS, 291, 514
\bibitem[Tendulkar et al.(2012)]{Tendulkar+2012} Tendulkar, S. P., Cameron, P. B., \& Kulkarni, S. R. 2012, \apj, 761, 76
\bibitem[Tendulkar et al.(2013)]{Tendulkar+13} Tendulkar, S. P. et al. 2013, \apj, 772, 31
\bibitem[Thompson \& Blaes(1998)]{TB98} Thompson, C. \& Blaes, O. 1998, Phys. Rev. D, 57, 3219
\bibitem[Thompson \& Duncan(1993)]{TD93} Thompson, C. \& Duncan, R. C. 1993, \apj, 408, 194
\bibitem[Thompson et al.(2012)]{Thompson+2012} Thompson, D. J., Baldini, L., \& Uchiyama, Y. (2012), APh, 39, 22 
\bibitem[Tian et al.(2007)]{Tian+2007} Tian, W. W. et al. 2007, \apj, 657, L25
\bibitem[Tian \& Leahy(2008)]{TL08} Tian, W. W. \& Leahy, D. A. 2008, MNRAS, 391, 54
\bibitem[Tian \& Leahy(2012)]{TL12} Tian, W. W. \& Leahy, D. A. 2012, MNRAS, 421, 2593
\bibitem[Tong(2016)]{Tong16} Tong, H. 2016, ArXiv:1605.00522
\bibitem[Tong et al.(2013)]{Tong+13} Tong, H. et al. 2013, \apj, 768, 144
\bibitem[Torii et al.(1998)]{Torii+98} Torii, K. et al. 1998, \apj, 503, 843
\bibitem[Turolla, Zane, Watts(2015)]{TZW15} Turolla, R., Zane, S., \& Watts, A. L. 2015, Rep. Prog. Phys., 78, 116901
\bibitem[Van der Swaluw et al.(2001)]{Swaluw+01} Van der Swaluw, E. et al. 2001, A\&A, 380, 309
\bibitem[Vasisht \& Gotthelf(1997)]{VG97} Vasisht, G. \& Gotthelf, E. V. 1997, \apj, 486, L129
\bibitem[Vasisht et al.(2000)]{Vasisht+00} Vasisht, G. et al. 2000, \apj, 542, L49
\bibitem[Watcher et al.(2004)]{Watcher+04} Watcher, S. et al. 2004, \apj, 615, 887
\bibitem[Woods et al.(2002)]{Woods+02} Woods, P. M. et al. 2002, \apj, 576, 381
\bibitem[Woods et al.(2007)]{Woods+07} Woods, P. M. et al. 2007, \apj, 654, 470
\bibitem[Younes et al.(2012)]{Y+12} Younes, G. et al. 2012, \apj, 757, 39
\bibitem[Younes et al.(2015)]{Y+15} Younes, G. et al. 2015, \apj, 809, 165
\bibitem[Younes et al.(2016)]{Y+16} Younes, G. et al. 2016, \apj, 824, 138
\end{thebibliography}



\onecolumn
\appendix
\section{Braking index with magnetic field decay}
\label{sec:appA}
Equation~(\ref{eq:ndef}) for the spin-down law can be generalized to allow for a dependence on the (generally time dependent) surface dipole field $B_s(t)$,
\begin{equation}
    \dot{\Omega} = -KB_s^2(t)\Omega^n \quad\Longleftrightarrow\quad \dot{P} = (2\pi)^{n-1}KB_s^2(t)P^{2-n}\ .
    \label{eq:sdlaw}
\end{equation}
This equation reduces to magnetic dipole spin-down for $n=3$ and $K = fR_{\rm NS}^6/(Ic^3)$, and it reduces to Equation~(\ref{eq:ndef}) for a constant $B_s$.
In this case the braking index that is inferred from observation is 
\begin{equation}
    n' \equiv \frac{\ddot\Omega\Omega}{\dot\Omega^2} = n + \frac{2\dot B_s\Omega}{B_s\dot\Omega}\equiv n + \Delta n 
\end{equation}
where the difference from the standard braking index is given by
\begin{equation}
    \Delta n = \frac{2\dot B_s\Omega}{B_s\dot\Omega} = - \frac{2\dot{B}_sP}{B_s\dot{P}} 
    = -\frac{d\log(B_s^2)}{d\log t}\left(\frac{d\log P}{d\log t}\right)^{-1}\ .
\end{equation}
Note the standard spin-down relation is recovered when $\dot B$ vanishes. Also, since $\dot{P}>0$, 
a decaying magnetic field ($\dot{B}_s<0$) yields $\Delta n>0$. However, since it is hard to measure $\ddot\Omega$ 
for the majority of pulsars and especially for magnetars due to large timing noise, $n'$ can't be measured directly. On the other hand, the standard 
braking index $n$ can be ascertained by comparing the characteristic spin-down age to that of the host SNR. 

Assuming a magnetic field time evolution of the form\footnote{This corresponds to the field decay parameterization of \citet{DallOsso2012} for $\alpha>0$ and $t_B\to\tau_{d,i}/\alpha$. They consider only a decay of the magnetic field, and for $\alpha\leq0$ they obtain a different functional form (exponential decay for $\alpha=0$ and decay to zero over a finite time for $\alpha<0$). For our parameterization $\alpha<0$ corresponds to a growth of the magnetic field, which was not considered by \citet{DallOsso2012}, as such a growth is not expected physically under most scenarios.}
\begin{equation}
    B_s(t) = B_0\left(1+\frac{t}{t_B}\right)^{-1/\alpha} = B_0\tau^{-1/\alpha}\ ,\quad\quad \frac{d\log(B_s^2)}{d\log\tau} = -\frac{2}{\alpha}\ ,
    \label{eq:B_evol}
\end{equation}
with some characteristic magnetic field decay timescale $t_B$, where we conveniently define $\tau\equiv 1+t/t_B$, for which
\begin{equation}
    \Delta n = -\frac{d\log(B_s^2)}{d\log \tau}\left(\frac{d\log P}{d\log\tau}\right)^{-1}\ .
\end{equation}
Integrating Equation~(\ref{eq:sdlaw}) over time gives
\begin{equation}
\frac{P(t)}{P_0} = \frac{\Omega_0}{\Omega(t)} = \left[1+\frac{\alpha t_B}{(\alpha-2)t_0}\left(\left\{1+\frac{t}{t_B}\right\}^\frac{\alpha-2}{\alpha}-1\right)\right]^{\frac{1}{n-1}}
    = \left[1+\frac{\alpha t_B}{(\alpha-2)t_0}\left(\tau^\frac{\alpha-2}{\alpha}-1\right)\right]^{\frac{1}{n-1}}\ ,
    \quad\quad\rm{for}~\alpha\neq2
    \label{eq:P_evol}
\end{equation}
where the initial spin-down time
\begin{equation}
    t_0 = \frac{1}{(n-1)KB_0^2}\fracb{P_0}{2\pi}^{n-1} = \frac{P_0}{(n-1)\dot{P}}\fracb{P}{P_0}^{2-n}\tau^{-\frac{2}{\alpha}}
    \xrightarrow{\: |\alpha| \to \infty \: } \frac{P_0}{(n-1)\dot{P}}\fracb{P}{P_0}^{2-n}\ .
\end{equation}
This implies the following spin-down luminosity,
\begin{equation}
    L_{\rm sd} = 4\pi^2I\frac{\dot{P}}{P^3} = L_0\fracb{P}{P_0}^{-n-1}\tau^{-\frac{2}{\alpha}} = 
    L_0\left\{1+\frac{t}{t_B}\right\}^{-\frac{2}{\alpha}}\left[1+\frac{\alpha t_B}{(\alpha-2)t_0}\left(\left\{1+\frac{t}{t_B}\right\}^\frac{\alpha-2}{\alpha}-1\right)\right]^{-\frac{n+1}{n-1}}
\end{equation}
where $L_0 = fR_{\rm NS}^6B_0^2\Omega_0^4/c^3$ for magnetic dipole braking ($n=3$), and more generally $L_0 = 2E_0/(n-1)t_0$.

\begin{figure}
\centering
\includegraphics[width=0.7\textwidth]{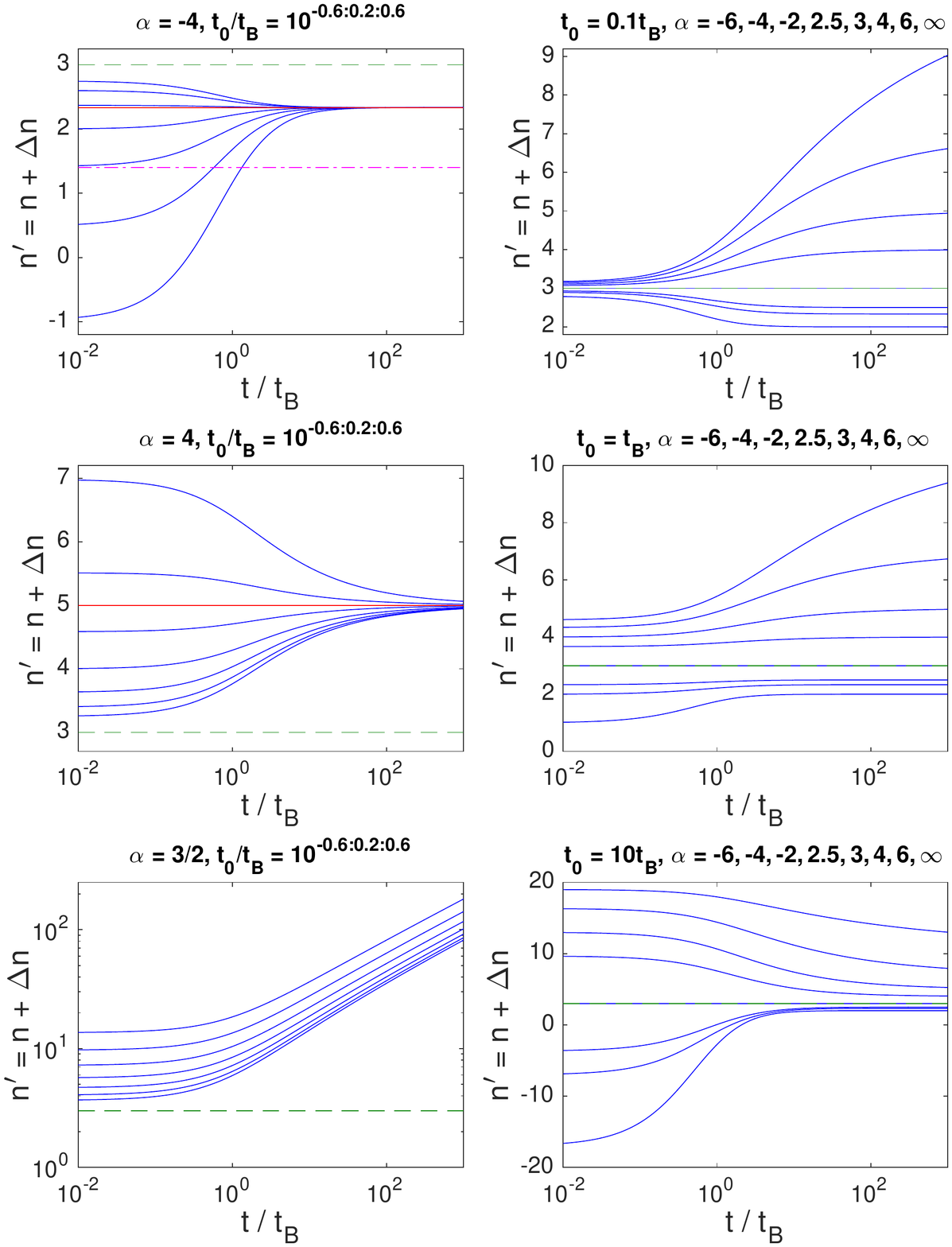}
\caption{The evolution of the true braking index $n'=n+\Delta n$ as a function of $t/t_B$ is shown for several representative cases,
with $n=3$ and varying $t_0/t_B$ and $\alpha$.}
\label{fig:n-prime}
\end{figure}

\subsection{Spin-Down Freezout}
Note that when the magnetic field decays rapidly enough, $\alpha<2$, the spin-down freezes out at late times, and the rotational period
approaches a constant asymptotic value \citep{DallOsso2012},
\begin{equation}
    P_\infty = P_0\left[1+\frac{\alpha t_B/t_0}{(2-\alpha)}\right]^{\frac{1}{n-1}} \quad\Longleftrightarrow\quad
    P_{\infty}^{n-1} = P_0^{n-1} +\frac{(n-1)KB_0^2\alpha t_B}{(2\pi)^{1-n}(2-\alpha)}\longrightarrow P_{\infty}^{2} 
    = P_{0}^{2} +\frac{8\pi^{2}fR_{\rm NS}^6B_0^2\alpha t_B}{(2-\alpha)Ic^3} \ .
\end{equation}
Equation~(\ref{eq:P_evol}) implies
\begin{equation}
\frac{d\log P}{d\log\tau} = \frac{t_B/t_0}{n-1}\,\tau^\frac{\alpha-2}{\alpha}
\left[1+\frac{\alpha t_B}{(\alpha-2)t_0}\left(\tau^\frac{\alpha-2}{\alpha}-1\right)\right]^{-1}\ ,
\quad\quad\rm{for}~\alpha\neq-2\ ,
\end{equation}
and a difference from the standard braking index of
\begin{equation}
    \Delta n = \frac{2\alpha\Omega^{1-n}}{KB_0^2t_B}\left(1+\frac{t}{t_B}\right)^{-(2\alpha+1)}
    = (n-1)\frac{2t_0}{\alpha t_B}\,\tau^\frac{2-\alpha}{\alpha}\left[1+\frac{\alpha t_B}{(\alpha-2)t_0}\left(\tau^\frac{\alpha-2}{\alpha}-1\right)\right]\ .
\end{equation}
Initially, at $t=0$, we have
\begin{equation}
    \Delta n_{i} = (n-1)\frac{2t_0}{\alpha t_B} \ ,\quad\quad n'_{i} = n+\Delta n_{i} = n +(n-1)\frac{2t_0}{\alpha t_B}\ . 
\end{equation}
For $\alpha < 2$ at late times the braking index grows as a power-law, $n\ll n'\approx \Delta n \propto t^{(2-\alpha)/\alpha}$.
For $\alpha > 2$ the following asymptotic value is approached at late times,
\begin{equation}
    \Delta n_\infty = \frac{2(n-1)}{\alpha-2}\ ,\quad\quad n'_{\infty} = n+\Delta n_\infty = \frac{n\alpha-2}{\alpha-2}\ .
\end{equation}
Equation~(\ref{eq:sdlaw}) corresponds to $\dot{P}\propto B^2 P^{2-n}$, and imposing a magnetic dipole braking spin-down rate for which 
$\dot{P}\propto B^2 P^{-1}$ corresponds to $n = 3$. Under this restriction, for $\alpha > 2$ one has
\begin{equation}
    n'_{\infty} = \frac{3\alpha-2}{\alpha-2} \quad\Longleftrightarrow\quad \alpha = \frac{2(n'_{\infty}-1)}{n'_{\infty}-3}\ ,
\end{equation}
which corresponds to the scaling of Equation~(\ref{eq:B-evol}) under the substitution $n\to n'_{\infty}$.
The evolution of the true braking index $n'=n+\Delta n$ for several representative cases is shown in Figure~\ref{fig:n-prime}.

\subsection{Solving for $\alpha$ and $t_B$}
\begin{figure}
    \centering
    \includegraphics{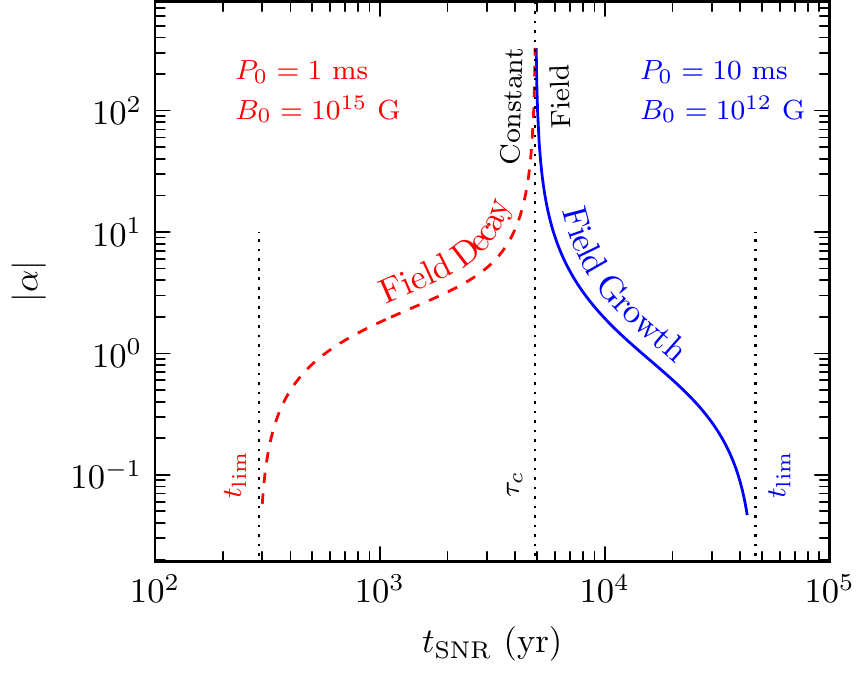}
    \includegraphics{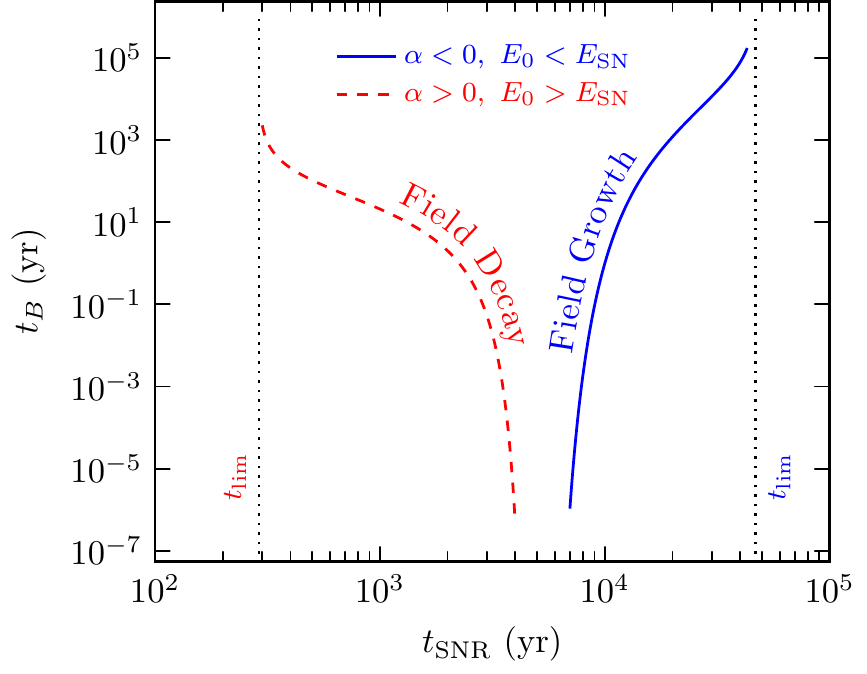}
    \caption{\textit{Left}: Magnetic field power law index as a function of the age of 
    the system $t_{\rm SNR}$ for $n = 3$ and $f=1$. (Solid-line) The magnetar is born as an ordinary 
    NS whose surface magnetic field grows over its lifetime to the currently 
    inferred value $B_s = 1.16\times10^{14}~{\rm G}$ from $P$ and $\dot P$. 
    Here $\alpha < 0$, but we plot $\vert\alpha\vert$ for clarity. 
    (Dashed-line) The magnetar is born as a fast rotator with $P_{0,-3} = 1$ 
    and strong surface field $B_0 = 10^{15}G$ which decays over its lifetime 
    to the current surface field $B_s$. In both scenarios, $\vert\alpha\vert\to\infty$ 
    at $t_{\rm SNR} = \tau_c = 4.9~{\rm kyr}$ for which the surface field remains constant at its 
    initial value, $B_s(t) = B_0$. \textit{Right}: Magnetic field decay/growth time $t_B$ as a function 
    of the age of the system $t_{\rm SNR}$ for the two cases shown on the left. 
    The vertical dashed lines show the limiting value of $t_{\rm SNR}$ for $|\alpha|\to0$ from 
    Equation~(\ref{eq:tlimit}).}
    \label{fig:alpha-tB-plot}
\end{figure}

\begin{figure}
\centering
\includegraphics[width=0.477\textwidth]{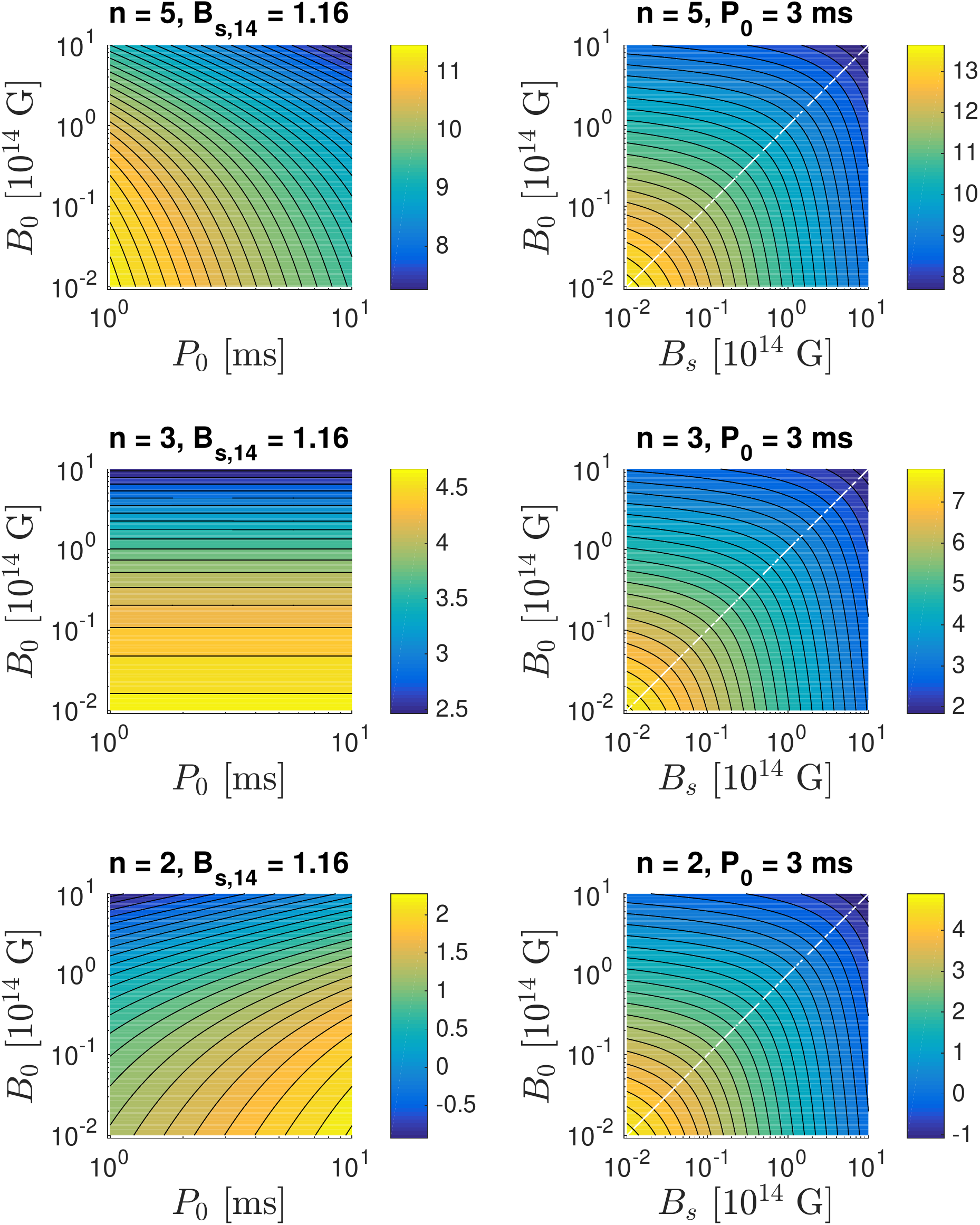}
\caption{Contour plots of $\log_{10}(t_{\rm lim}~[{\rm yr}])$ for the 
limiting time given in Equation~(\ref{eq:tlimit}) for different values 
of the relevant parameters. The remaining parameters are fixed to their 
fiducial values from the main text.}
\label{fig:t_lim}
\end{figure}

A useful constraint on the magnetic field decay index $\alpha$ can be obtained by using the present spin period and inferred surface 
magnetic field of Swift J1834. By using the magnetic field time evolution from Equation~(\ref{eq:B_evol}) and the time evolution of the 
spin period, Equation~(\ref{eq:P_evol}) can be cast into the following form
\begin{equation}
    \frac{P(t)}{P_0} = \left[1+\left\{\fracb{B_s}{B_0}^{-\alpha}-1\right\}^{-1}\frac{\alpha t}{(\alpha-2)t_0}
    \left\{\fracb{B_s}{B_0}^{2-\alpha}-1\right\}\right]^\frac{1}{n-1}\ .
    \label{eq:PB_evol}
\end{equation}
Here we look at two scenarios and assume standard magnetic dipole braking with $n=3$: 
(i) the magnetar begins its life as a regular pulsar with initial spin period $P_0 \sim 10~{\rm ms}$ 
and surface magnetic field $B_0 \sim 10^{12}~{\rm G}$, but experiences a growth in surface field by the time $t = t_{\rm SNR}$, 
or (ii) the proto-NS is a rapid rotator with $P_0 \sim 1~{\rm ms}$ and quickly ramps up its surface magnetic field under the action 
of the $\alpha$-$\omega$ dynamo mechanism, so that $B_0 \sim 10^{15}~{\rm G}$, which then decays over the age of the system to its 
current value. In Figure~\ref{fig:alpha-tB-plot} we solve Equation~(\ref{eq:PB_evol}) for $\alpha$ given that the current spin period $P = 2.48~{\rm s}$ and the surface magnetic field is that inferred from $P$ and $\dot P$, $B_s = 1.16\times10^{14}f^{-1/2}~{\rm G}$.
In both cases, as $|\alpha|\ll1$, the age of the system $t_{\rm SNR}$ in Equation~(\ref{eq:PB_evol}) approaches a limiting value for a given set of 
($P_0$, $P$, $B_0$, $B_s$, $n$) parameters
\begin{equation}
    t_{\rm lim} = t_{\rm SNR}(|\alpha|\to 0) = 2t_0\ln\fracb{B_s}{B_0}\left[\fracb{B_s}{B_0}^2-1\right]^{-1}\left[\fracb{P}{P_0}^{n-1}-1\right]\ ,
    \label{eq:tlimit}
\end{equation}
which corresponds to $t_B\rightarrow\infty$. In the opposite limit, as $\alpha\rightarrow\pm\infty$, which implies constant surface field $B_s(t)=B_0$, 
$t_B\rightarrow 0$ and the age of the system approaches the characteristic age 
$t_{\rm SNR}\to\tau_c=4.9~{\rm kyr}$.
We show the values of $t_{\rm lim}$ for different parameter values in Figure~\ref{fig:t_lim}.

The constraint on $\alpha$ can be further used to constrain the characteristic field decay/growth time $t_B$, as shown 
in Fig.(\ref{fig:alpha-tB-plot}), using
\begin{equation}
    \frac{t_B}{t_{\rm SNR}} = \left[\fracb{B_s}{B_0}^{-\alpha}-1\right]^{-1}\ .
\end{equation}
For $\alpha > 0$, the corresponding field decay time can be much shorter than $10^3\;$yr, however, 
such short decay times are unphysical. Field decay in magnetars is believed to be occuring in their crust due to Hall drift or Ohmic 
decay modes for which the characteristic decay time is $t_B\sim10^{3}-10^5~{\rm yr}$ \citep[e.g.,][]{GR92}.

\section{Magnetar Wind Nebula Dynamical Evolution}
\label{sec:appB}
We consider the expansion of the MWN inside the freely expanding SNR and its interaction with the ISM. 
This depends on the density profile of the unshocked SN ejecta and on the details of how energy is exchanged 
between it and the shocked magnetar wind. Radiative losses play an important role, however for simplicity we only consider the adiabatic expansion of the nebula. Following 
\citet{BCF01}, the density can be modeled as two components, separated by a transition radius 
$r_t = v_t t$, with a spatially flat inner profile given by
\be
\rho_{\rm ej}(r,t) = \frac{5q-25}{2\pi q}E_{\rm SN}v_t^{-5}t^{-3}\quad\quad{\rm for}\quad r < v_t t
\ee
and a steep outer profile with $\rho_{\rm ej} \propto r^{-q}t^{q-3}$; we use $q = 10$ as a fiducial index. The ejecta expands 
ballistically with transition velocity
\be
v_t = \left(\frac{10q-50}{3q-9}\frac{E_{\rm SN}}{M_{\rm ej}}\right)^{1/2} = 6.3\times10^8E_{\rm SN,51}^{1/2}M_3^{-1/2}~{\rm cm~s}^{-1}
\ee
The kinetic energy of the expanding density core
\be
E_c = \frac{1}{2}M_cv_t^2 = \fracb{q-5}{q}E_{\rm SN} = \frac{1}{2}E_{\rm SN}
\ee
delineates the two scenarios in which the magnetar is either energetically dominant ($E_0 > E_c$) and significantly 
alters the dynamical evolution of the SNR or sub-dominant ($E_0 < E_c$) and mirrors the standard case of PWNe. 
\subsection{$E_0 > E_c$}
The magnetar continues to inject energy for $t < t_0$, and the initial expansion of the MWN is the same for 
both cases. Ignoring radiative losses, the expansion of the MWN in the \textit{thin-shell} 
approximation is governed by \citep[e.g.][]{OG71,RC84}
\begin{eqnarray}
    LR &=& \frac{d}{dt}(4\pi pR^4)\\
    M_s \ddot R &=& 4\pi R^2[p - \rho_{\rm ej}(\dot R-v_{\rm ej})^2]
\label{eq:pwnR}
\end{eqnarray}
where $R$ and $p$ are the radius and pressure inside the MWN, $M_s$ is the swept up ejecta mass, and 
$L = L_0$ for $t < t_0$ is the power injected by the magnetar. With radius expanding as a power law in time 
$R \propto t^a$, the internal energy of the MWN increases linearly with time
\begin{equation}
    U = 4\pi R^3p = \frac{L_0 t}{a+1}
\end{equation}
Using this result in Equation~(\ref{eq:pwnR}) and noting that $v_{\rm ej} = R/t$ for $v_{\rm ej} < v_t$, 
gives $a = 6/5$ \citep{RC84,BCF01}. The spinning down magnetar injects most of its rotational energy 
$E_0$ at $t \sim t_0$, and if $E_0 \gtrsim E_c$, the evolution of the MWN will 
be significantly different from what's generally observed for PWNe. Whether this is indeed the case 
can be learned by comparing $t_c$, the time at which the MWN reaches the edge of the density 
core, to $t_0$, the time after which the rate of energy injection by the magnetar decreases significantly. 
Then, for $R(t) = R_t = v_t t$, the core crossing time and radius is \citep{BCF01}
\begin{eqnarray}\label{eq:tc_over_to}
\frac{t_c}{t_0} &&= 2.64\fracb{q-5}{q}\fracb{n-1}{2}\frac{E_{\rm SN}}{E_0} = 0.07\fracb{n-1}{2}P_{0,-3}^2E_{\rm SN,51} \\
R_c &&=v_tt_c = 2.64\fracb{n-1}{2}\fracb{q-5}{q}
\left(\frac{10q-50}{3q-9}\frac{E_{\rm SN}^3}{M_{\rm ej}}\right)^{1/2}\frac{t_0}{E_0} = 
1.4\times10^{12}P_{0,-3}^4E_{\rm SN,51}^{3/2}f^{-1}B_{14}^{-2}M_3^{-1/2}~{\rm cm}
\end{eqnarray}
For $t < t_c$, the ejecta mass will accumulate in a thin shell at the contact discontinuity that 
separates the relativistically hot MWN gas and the unshocked SNR ejecta, with its radius growing as
\begin{equation}
    R(t) = R_c\fracb{t}{t_c}^{6/5}
\end{equation}
For $t_0 > t > t_c$, the expansion of the MWN will accelerate down the steep density gradient while it is still 
being energized by the magnetar. At this point the swept up mass $\sim M_{\rm ej}$ and it can be shown that the 
radius of the MWN will grow as \citep{RC84}
\begin{equation}
    R(t) = R_c\fracb{t}{t_c}^{3/2}
\end{equation}
At this point, the dynamical evolution of the SNR differs from the canonical case of a point explosion with 
no further energy injection. Here the SNR volume is replaced by that of the MWN and the forward blast wave 
continues to accelerate in the ISM until $t = t_0$. Here, it should be noted that the contact discontinuity 
between the relativistically hot bubble and the unshocked ISM is highly susceptible to its fragmentation by the 
Rayleigh-Taylor instability. Consequently, the hot gas will escape through gaps in the fragmented shell and 
directly interact with the cold ISM. The treatment of this phase is out of the scope of this work and is 
left for future study.

The MWN blast wave will begin to coast at a constant velocity for $t > t_0$ until the onset of the Sedov-Taylor phase, with 
\begin{equation}
    R(t) = R_0\fracb{t}{t_0}
\end{equation}
where
\begin{equation}
    R_0 = \left[\frac{1}{2.64}\fracb{2}{n-1}\fracb{q}{q-5}\fracb{10q-50}{3q-9}\frac{E_0}{M_{\rm ej}}\right]^{1/2}t_0 
    = 8.3\times10^{13}f^{-1}B_{14}^{-2}P_{0,-3}M_3^{-1/2}~{\rm cm}~.
\end{equation}
\subsection{$E_0 < E_c$}
In this case, the energy injected by the magnetar is not large enough and the MWN never crosses the 
entire density core, such that $t_0 < t_c$, which translates into a lower bound on the initial 
spin period and an upper bound on $E_0$
\begin{eqnarray}
    P_0 &&> 4.1\left[\frac{n-1}{2}E_{\rm SN,51}\right]^{-1/2}~{\rm ms} \\
    E_0 &&< 1.2\times10^{51}\left(\frac{n-1}{2}\right)E_{\rm SN,51}~{\rm erg}
\end{eqnarray}
This limit is larger (unless $n > 4$) than the upper bound of $P_0 \lesssim 3~{\rm ms}$ conjectured by
\citep{DT92,TD93} for 
the formation of a magnetar by magnetic field amplification through the action of an $\alpha-\omega$ 
dynamo, which would operate in the convective and differentially rotating cores of rapidly spinning proto-NSs. 
However, as shown in Appendix.(\ref{sec:appA}), magnetic field decay, which powers 
the high quiescent luminosity and 
bursting activity in magnetars, tends to produce high braking indices, $n > 3$. On the other hand, if $n < 3$ 
then it's unclear how the pulsar evolves into a magnetar. In this case, the wind nebula and its interaction 
with the SNR ejecta would proceed in a way much similar to the canonical case of PWNe, where its size would 
grow as
\begin{equation}
    R(t) = R_0\fracb{t}{t_0}^a
\end{equation}
where $a = 6/5$ for $t < t_0$ and $a = 1$ thereafter, until it is crushed by the reverse shock \citep{BCF01}, with
\begin{equation}
    R_0 = 1.50\fracb{2 q}{(q-5)(n-1)}^\frac{1}{5}\fracb{q-5}{q-3}^\frac{1}{2}\fracb{E_{\rm SN}^3E_0^2}{M_{\rm ej}^5}^\frac{1}{10}t_0 = 
    1.5\times10^{15}E_{\rm SN,51}^\frac{3}{10}M_3^{-\frac{1}{2}}f^{-1}B_{14}^{-2}P_{0,-2}^{8/5}~{\rm cm}
\end{equation}
where $R_0 < v_t t_0$.
\section{Adiabiatic and Radiative Energy Losses}
\label{sec:appC}
\begin{figure}
    \centering
    \includegraphics{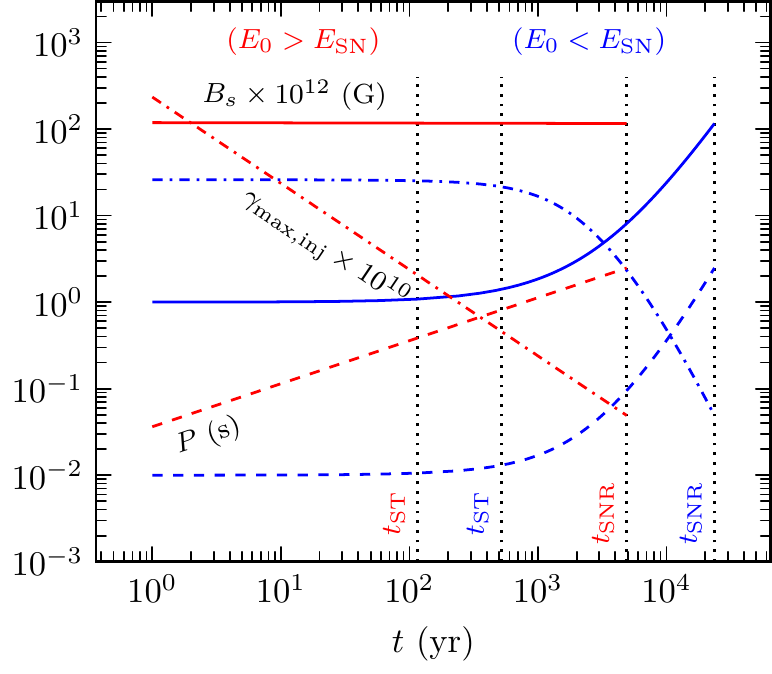}
    \caption{Evolution of the surface magnetic field $B_s$ (solid), spin period $P$ (dashed), 
    and maximum Lorentz factor of injected electrons $g_{\rm max}$ (dot-dashed).
    The \textit{red} lines correspond 
    to the case where $E_0 > E_{\rm SN}$ with initial $P_0 = 1~{\rm ms}$, constant surface 
    field $B_s = B_0 = 1.16\times10^{14}~{\rm G}$ as inferred from the measured $P$ and $\dot P$, 
    and system age $t_{\rm SNR} = \tau_c = 4.9~{\rm kyr}$. The \textit{blue} lines are for the 
    case where $E_0 < E_{\rm SN}$ with initial $P_0 = 10~{\rm ms}$, initial surface field 
    $B_0 = 10^{12}~{\rm G}$ which then grows to the current surface field over the system 
    age $t_{\rm SNR} = 23.6~{\rm kyr}$ (see Eq. \ref{eq:tSNR}). The dotted lines show the 
    Sedov-Taylor times $t_{\rm ST}$ and system ages for the two cases. Other assumed parameters are: 
    $n = 3$, $n_{\rm ext} = 1~{\rm cm}^{-3}$, $M_{\rm ej} = 3M_\odot$, $E_{\rm SN} = 10^{51}~{\rm erg}$, $f=1$.}
    \label{fig:gmax_plot}
\end{figure}
Electrons injected at the termination shock $R_{\rm TS}$ with energy $\gamma_{e,i}m_ec^2$ at time $t_i$ lose energy to adiabatic expansion 
of the nebular volume and synchrotron radiation. Their energy evolves in time according to 
Equation~(\ref{eq:gamma_t}), which can be written in
terms of the adiabatic-expansion or radius-doubling time, $t_{\rm ad}(t)=t/a$, and the synchrotron cooling time, 
$t_{\rm syn}(\gamma_e,t) = t_{c,0}(t)/\gamma_e$ where $t_{c,0}(t)=1/[bB^2(t)]=6\pi m_ec/[\sigma_T B^2(t)]$,
\begin{equation}
    \frac{d\ln\gamma_e}{dt} = -\frac{1}{t_{\rm ad}(t)} - \frac{1}{t_{\rm syn}(\gamma_e,t)} = -\frac{a}{t} - \frac{\gamma_e}{t_{c,0}(t)}\ .
    \label{eq:dgamma_e_dt}
\end{equation}
After being compressed by the reverse shock and after establishing pressure equilibrium with the SNR, the MWN re-expands slowly as a 
power law in time given by Equation~(\ref{eq:Rt_ad}), with $a=3/2(5-k)$, i.e. 
$R(t>t_{\rm ST})/R_f = (t/t_{\rm ST})^a$.
Therefore, the magnetic field in the nebula for a constant $\sigma$ scales as its energy density, 
$B^2 \propto E/R^3 \propto R^{-4}\propto t^{-4a}$ so that $B^2(t)=B_{\rm ST}^2(t/t_{\rm ST})^{-4a}$ where $B_{\rm ST}=B(t_{\rm ST})$,
and $t_{c,0}(t)=t_{c,0}(t_{\rm ST})(t/t_{\rm ST})^{4a}$.
For injection times $t_i>t_{\rm ST}$ Equation~(\ref{eq:dgamma_e_dt}) can be solved analytically by switching variables to $y(t) = \gamma_e(t)^{-1}$,
\begin{equation}
    \frac{dy}{dt} = \frac{a}{t}y + \frac{1}{t_{c,0}(t)}\ .
\end{equation}
The solution to the above equation for an initial value $y(t_i)=y_i = 1/\gamma_i$ is
\begin{equation}
    y(t_i,t,y_i) = t^a\left[y_i t_i^{-a}+I(t_i,t)\right]\ ,
\end{equation}
where $\gamma_i$ is the initial Lorentz factor of the electron injected into the nebula at $R_{\rm TS}$ at 
injection time $t_i$, and
\begin{equation}
    I(t_i,t) = \int_{t_i}^{t}\frac{t^{\prime\,-a}dt'}{t_{c,0}(t')} = \frac{t_{\rm ST}^{4a}}{t_{c,0}(t_{\rm ST})}\fracb{t_i^{1-5a}-t^{1-5a}}{5a-1}\ ,\quad\quad
    t^{a}I(t_i,t) = \frac{t}{t_{c,0}(t)}\left[\frac{\fracb{t}{t_i}^{5a-1}-1}{5a-1}\right]\ .
\end{equation}
This solution can be expressed back in terms of $\gamma_e$,
\begin{equation}
    \gamma_e(t_i,t,\gamma_i) = \gamma_i\fracb{t}{t_i}^{-a}\left[1+\frac{t_i}{t_{\rm syn}(\gamma_i,t_i)}\fracb{1-\fracb{t}{t_i}^{1-5a}}{5a-1}\right]^{-1}
    = \gamma_i\left[\fracb{t}{t_i}^{a}+\frac{t}{t_{\rm syn}(\gamma_i,t)}\fracb{\fracb{t}{t_i}^{5a-1}-1}{5a-1}\right]^{-1} \ .
\end{equation}
It is of relevance here to only look at the cooling evolution of the maximum energy electrons that were injected at time $t_i>t_{\rm ST}$. 
The maximum injection energy depends on the strength of the surface magnetic field and the spin period at time of injection, 
as given in Equation~(\ref{eq:gmax}),
\begin{equation}
    \gamma_{\rm max} = 2.6\times10^9B_{14}P_{\rm sec}^{-2}
\end{equation}

In Fig.(\ref{fig:gmax_plot}), we show the evolution of the surface magnetic field, 
spin period, and maximum Lorentz factor of electrons injected into the nebula after 
the crushing and during the re-expansion phase for $t>t_{\rm ST}$. We show the 
spin-down evolution for two cases. (a) When $E_0 > E_{\rm SN}$, the magnetar is 
assumed to have been born rotating very fast with initial spin period 
$P_0 = 1~{\rm ms}$ and initial surface field $B_0 = B_s = 1.16\times10^{14}~{\rm G}$ 
(assuming $f = 1$). From Fig.(\ref{fig:alpha-tB-plot}), it is clear that any field decay 
scenario predicts $t_{\rm SNR} < \tau_c$, which implies that the object has to be very 
young. Moreover, physical field decay timescales $t_B \gtrsim 10^3$ are only obtained when 
$t_{\rm SNR}\lesssim 400~{\rm yr}$. However, this age estimate is in strong contradiction with 
the age estimate derived in Eq. \ref{eq:tSNR} based on the current size of the radio SNR. 
Therefore, one is forced to consider a constant surface field scenario (with the alternative 
being the field growth case) with $t_{\rm SNR} = \tau_c = 4.9~{\rm kyr}$. (b) When 
$E_0 < E_{\rm SN}$, we consider the field growth scenario where the magnetar has 
initial spin period $P_0 = 10~{\rm ms}$ and surface field $B_0 = 10^{12}~{\rm G}$. 
Over the age of the system $t_{\rm SNR} = 23.6~{\rm kyr}$ (see Eq. (\ref{eq:tSNR}) for 
the choice of age), the surface field and spin period of the magnetar grow to the 
currently measured values.


\bsp	
\label{lastpage}
\end{document}